\documentclass[journal]{IEEEtran}

\usepackage{graphicx}
\usepackage{verbatim}

\usepackage{amsmath}
\usepackage{amsfonts}
\usepackage{amssymb}
\usepackage{mathrsfs} 

\usepackage{multirow}
\usepackage{multirow}
\usepackage{tabularx, booktabs}
\newcolumntype{Y}{>{\centering\arraybackslash}X}
\usepackage{array}

\newtheorem{theorem}{Theorem}
\newtheorem{lemma}[theorem]{Lemma}
\newtheorem{remark}{Remark}

\def\R{R}
\def\I{I}
\def\J{J}
\def\U{U}
\def\V{V}
\def\S{S}
\def\B{B}
\def\T{T}
\def\L{L}
\def\n{n}
\def\C{C}
\def\X_l{\mathbf{X}_\ell}
\def\Y_l{\mathbf{Y}_\ell}
\def\XL{\mathbf{X}^\L}
\def\YL{\mathbf{Y}^\L}
\def\k{\mathsf{k}}
\def\s{\mathsf{s}}
\def\d{\mathsf{d}}

\def\ceta{\mathsf{c}}
\def\K{\mathsf{K}}
\def\A{\mathsf{A}}
\def\pow{\alpha}
\def\H{\mathsf{H}}

\begin{document}

\title{On Single-Antenna Rayleigh Block-Fading Channels at Finite Blocklength}

\author{
\IEEEauthorblockN{Alejandro Lancho,~\IEEEmembership{Member,~IEEE,} Tobias Koch,~\IEEEmembership{Senior Member,~IEEE,}  and Giuseppe~Durisi,~\IEEEmembership{Senior~Member,~IEEE} }

\thanks{A.~Lancho and T.~Koch have received funding from the Spanish Ministerio de Economia y Competitividad under Grants TEC2013-41718-R and TEC2016-78434-C3-3-R (AEI/FEDER, EU), from the European Research Council (ERC) under the European Union's Horizon 2020 research and innovation programme (grant agreement number 714161), and from the Comunidad de Madrid under Grant S2103/ICE-2845. A.~Lancho has also received funding from an FPU fellowship from the Spanish Ministerio de Educaci\'on, Cultura y Deporte under Grant FPU14/01274. T.~Koch has also received funding from the Spanish Ministerio de Economia y Competitividad under Grant RYC-2014-16332 and from the 7th European Union Framework Programme under Grant 333680. G.~Durisi has been supported by the Swedish Research Council under Grants 2012-4571 and 2016-03293. The material in this paper was presented in part at the 2017 IEEE International Symposium on Information Theory (ISIT), Aachen, Germany, June 2017, and at the 52nd Annual Conference on Information Sciences and Systems (CISS), Princeton, NJ, USA, March 2018.
}
\thanks{A.~Lancho was with the Signal Theory and Communications Department, Universidad Carlos III de Madrid, 28911, Legan\'es, Spain and also with the Gregorio Mara\~n\'on Health Research Institute, 28007, Madrid, Spain. He is now with the Department of Electrical Engineering, Chalmers University of Technology, Gothenburg, 41296, Sweden (e-mail: \mbox{lancho@ieee.org}).}
\thanks{T.~Koch is with the Signal Theory and Communications Department, Universidad Carlos III de Madrid, 28911, Legan\'es, Spain and also with the Gregorio Mara\~n\'on Health Research Institute, 28007, Madrid, Spain (e-mail: \mbox{koch@tsc.uc3m.es}).}
\thanks{G.~Durisi is with the Department of Electrical Engineering, Chalmers University of Technology, Gothenburg 41296, Sweden (e-mail: \mbox{durisi@chalmers.se}).}
\thanks{Copyright (c) 2019 IEEE. Personal use of this material is permitted.  However, permission to use this material for any other purposes must be obtained from the IEEE by sending a request to \mbox{pubs-permissions@ieee.org}.}}

\maketitle

\begin{abstract}
This paper concerns the maximum coding rate at which data can be transmitted over a noncoherent, single-antenna, Rayleigh block-fading channel using an error-correcting code of a given blocklength with a block-error probability not exceeding a given value. A high-SNR normal approximation of the maximum coding rate is presented that becomes accurate as the signal-to-noise ratio (SNR) and the number of coherence intervals $L$ over which we code tend to infinity. Numerical analyses suggest that the approximation is accurate at SNR values above $15$\,dB and when the number of coherence intervals is $10$ or more.
\end{abstract}


\section{Introduction}

\IEEEPARstart{T}{here} exists an increasing interest in the problem of transmitting short packets in wireless communications. For example, the vast majority of wireless connections in the next generations of cellular systems will most likely be originated by autonomous machines and devices, which predominantly exchange short packets. It is also expected that enhanced mobile-broadband services will be complemented by new services that target systems requiring reliable real-time communication with stringent requirements on latency and reliability. For more details see \cite{Durisi_Koch_Popovski_2015} and references therein. While in the absence of latency constraints, \emph{capacity} and \emph{outage capacity} provide accurate benchmarks for the throughput achievable in wireless communication systems, for low-latency wireless communications a more refined analysis of the maximum coding rate as a function of the blocklength is needed. Such an analysis is provided in this paper.

Let $\R^*(\n,\epsilon)$ denote the maximum coding rate at which data can be transmitted using an error-correcting code of a determined length $\n$ with a block-error probability no larger than $\epsilon$. Building upon Dobrushin's and Strassen's asymptotic results,  Hayashi \cite{Hayashi_2009} and Polyanskiy, Poor and Verd\'{u} \cite{Polyanskiy_Poor_Verdu} showed that for various channels with a positive capacity $\C$, the maximum coding rate can be tightly approximated by
\begin{equation}\label{eq:rate_ppv}
	\R^*(\n,\epsilon)=\C-\sqrt{\frac{\V}{\n}}Q^{-1}(\epsilon)+\mathcal{O}\biggl(\frac{\log\n}{\n}\biggr)
\end{equation}
where $\V$ denotes the channel dispersion \cite[Def.~1]{Polyanskiy_Poor_Verdu}, $Q^{-1}(\epsilon)$ denotes the inverse of the Gaussian $Q$-function
\begin{equation}
Q(x) \triangleq \int_x^{\infty} \frac{1}{\sqrt{2\pi}} e^{-\frac{t^2}{2}} \mathrm{d} t
\end{equation}
and $\mathcal{O}(\log\n/\n)$ comprises terms that decay no slower than $\log\n/\n$. The approximation that follows from \eqref{eq:rate_ppv} by ignoring the $\mathcal{O}(\log\n/\n)$ term is sometimes referred to as \emph{normal approximation}.

The work by Polyanskiy \emph{et al.} \cite{Polyanskiy_Poor_Verdu} has been generalized to some wireless communication channels. For instance, the channel dispersion of coherent fading channels---where the receiver has perfect knowledge of the realizations of the fading coefficients---was studied by Polyanskiy and Verd\'u for the single-antenna case \cite{Polyanskiy_2011}, and by Collins and Polyanskiy for the multiple-input single-output (MISO) \cite{Collins_Polyanskiy_2014} and the multiple-input multiple-output (MIMO) case \cite{Collins_Polyanskiy_MIMO_coh_BFC,Collins_Polyanskiy_2018}. The channel dispersion of single-antenna quasistatic fading channels when both transmitter and receiver have perfect knowledge of the realization of the fading coefficients and the transmitter satisfies a long-term power constraint was obtained by Yang \emph{et al.} \cite{Yang_LongTerm}. In the noncoherent setting---where neither the transmitter nor the receiver have \emph{a priori} knowledge of the realizations of the fading coefficients---the channel dispersion is only known in the quasistatic case, where it is zero \cite{Yang_2014,MolavianJazi_2013}. Upper and lower bounds on the second-order coding rate of quasistatic MIMO Rayleigh-fading channels have further been reported in \cite{Hoydis_2015} for the asymptotically-ergodic setup where the number of antennas grows linearly with the blocklength. For noncoherent Rayleigh block-fading channels, nonasymptotic bounds on the maximum coding rate were presented by Yang \emph{et al.} for the single-antenna case \cite{finte_block_length_2012} and by \"Ostman \emph{et al.} for the MIMO case \cite{Ostman_2014, Short_Packets_Durisi_Koch_TCOM_2015}. For further references see \cite{Durisi_Koch_Popovski_2015}.

In a nutshell, in the noncoherent setting the channel dispersion is only known in the quasistatic case. For general block-fading channels, the maximum coding rate needs to be assessed by means of nonasymptotic bounds, whose evaluation is often computationally demanding. Obtaining an expression for the channel dispersion of noncoherent block-fading channels is difficult because for such channels the capacity-achieving input distribution is in general unknown. Thus, the standard approach of obtaining expressions of the form \eqref{eq:rate_ppv}, which consists of first evaluating nonasymptotic upper and lower bounds on $\R^*(\n,\epsilon)$ for the capacity-achieving input and output distributions and then analyzing these bounds in the limit as $\n\to\infty$, cannot be followed. However, the behavior of capacity at high signal-to-noise ratio (SNR) is well understood. Indeed, it was demonstrated that an input distribution called \emph{unitary space-time modulation} (USTM) achieves a lower bound on the capacity that is asymptotically tight \cite{Hochwald_2000,Zheng_2002, Yang_2013}. Thus, a characterization of the channel dispersion at high SNR may be feasible.

In this paper, we present an expression similar to \eqref{eq:rate_ppv} of the maximum coding rate $\R^*(\L,\T,\epsilon,\rho)$ achievable over noncoherent, single-antenna, Rayleigh block-fading channels using error-correcting codes that span $\L$ coherence intervals of length $\T$, have a block-error probability no larger than $\epsilon$, and satisfy the power constraint $\rho$. By replacing the capacity and channel dispersion by asymptotically tight approximations, we obtain a high-SNR normal approximation of $\R^*(\L,\T,\epsilon,\rho)$. The obtained normal approximation is useful in two ways. On the one hand, it complements the nonasymptotic bounds provided in \cite{finte_block_length_2012, Ostman_2014, Short_Packets_Durisi_Koch_TCOM_2015}. On the other hand, it allows for a mathematical analysis of $\R^*(\L,\T,\epsilon,\rho)$.

The rest of this paper is organized as follows. Section~\ref{sec:notation} introduces the notation used in this paper. Section~\ref{sec:system} presents the system model. Section~\ref{sec:preliminaries} introduces the most important quantities used in this paper. Section~\ref{sec:results} is divided into three subsections. The first subsection presents the main result of the paper: a high-SNR normal approximation of $\R^*(\L,\T,\epsilon,\rho)$. The second subsection discusses the accuracy of the normal approximation by means of numerical evaluations. The third subsection discusses some applications of our normal approximation. Section~\ref{sec:proof} contains the proof of the main result. Section~\ref{sec:conclusion} concludes the paper with a discussion of the presented results. Some of the proofs are deferred to the appendices.

\section{Notation}
\label{sec:notation}
We denote scalar random variables by upper case letters such as $X$, and their realizations by lower case letters such as $x$. Likewise, we use boldface upper case letters to denote random vectors, \emph{i.e.}, $\mathbf{X}$, and we use boldface lower case letters such as $\mathbf{x}$ to denote their realizations. We use upper case letters with the standard font to denote distributions, and lower case letters with the standard font to denote probability density functions (pdfs). The serif font is used to denote constants independent of $L$ and $\rho$, except $\mathsf{E}[\cdot]$, which denotes the expectation operator, and $\mathsf{P}[\cdot]$, which is used for probabilities.  
The superscripts $(\cdot)^\text{T}$ and $(\cdot)^\H$ denote transposition and Hermitian transposition, respectively. The complement of a set $\mathscr{A}$ is denoted as $\mathscr{A}^\mathsf{c}$. We use ``$\stackrel{\mathscr{L}}{=}$'' to denote equality in distribution.

We denote by $\log(\cdot)$ the natural logarithm,  by $\text{I}\{\cdot\}$ the indicator function, by $\Gamma(\cdot)$ the Gamma function \cite[Sec. 6.1.1]{Abramowitz}, by $\tilde{\gamma}(\cdot, \cdot)$ the regularized lower incomplete gamma function \cite[Sec. 6.5]{Abramowitz}, by $\psi(\cdot)$ the digamma function \cite[Sec. 6.3.2]{Abramowitz}, by $_2F_1(\cdot,\cdot;\cdot;\cdot)$ the Gauss hypergeometric function \cite[Sec. 9.1]{table_integrals}, by $\text{E}_1(\cdot)$ the exponential integral function \cite[Sec.~5.1.1]{Abramowitz}, and by $\zeta(z,q)$ Riemann's zeta function \cite[Sec.~9.511]{table_integrals}. The gamma distribution with parameters $z$ and $q$ is denoted by $\text{Gamma}(z,q)$. We denote by $\gamma\approx 0.5772$ Euler's constant. 

Double limits such as
\begin{equation}
\lim\limits_{\begin{subarray}{c} \L\to\infty,\\ \rho\to\infty\end{subarray}} f(\L,\rho) = \K
\end{equation}
indicate that for every $\epsilon>0$ there exists a pair $(\L_0,\rho_0)$ independent of $(L,\rho)$ such that for every $\L\geq\L_0$ and $\rho\geq\rho_0$ we have $|f(\L,\rho)-\K| \leq \epsilon$. We denote by $\varliminf$ the \emph{limit inferior} and by $\varlimsup$ the \emph{limit superior}. \emph{Double limit inferiors} and \emph{double limit superiors} are defined accordingly using the above definition of a double limit. For example,
\begin{equation}
\varliminf_{\begin{subarray}{c} \L\to\infty,\\ \rho\to\infty\end{subarray}}f(\L,\rho) = \lim\limits_{\begin{subarray}{c} L_0\to\infty,\\ \rho_0\to\infty\end{subarray}} \inf_{\L\geq L_0}\inf_{\rho\geq\rho_0} f(\L,\rho).
\end{equation}
\section{System Model}
\label{sec:system}
We consider a single-antenna Rayleigh block-fading channel with coherence interval $\T>2$. For this channel model, the input-output relation within the $\ell$-th coherence interval is given by
\begin{equation}
		\label{ch_model}
			\Y_l = H_{\ell}\X_l + \mathbf{W}_{\ell}
	\end{equation}
where $\X_l$ and $\Y_l$ are $\T$-dimensional, complex-valued, random vectors containing the input and output signals, respectively; $\mathbf{W}_{\ell}$ is the additive noise, which is assumed to be a random vector with independent and identically distributed (i.i.d.), zero-mean, unit-variance, circularly-symmetric, complex Gaussian entries; and $H_{\ell}$ is Rayleigh fading, i.e., it is a zero-mean, unit-variance, circularly-symmetric, complex Gaussian random variable. We assume that $H_{\ell}$ and $\mathbf{W}_{\ell}$ are independent and take on independent realizations over successive coherence intervals. We further assume that the joint law of $(H_{\ell},\mathbf{W}_{\ell})$ does not depend on the channel inputs. We consider a noncoherent setting where transmitter and receiver are aware of the distribution of $H_{\ell}$ but not of its realization.

We next introduce the notion of a channel code. For simplicity, we shall restrict ourselves to codes whose blocklength $\n$ satisfies $\n=\L\T$, where $\L$ denotes the number of coherence intervals of length $\T$ needed to transmit the whole code. An $(M,\L,\T,\epsilon,\rho)$ code for the channel \eqref{ch_model} consists of the following:
\begin{enumerate}
	\item An encoder $f$: $\{1,\dots,M\}\rightarrow\mathbb{C}^{\L\T}$ that maps the message $A$, taking value in $\{1,\dots,M\}$, to a codeword \mbox{$\XL=[\mathbf{X}_1,\dots,\mathbf{X}_\L]$}. The codewords are assumed to satisfy the power constraint\footnote{In the information theory literature, it is more common to impose a power constraint per codeword $\XL$. However, practical systems typically require a per-coherence-interval constraint. Note that, in contrast to \cite{Short_Packets_Durisi_Koch_TCOM_2015}, where the power constraint \eqref{power_const} is assumed to hold with equality, here we consider the more general case where the power constraint may also be satisfied with strict inequality.}
	\begin{equation}
		\label{power_const}
		\|\X_l\|^2\leq \T\rho, \quad \ell=1,\dots,\L.
	\end{equation}
	Since the variance of $H_{\ell}$ and of the entries of $\mathbf{W}_{\ell}$ are normalized to one, $\rho$ in \eqref{power_const} can be interpreted as the average SNR at the receiver.
	\item A decoder $g$: $\mathbb{C}^{\L\T}\rightarrow\{1,\dots,M\}$ satisfying the maximum error probability constraint
	\begin{equation}\label{eq:max_err_prob}
		\max_{1\leq a\leq M}{\text{P}\bigl[g(\YL)\neq A\big|A=a\bigr]}\leq\epsilon
	\end{equation}
	where $\YL=[\mathbf{Y}_1,\dots,\mathbf{Y}_\L]$ is the channel output induced by the transmitted codeword $\XL  = f(a)$ according to \eqref{ch_model}.	
\end{enumerate}
The \emph{maximum coding rate} $\R^*(\L,\T,\epsilon,\rho)$ is defined as the largest rate $\log M/(\L\T)$ (in nats per channel use) for which there exists an $(M,\L,\T,\epsilon,\rho)$ code, i.e.,
\begin{equation}
	\R^*(\L,\T,\epsilon,\rho)\triangleq\sup\biggl\{\frac{\log M}{\L\T}\,:\,\exists(M,\L,\T,\epsilon,\rho) \text{ code}\biggr\}.
\end{equation}

\section{Preliminaries}
\label{sec:preliminaries}
We next introduce some preliminary results that will be helpful in the remainder of the paper.

Conditioned on $\XL=\mathbf{x}^\L$, the output vector $\YL$ is blockwise i.i.d.\ Gaussian. Thus, the conditional pdf of $\Y_l$ given $\X_l=\mathbf{x}$ is independent of $\ell$ and satisfies
\begin{equation}
		 \label{cond_pdf}
	\text{p}_{\mathbf{Y}|\mathbf{X}}(\mathbf{y}|\mathbf{x})={\frac{1}{\pi^\T(1+\|\mathbf{x}\|^2)}\exp\biggl\{-\|\mathbf{y}\|^2+\frac{|\mathbf{y}^\H\mathbf{x}|^2}{1+\|\mathbf{x}\|^2}\biggr\}}
\end{equation}
for $\mathbf{x},\mathbf{y}\in\mathbb{C}^\T$. Here and throughout the paper, we omit the subscript $\ell$ when immaterial. We shall refer to the distribution $\text{P}_{\XL}^{(\text{U})}$, according to which $\XL=\sqrt{\T\rho}\mathbf{U}^\L$ (where $\mathbf{U}^\L=[\mathbf{U}_1,\dots,\mathbf{U}_\L]$ and $\mathbf{U}_1,\dots,\mathbf{U}_\L$ are i.i.d. and uniformly distributed on the unit sphere in $\mathbb{C}^\T$), as USTM \cite{Hochwald_2000}. This distribution is relevant because it gives rise to a lower bound on capacity that is asymptotically tight at high SNR \cite{Zheng_2002, Yang_2013}. In fact, it can be shown that this lower bound accurately approximates capacity already for intermediate SNR values. For example, \cite[Fig.~1]{finte_block_length_2012} illustrates that the lower bound is indistinguishable from the upper bound on capacity given in \cite[Eq. (17)]{finte_block_length_2012} for $\rho\geq 10$\,dB. 

The outputs $\YL$ induced by the USTM input distribution have the pdf
\begin{equation}\label{output_pdf}
\text{q}_{\mathbf{Y}^\L}^{(\text{U})}(\mathbf{y}^\L) = \prod_{\ell=1}^\L \text{q}_{\mathbf{Y}}^{(\text{U})}(\mathbf{y}_{\ell}), \quad \mathbf{y}^L=[\mathbf{y}_1,\ldots,\mathbf{y}_L]\in\mathbb{C}^{\L\T}
\end{equation}
where \cite[Eq. (18)]{finte_block_length_2012}
\begin{IEEEeqnarray}{lCl}
	\label{marginal_output_pdf} 
	\text{q}_{\mathbf{Y}}^{(\text{U})}(\mathbf{y}) &=& \frac{e^{-\|\mathbf{y}\|^2/(1+\T\rho)}\|\mathbf{y}\|^{2(1-\T)}\Gamma(\T)}{\pi^\T(1+\T\rho)}\nonumber\\ &&{}\times\tilde{\gamma}\biggl(\T-1,\frac{\T\rho\|\mathbf{y}\|^2}{1+\T\rho}\biggr)\biggl(1+\frac{1}{\T\rho}\biggr)^{\T-1}
\end{IEEEeqnarray}
for $\mathbf{y}\in\mathbb{C}^\T$. Observe that the expression of $\text{q}_{\mathbf{Y}}^{(\text{U})}$ contains a regularized lower incomplete gamma function. The following lemma presents an upper and lower bound on the logarithm of this function, which we shall use throughout the paper.
\begin{lemma}\label{lm:inc_gamma}
The logarithm of the regularized lower incomplete gamma function $\tilde{\gamma}(\T-1,x)$, $x>0$ can be bounded as
 	\begin{equation}\label{eq:part1_inc_gamma}
 		0 \leq \log\frac{1}{\tilde{\gamma}(\T-1,x)}
 		 \leq (\T-1) \log\Biggl(1+\frac{\Gamma(\T)^{\frac{1}{\T-1}}}{x}\Biggr).
 	\end{equation}	 
\end{lemma}
\begin{IEEEproof}
See Appendix~\ref{app:inc_gamma}.
\end{IEEEproof}

Throughout this paper, we shall denote by $\YL$ a blockwise i.i.d. random vector whose conditional pdf, conditioned on $\XL = \mathbf{x}^{\L}$, is given by $\prod_{\ell=1}^\L \text{p}_{\mathbf{Y}|\mathbf{X}}(\mathbf{y}_\ell|\mathbf{x}_\ell)$ with $\text{p}_{\mathbf{Y}|\mathbf{X}}$ as in \eqref{cond_pdf}. We shall denote by $\tilde{\mathbf{Y}}^{\L}$ a blockwise i.i.d. random vector that is independent of $\XL$ and has pdf $\text{q}_{\mathbf{Y}^\L}^{(\text{U})}$.

Conditioned on $\|\X_l\|^2=\T\pow_{\ell}$, the random variables $|\Y_l^\H\X_l|^2$ and $\|\Y_l\|^2$ can be written as
\begin{IEEEeqnarray}{rCl}\label{eq:distributions_cond_x}
	|\Y_l^\H\X_l|^2&\stackrel{\mathscr{L}}{=}&|H_\ell^*\T\pow_{\ell}+\boldsymbol{W}_\ell^*(1)\sqrt{\T\pow_{\ell}}|^2\IEEEnonumber\\
	&\stackrel{\mathscr{L}}{=}&\T\pow_{\ell}(1+\T\pow_{\ell})Z_{1,\ell}\label{eq:norm_input_output_final}\\
	\|\Y_l\|^2&\stackrel{\mathscr{L}}{=}&\|H_\ell\sqrt{\T\pow}\mathbf{e}_1+\boldsymbol{W}_\ell\|^2\IEEEnonumber\\
	&\stackrel{\mathscr{L}}{=}&(1+\T\pow_{\ell})Z_{1,\ell}+Z_{2,\ell}\label{eq:norm_output}
\end{IEEEeqnarray}
where $\mathbf{e}_1$ is the length-$\T$ unitary vector $ [1,0,\dots,0]^{\text{T}}$, $\{Z_{1,\ell},\ell\in\mathbb{Z}\}$ is a sequence of i.i.d. $\text{Gamma}(1,1)$-distributed random variables, and $\{Z_{2,\ell},\ell\in\mathbb{Z}\}$ is a sequence of i.i.d. $\text{Gamma}(\T-1,1)$-distributed random variables.

Conditioned on $\|\X_l\|^2=\T\pow_{\ell}$, the random variables $|\tilde{\mathbf{Y}}_\ell^\H\X_l|^2$ and $\|\tilde{\mathbf{Y}}_\ell\|^2$ can be written as
\begin{IEEEeqnarray}{rCl}
	|\tilde{\mathbf{Y}}_\ell^\H\X_l|^2&\stackrel{\mathscr{L}}{=}&|(H_\ell^*\sqrt{\T\rho}\mathbf{U}_\ell(1)+\boldsymbol{W}_\ell^*(1))\sqrt{\T\pow_{\ell}}|^2\IEEEyesnumber\label{eq:norm_input_output_initial_Q}\\
	\|\tilde{\mathbf{Y}}_\ell\|^2&\stackrel{\mathscr{L}}{=}&\|H_\ell\sqrt{\T\rho}\mathbf{U}_\ell+\boldsymbol{W}_\ell\|^2.\label{eq:norm_output_Q}
\end{IEEEeqnarray}
In \eqref{eq:distributions_cond_x}--\eqref{eq:norm_output_Q}, the parameter $\pow_{\ell}$ lies in the interval $[0,\rho]$ and can be thought of as the power allocated over the coherence interval $\ell$.

The \emph{information density} between the random vectors $\XL$ and $\YL$ is defined as
\begin{equation}
\label{inf_dens_original}
i(\XL;\YL)\triangleq\log\Biggl({\frac{\text{p}_{\YL|\XL}\bigl(\YL\bigm|\XL\bigr)}{\text{p}_{\YL}\bigl(\YL\bigr)}}\Biggr)
\end{equation}
where $\text{p}_{\YL}$ is the output pdf induced by the input distribution.\footnote{The existence of the conditional pdf $\text{p}_{\YL|\XL}$ implies that the output pdf $\text{p}_{\YL}$ exists for every input distribution.} When the input distribution is USTM, the information density $i(\XL;\YL)$ can be expressed as
\begin{equation}
	i(\XL;\YL) = \sum_{\ell=1}^{\L}{i_{\ell}(\T,\rho)}
\end{equation}
where
\begin{IEEEeqnarray}{lCl}
	i_{\ell}(\T,\rho)& \triangleq &(\T-1)\log(\T\rho)-\log\Gamma(\T)-\frac{\T\rho Z_{2,\ell}}{1+\T\rho}\IEEEnonumber*\\
	&&{}+(\T-1)\log\biggl(\frac{(1+\T\rho)Z_{1,\ell}+Z_{2,\ell}}{1+\T\rho}\biggr)\\
	&&{}-\log\tilde{\gamma}\biggl(\T-1,\frac{\T\rho((1+\T\rho)Z_{1,\ell}+Z_{2,\ell})}{1+\T\rho}\biggr).\IEEEeqnarraynumspace\IEEEyesnumber\label{eq:i_def_2}
\end{IEEEeqnarray}
Using the left-most inequality in Lemma \ref{lm:inc_gamma}, we can lower-bound \eqref{eq:i_def_2} by
\begin{IEEEeqnarray}{lCl}
	\underline{i}_\ell(\T,\rho)&\triangleq&(\T-1)\log(\T\rho)-\log\Gamma(\T)-\frac{\T\rho Z_{2,\ell}}{1+\T\rho}\nonumber\\
	&&{}+(\T-1)\log\biggl(\frac{(1+\T\rho)Z_{1,\ell}+Z_{2,\ell}}{1+\T\rho}\biggr).\label{eq:i_lower_def}
\end{IEEEeqnarray}

The expected value of \eqref{eq:i_def_2}, denoted by $\I(\T,\rho)$, is given by
\begin{IEEEeqnarray}{lCl}
	\I(\T,\rho) & \triangleq & \mathsf{E}\big[i_\ell(\T,\rho)] \IEEEnonumber*\\
	& = & (\T-1)\log(\T\rho) -\log\Gamma(\T)-\frac{(\T-1)\T\rho}{1+\T\rho}\\
	&&{}(\T-1)\mathsf{E}\biggl[\log\biggl(\frac{(1+\T\rho)Z_{1}+Z_{2}}{1+\T\rho}\biggr)\biggr] \nonumber\\
	&&{}+\mathsf{E}\biggl[\log\tilde{\gamma}\biggl(\T-1,\frac{\T\rho((1+\T\rho)Z_{1}+Z_{2})}{1+\T\rho}\biggr)\biggr].\nonumber\\\IEEEyesnumber\label{eq:I_def}
\end{IEEEeqnarray}
Likewise, the expected value of \eqref{eq:i_lower_def}, denoted by $\underline{\I}(\T,\rho)$, can be computed as
\begin{IEEEeqnarray}{lCl}
	\underline{\I}(\T,\rho) & \triangleq & \mathsf{E}\big[\underline{i}_\ell(\T,\rho)] \IEEEnonumber*\\
	& = & (\T-1)\log(\T\rho) -\log\Gamma(\T)-\frac{(\T-1)\T\rho}{1+\T\rho} \\
		&&{}-(\T-1)\log(1+\T\rho)\\
		&&{}+(\T-1)\mathsf{E}\bigl[\log((1+\T\rho)Z_1+Z_2)\bigr]\IEEEnonumber*\\
	&=&{}(\T-1)\log(\T\rho)-\log\Gamma(\T) \\
	&&{}-(\T-1)\biggl[\log(1+\T\rho)+ \frac{\T\rho}{1+\T\rho} - \psi(\T-1)\biggr]\nonumber\\
	& & {} +{_2F_1\biggl(1,\T-1;\T;\frac{\T\rho}{1+\T\rho}\biggr)}\IEEEeqnarraynumspace\IEEEyesnumber\label{eq:sec:first_first_moment_closed}
\end{IEEEeqnarray}
where the expected value in the fourth line has been solved using \cite[Sec. 4.337-1]{table_integrals} to integrate with respect to $Z_1$ and \cite[Sec. 4.352-1]{table_integrals}, \cite[Sec. 3.381-4]{table_integrals}, and \cite[Sec. 4.2.20]{table_Ei} to integrate with respect to $Z_2$. Clearly, 
\begin{equation}\label{eq:low_bound_I}
\I(\T,\rho) \geq \underline{\I}(\T,\rho).
\end{equation}

We define the \emph{mismatched information density}\footnote{We use the word ``mismatched'' to indicate that the output distribution $\text{q}_{\YL}^{(\text{U})}$ in the denominator in \eqref{inf_dens_def} is not the one induced by the input distribution and the channel.} between the random vectors $\XL$ and $\YL$ as
\begin{equation}
	\label{inf_dens_def}
	j(\XL;\YL)\triangleq\log\Biggl({\frac{\text{p}_{\YL|\XL}\bigl(\YL\bigm|\XL\bigr)}{\text{q}_{\YL}^{(\text{U})}\bigl(\YL\bigr)}}\Biggr).
\end{equation}
Using this definition together with \eqref{cond_pdf}, \eqref{output_pdf}, and \eqref{marginal_output_pdf}, the mismatched information density $j(\XL;\YL)$ can be written as
\begin{equation}
	\label{inf_dens_def_2}
	j(\XL;\YL)=\sum_{\ell=1}^\L{j_\ell(\X_l;\Y_l)}
\end{equation}
where
\begin{IEEEeqnarray}{lCl}
	j_\ell(\X_l;\Y_l) & \triangleq & \log\biggl(\frac{1+\T\rho}{\Gamma(\T)}\biggr)+\frac{|\boldsymbol{Y}_\ell^\H\X_l|^2}{1+\|\X_l\|^2}-\frac{\T\rho\|\Y_l\|^2}{1+\T\rho}\nonumber\\
	&&{}+(\T-1)\log\biggl(\frac{\T\rho\|\Y_l\|^2}{1+\T\rho}\biggr)
	-\log(1+\|\X_l\|^2)\nonumber\\
	&&{}-\log\tilde{\gamma}\biggl(\T-1,\frac{\T\rho\|\Y_l\|^2}{1+\T\rho}\biggr).\IEEEyesnumber\label{eq:inf_density_def}
\end{IEEEeqnarray}
By \eqref{eq:norm_input_output_final} and \eqref{eq:norm_output}, $j(\X_l;\Y_l)$ depends on $\X_l$ only via its magnitude. We can thus express $j(\X_l;\Y_l)$, conditioned on $\|\X_l\|^2=\T\pow_{\ell}$, as
\begin{IEEEeqnarray}{lCl}
	j_{\ell}(\T,\pow_{\ell})& \triangleq &(\T-1)\log(\T\rho) -\log\Gamma(\T)-\frac{(\T\rho-\T\pow_{\ell})Z_{1,\ell}}{1+\T\rho}\nonumber\\
	&&{}-\frac{\T\rho Z_{2,\ell}}{1+\T\rho}+\log\biggl(\frac{1+\T\rho}{1+\T\pow_{\ell}}\biggr)\IEEEnonumber*\\
	&&{}+(\T-1)\log\biggl(\frac{(1+\T\pow_{\ell})Z_{1,\ell}+Z_{2,\ell}}{1+\T\rho}\biggr)\\
	&&{}-\log\tilde{\gamma}\biggl(\T-1,\frac{\T\rho((1+\T\pow_{\ell})Z_{1,\ell}+Z_{2,\ell})}{1+\T\rho}\biggr).\nonumber\\\IEEEyesnumber\label{eq:j_def}
\end{IEEEeqnarray}
Note that the information densities $i(\mathbf{X}^L;\mathbf{Y}^L)$ and $j(\mathbf{X}^L;\mathbf{Y}^L)$ only differ in the output densities $\text{p}_{\textbf{Y}^L}$ and $\text{q}_{\textbf{Y}^L}^{(\text{U})}$ in the denominators of \eqref{inf_dens_original} and \eqref{inf_dens_def}, respectively. Consequently, for USTM inputs, where $\text{p}_{\textbf{Y}^L}=\text{q}_{\textbf{Y}^L}^{(\text{U})}$, we have $i_{\ell}(\T,\rho)=j_{\ell}(\T,\rho)$.

Define $\beta(\T,\rho) \triangleq \Gamma(\T)^{\frac{1}{\T-1}}\frac{1+\T\rho}{\T\rho}$, and let
\begin{IEEEeqnarray}{lCl}
	\IEEEeqnarraymulticol{3}{l}{\bar{j}_\ell(\T,\pow_{\ell})} \IEEEnonumber*\\
	\quad &\triangleq&(\T-1)\log(\T\rho) -\log\Gamma(\T)-\frac{(\T\rho-\T\pow_{\ell})Z_{1,\ell}}{1+\T\rho}\\
	&&{}-\frac{\T\rho Z_{2,\ell}}{1+\T\rho}+\log\biggl(\frac{1+\T\rho}{1+\T\pow_{\ell}}\biggr)\\
	&&{}+(\T-1)\log\biggl(\frac{(1+\T\pow_{\ell})Z_{1,\ell}+Z_{2,\ell}}{1+\T\rho}\biggr)\\
	&&{}+(\T-1)\log\biggl(1+\frac{\beta(\T,\rho)}{(1+\T\pow_{\ell})Z_{1,\ell}+Z_{2,\ell}}\biggr).\IEEEyesnumber\IEEEeqnarraynumspace\label{eq:j_upper_def}
\end{IEEEeqnarray}
By Lemma \ref{lm:inc_gamma}, we have that, with probability one,
\begin{equation}\label{eq:j_u_bound}
 j_{\ell}(\T,\pow_{\ell})\leq\bar{j}_\ell(\T,\pow_{\ell}), \quad \pow_{\ell}\in[0,\rho].
 \end{equation}
 
Let $\J(\T,\pow_{\ell})\triangleq\mathsf{E}[j_{\ell}(\T,\pow_{\ell})]$ and $\bar{\J}(\T,\pow_{\ell})\triangleq\mathsf{E}[\bar{j}_\ell(\T,\pow_{\ell})]$ denote the conditional expected values of \eqref{eq:j_def} and \eqref{eq:j_upper_def} given $\|\X_l\|^2=\T\pow_{\ell}$. The latter expected value can be evaluated as
\begin{IEEEeqnarray}{lCl}
	\bar{\J}(\T,\pow_{\ell})& = &(\T-1)\log(\T\rho) -\log\Gamma(\T)-\frac{(\T-1)\T\rho}{1+\T\rho}\IEEEnonumber*\\
		&&{}-(\T-1)\log(1+\T\rho)\\
		&&{}+(\T-1)\mathsf{E}\bigl[\log((1+\T\pow_{\ell})Z_1+Z_2+\beta(\T,\rho))\bigr]\\
		&&{}-\frac{\T\rho-\T\pow_{\ell}}{1+\T\rho}+\log\biggl(\frac{1+\T\rho}{1+\T\pow_{\ell}}\biggr).\IEEEyesnumber\label{eq:J_def}
\end{IEEEeqnarray}
Clearly,
\begin{equation}
\label{eq:tobi_JbarJ}
\J(\T,\pow_{\ell}) \leq \bar{\J}(\T,\pow_{\ell}), \quad \pow_{\ell} \in [0,\rho].
\end{equation}

It can be shown that $\bar{\J}(\cdot)$ and $\underline{\I}(\cdot)$ bound the capacity \cite{Biglieri}
\begin{equation} 
\C(T,\rho) = \sup\limits_{\text{P}_{\XL}:\,\mathsf{E}[\|\X_l\|^2] \leq \T\rho}{\frac{\mathsf{E}[i(\X_l;\Y_l)]}{\T}}.
\end{equation}
Indeed, on the one hand we have
\begin{equation} 
\label{eq:capacity_UB}
\C(T,\rho) \leq\sup\limits_{0\leq\pow\leq \rho}{\frac{ \J(\T,\pow)}{\T}}\leq\sup\limits_{0\leq\pow\leq \rho}{\frac {\bar{\J}(\T,\pow)}{\T}}
\end{equation}
where the first inequality follows from \cite[Th. 5.1]{Lapidoth_Moser_duality} and the second inequality follows from \eqref{eq:tobi_JbarJ}. On the other hand,
\begin{equation}
\label{eq:capacity_LB}
\C(T,\rho) \geq \frac{\I(\T,\rho)}{\T}\geq\frac{\underline{\I}(\T,\rho)}{\T}
\end{equation}
where the first inequality follows because USTM is a valid input distribution and the second inequality follows from \eqref{eq:low_bound_I}. Note that $\J(\T,\rho)=\I(\T,\rho)$ when the input distribution is USTM. It can be further shown that
\begin{equation}
\lim\limits_{\rho\to\infty}\biggl\{\sup\limits_{0\leq\pow\leq\rho}\bar{\J}(\T,\pow) - \underline{\I}(\T,\rho)\biggr\} = 0.
\end{equation}
Thus, USTM yields an asymptotically tight lower bound on capacity, as already mentioned before.

Let
\begin{IEEEeqnarray}{rCl}
\U(\T,\rho) & \triangleq & \mathsf{E}\Bigl[\bigl(i_{\ell}(\T,\rho)-\I(\T,\rho)\bigr)^2\Bigr] \label{eq:second_def}\\
\bar{\V}_\rho(\T,\pow) & \triangleq & \mathsf{E}\Bigl[\bigl(\bar{j}_{\ell}(\T,\pow)-\bar{\J}(\T,\pow)\bigr)^2\Bigr] \label{eq:second_def_bar}
\end{IEEEeqnarray}
where the subscript $\rho$ in $\bar{\V}_\rho(\T,\pow)$ is introduced to highlight that $\bar{\V}_\rho(\T,\pow)$ depends both on $\pow$ and $\rho$, but it is omitted when $\pow=\rho$. In Lemma~\ref{lm:first_asy} (Appendix~\ref{app:first_asymp}) and Lemma~\ref{lm:var_asy} (Appendix~\ref{app:second_asymp}), we show that $\I(\T,\rho)$, $\U(\T,\rho)$, $\bar{\J}(\T,\rho)$, and $\bar{\V}_\rho(\T,\rho)$ can be approximated as
\begin{IEEEeqnarray}{lCl}
\I(\T,\rho) & = & \underline{\I}(\T,\rho) + \K_{\I}(\T,\rho) \label{eq:underI}\\
\U(\T,\rho) & = & \tilde{\U}(\T) + \K_{\U}(\T,\rho) \label{eq:underU}\\
\bar{\J}(\T,\rho) & = & \underline{\I}(\T,\rho) + \K_{\bar{\J}}(\T,\rho) \label{eq:barJ}\\
\bar{\V}(\T,\rho) & = & \tilde{\U}(\T) + \K_{\bar{\V}}(\T,\rho) \label{eq:barV}
\end{IEEEeqnarray}
where $\K_{\xi}(\T,\rho)$, $\xi=\{\I,\U,\bar{\J},\bar{\V}\}$ are functions of $\T$ and $\rho$ that satisfy
\begin{equation}
\label{eq:former_36}
\lim_{\rho\to\infty} \K_\xi(\T,\rho) = 0, \quad \T>2.
\end{equation}
A closed form expression of $\underline{\I}(\T,\rho)$ is given in \eqref{eq:sec:first_first_moment_closed}. Moreover, $\tilde{\U}(\T)$ in \eqref{eq:underU} and \eqref{eq:barV} is defined as
\begin{equation}
\label{eq:second_second_moment_high_SNR}
\tilde{\U}(\T) \triangleq (\T-1)^2\frac{\pi^2}{6}+(\T-1).
\end{equation}
\section{Main Result}
\label{sec:results}
The main result of this paper is a high-SNR normal approximation on $\R^*(\L,\T,\epsilon,\rho)$ presented in Section~\ref{sec:high_SNR}. In Section~\ref{sec:num_results}, we assess the accuracy of this approximation by means of numerical examples. Possible applications are discussed in Section~\ref{sub:wisdom}.

\subsection{A High-SNR Normal Approximation}\label{sec:high_SNR}
\begin{theorem}\label{th:high_SNR}
Assume that $\T>2$ and $0<\epsilon<\frac{1}{2}$. Then, the maximum coding rate $R^*(\L,\T,\epsilon,\rho)$ can be expanded as
\begin{IEEEeqnarray}{lCl}\label{eq:comb_for_asymp}
	\R^*(\L,\T,\epsilon,\rho)&=&\frac{\underline{\I}(\T,\rho)}{\T}-\sqrt{\frac{\tilde{\U}(\T)+\K_{\tilde{\U}}(\T,\rho)}{\L\T^2}}Q^{-1}(\epsilon)\nonumber\\
	&&{}+\K_{\underline{\I}}(\T,\rho)+\K_{L}(L,\T,\rho)
\end{IEEEeqnarray}
where $\underline{\I}(\T,\rho)$ and $\tilde{\U}(\T)$ are defined in \eqref{eq:sec:first_first_moment_closed} and \eqref{eq:second_second_moment_high_SNR}, respectively, and $\K_{\xi}(\T,\rho)$, $\xi=\{\underline{\I},\tilde{\U}\}$ are functions of $\T$ and $\rho$ that satisfy
\begin{equation}
\lim_{\rho\to\infty}\K_\xi(\T,\rho) = 0, \quad \T>2.
\end{equation}
Similarly, $\K_{L}(L,\T,\rho)$ is a function of $T$, $L$, and $\rho$ that satisfies
\begin{equation}
\sup_{\rho\geq\rho_0}\bigl|\K_{L}(L,\T,\rho)\bigr|\leq \A\frac{\log L}{L},\quad L\geq\L_0
\end{equation}
for every $\T>2$ and some $\A$, $\L_0$, and $\rho_0$ independent of $L$ and $\rho$.
\end{theorem}
\begin{IEEEproof}
See Section~\ref{sec:Proofs}.
\end{IEEEproof}

\begin{remark}
The assumption that $0<\epsilon<1/2$ is required to ensure that $Q^{-1}(\epsilon)$ is nonnegative, which simplifies the manipulations of the channel dispersion. Treating the case $1/2<\epsilon<1$ would require a separate analysis. For the sake of compactness, we decided to omit such an analysis, since we believe that $0<\epsilon<1/2$ covers all cases of practical interest.
\end{remark}

Ignoring the $\K_{\underline{\I}}(\T,\rho)$, $\K_{\tilde{\U}}(\T,\rho)$, and $\K_{L}(L,\T,\rho)$ terms in \eqref{eq:comb_for_asymp}, we obtain the high-SNR normal approximation
	\begin{equation}
		\label{eq:asymp}
		\R^*(\L,\T,\epsilon,\rho)\approx \frac{\underline{\I}(\T,\rho)}{\T}-\sqrt{\frac{\tilde{\U}(\T)}{\L\T^2}}Q^{-1}(\epsilon).
	\end{equation}
	
The closed form expression for $\underline{\I}(\T,\rho)$ in \eqref{eq:sec:first_first_moment_closed} contains a hypergeometric function, which is difficult to analyze mathematically. We therefore present also a simplified expression that is less accurate than~\eqref{eq:sec:first_first_moment_closed} but easier to analyze. Specifically, it follows from Lemma~\ref{lm:first_asy}  (Appendix~\ref{app:first_asymp}) that
\begin{multline}\label{eq:sec:first_first_moment}
	\underline{\I}(\T,\rho) = (\T-1)\log(\T\rho)-\log\Gamma(\T)\\
	{}-(\T-1)(1+\gamma)+\K'_{\underline{\I}}(\T,\rho)
\end{multline}
where $\K'_{\underline{\I}}(\T,\rho)$ is a function of $\T$ and $\rho$ that satisfies
\begin{equation}
\lim_{\rho\to\infty} \K'_{\underline{\I}}(\T,\rho) = 0, \quad \T>2.
\end{equation}

The quantity $\underline{\I}(\T,\rho)/\T$ is a high-SNR approximation of the information rate achievable with i.i.d.\ USTM inputs; cf.\ \cite[Eq. (12)]{Marzetta_99} (see also \cite[Eq. (5)]{finte_block_length_2012}). It is shown in \cite[Th. 4]{Hochwald_2000} that $\underline{\I}(\T,\rho)/\T$ is also an asymptotically-tight lower bound on the capacity $\C(T,\rho)$ in the sense that
\begin{equation}
\lim_{\rho\to\infty}\biggl\{\C(T,\rho)-\frac{\underline{\I}(\T,\rho)}{\T}\biggr\}=0.
\end{equation}
According to Theorem \ref{th:high_SNR},
\begin{equation}
\label{eq:disp_noncoh_yeah}
\frac{\tilde{\U}(\T)}{\T^2} = \frac{(\T-1)^2}{\T^2} \frac{\pi^2}{6} + \frac{\T-1}{\T^2}
\end{equation}
can be viewed as a high-SNR approximation of the channel dispersion.

For comparison, the capacity and dispersion of the coherent Rayleigh block-fading channel---where the receiver has perfect knowledge of the realizations of the fading coefficients---are given by \cite{Polyanskiy_2011,Collins_Polyanskiy_2018,Ericson70}
\begin{subequations}
\begin{IEEEeqnarray}{rCl}
\C_{c}(\rho)  & = & \mathsf{E}\bigl[\log(1+\rho Z_1)\bigr] \label{eq:capacity_coh}\\
 \V_{c}(\T,\rho) & = & \text{Var}\bigl[\log(1+\rho Z_1)\bigr] + \frac{1}{\T}-\frac{1}{\T}\mathsf{E}\biggl[\frac{1}{1+\rho Z_1}\biggr]^2.\IEEEeqnarraynumspace \label{eq:dispersion_coh}
\end{IEEEeqnarray}
\end{subequations}
Note that
\begin{equation}\label{eq:capacity_coh_hsnr}
\lim_{\rho\to\infty} \bigl\{\C_{c}(\rho) - (\log \rho - \gamma) \bigr\} = 0
\end{equation}
and, for every $\T$,
 \begin{equation}
 \label{eq:disp_coh_yeah}
 \lim_{\rho\to\infty}  \V_{c}(\T,\rho) =  \frac{\pi^2}{6} + \frac{1}{\T}.
 \end{equation} 
Furthermore, for the noncoherent channel the high-SNR capacity $\underline{\I}(\T,\rho)/\T$ satisfies (cf.\ \eqref{eq:sec:first_first_moment})
\begin{equation}\label{eq:capacity_n_coh}
\lim_{\T\to\infty}\lim_{\rho\to\infty}\Bigl\{\frac{\underline{\I}(\T,\rho)}{\T} - \frac{\T-1}{\T} \Bigl[\log(\rho)-\gamma\Bigr]\Bigr\} = 0.
\end{equation}
By comparing \eqref{eq:capacity_coh_hsnr} and \eqref{eq:capacity_n_coh}, we see that $\underline{\I}(\T,\rho)/\T$ is, up to terms that vanish as $\rho\to\infty$ and $\T\to\infty$, equal to $(1-1/\T) \C_{c}(\rho)$. Similarly, by comparing \eqref{eq:disp_noncoh_yeah} and \eqref{eq:disp_coh_yeah}, we observe that $\tilde{\U}(\T)/\T^2$ corresponds to the dispersion one obtains by transmitting one pilot symbol per coherence block to estimate the fading coefficient and by then transmitting $\T-1$ symbols per coherence block over a coherent fading channel. This suggests the heuristic that, at high SNR, one pilot symbol per coherence block should be transmitted to achieve both capacity and channel dispersion. However, this heuristic may be misleading since it is \emph{prima facie} unclear whether one pilot symbol per coherence block suffices to obtain a fading estimate of sufficient accuracy. A more refined analysis of the maximum coding rate achievable with pilot-assisted transmission has been recently performed by \"Ostman \emph{et al.} \cite{Ostman_2018}.

\subsection{Numerical Examples}\label{sec:num_results}

\begin{figure}
\begin{flushright}
\includegraphics[width=\columnwidth] {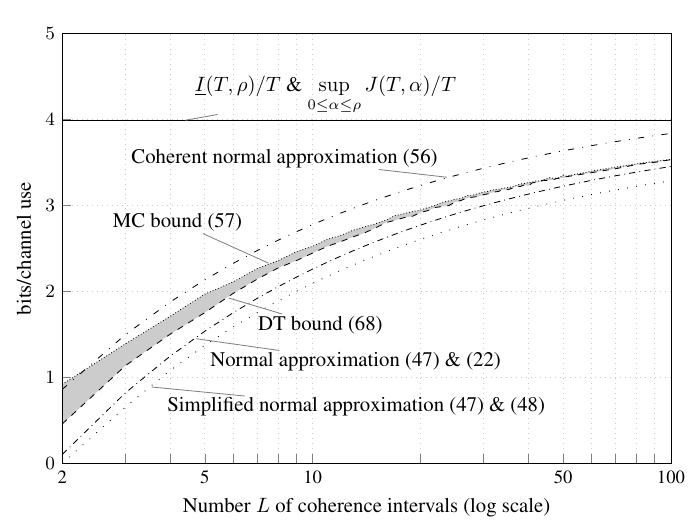}
\end{flushright}
\caption{Bounds on $\R^*(\L,\T,\epsilon,\rho)$ for $\rho=15$\,dB, $\T=20$, $\epsilon=10^{-3}$. The shaded area indicates the area in which $\R^*(\L,\T,\epsilon,\rho)$ lies.}
\label{fig:Rne_10db}
\end{figure}

\begin{figure}
\begin{flushright}
  \includegraphics[width=\columnwidth] {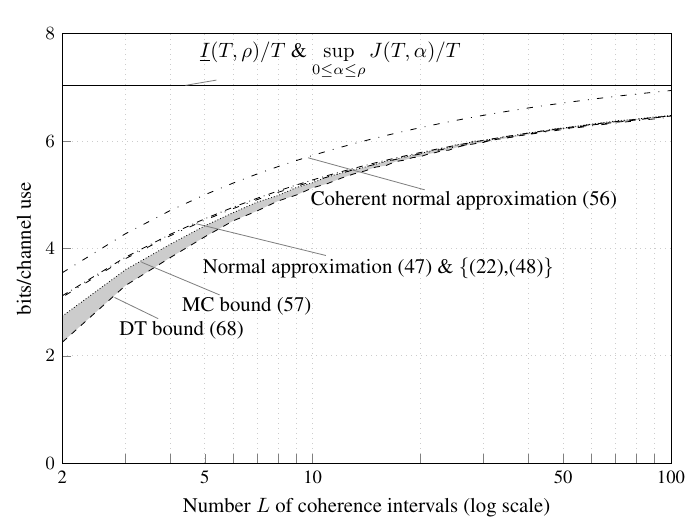}
\end{flushright}
\caption{Bounds on $\R^*(\L,\T,\epsilon,\rho)$ for $\rho=25$\,dB, $\T=20$, $\epsilon=10^{-3}$. The shaded area indicates the area in which $\R^*(\L,\T,\epsilon,\rho)$ lies.}
\label{fig:Rne_25db}
\end{figure}

We illustrate the accuracy of the high-SNR normal approximation \eqref{eq:asymp} by means of numerical examples.
In Figs.~\ref{fig:Rne_10db} and \ref{fig:Rne_25db}, we show the approximation \eqref{eq:asymp} as a function of $\L=\n/\T$ for a fixed coherence interval $\T$ and for different SNR values. In the normal approximation, we evaluate $\underline{\I}(\T,\rho)$ using both the exact expression \eqref{eq:sec:first_first_moment_closed} as well as the approximation \eqref{eq:sec:first_first_moment}. For comparison, we also plot the normal approximation of the coherent Rayleigh block-fading channel
\begin{equation}
\label{eq:NA_coh}
\R^*(\L,\T,\epsilon,\rho)\approx \C_c(\rho)-\sqrt{\frac{\V_c(\T,\rho)}{\L}}Q^{-1}(\epsilon)
\end{equation}
where $\C_c(\rho)$ and $\V_c(\T,\rho)$ are defined in \eqref{eq:capacity_coh} and \eqref{eq:dispersion_coh}, respectively. We further plot a nonasymptotic (in $\rho$ and $\L$) lower bound on $\R^*(\L,\T,\epsilon,\rho)$ that is based on the dependence testing (DT) lower bound \cite[Th. 22]{Polyanskiy_Poor_Verdu}  with USTM channel inputs (see \eqref{DT_bound_eq} below) and computed by Monte Carlo simulations. Similarly, we plot a nonasymptotic (in $\rho$ and $\L$) upper bound on $\R^*(\L,\T,\epsilon,\rho)$ that is based on the meta converse (MC) upper 
bound \cite[Th.~31]{Polyanskiy_Poor_Verdu} with auxiliary output pdf \eqref{output_pdf} (see \eqref{eq:MC} below).\footnote{The MC~bound appeared in the form used in this paper (cf.~\eqref{eq:MC}) in \cite[Th.~31]{Polyanskiy_Poor_Verdu}. It can also be obtained by particularizing the quantum result of Nagaoka~\cite{nagaoka2001} to the classical case.} More precisely, we plot the weakened version\footnote{The relaxation \eqref{eq:MC_numerical} of the MC bound coincides with the Verd\'{u}-Han bound \cite[Th. 4]{verdu_han} replacing the true output distribution $\text{P}_{\boldsymbol{Y}}$ by an arbitrary output distribution $\text{Q}_{\boldsymbol{Y}}$. This bound for an arbitrary output distribution $\text{Q}_{\boldsymbol{Y}}$ is a particularization of the Hayashi-Nagaoka lemma for classical quantum channels \cite[Lemma~4]{hayashi-nagaoka}.}
\begin{multline}
   \label{eq:MC_numerical}
   	\R^*(\L,\T,\epsilon,\rho) \leq\inf_{\xi>0}\Biggl\{\frac{\log\xi}{\L\T}\\
   	{} - \inf_{\boldsymbol{\pow}\in[o,\rho]^\L}\frac{\log\Bigl(1-\epsilon-\mathsf{P}\bigl[\sum_{\ell=1}^\L{j_\ell(\pow_\ell)}\geq\log\xi\bigr]\Bigr)}{\L\T}\Biggr\}
   \end{multline}
which is obtained by using \cite[Eq. (102)]{Polyanskiy_Poor_Verdu} and was evaluated by Monte Carlo simulations. In \eqref{eq:MC_numerical}, $\boldsymbol{\pow}=(\alpha_1,\ldots,\alpha_\L)$ denotes the vector of power allocations. We finally plot $\underline{\I}(\T,\rho)/\T$ given in \eqref{eq:sec:first_first_moment_closed}  and $\sup_{0\leq\alpha\leq\rho} J(\T,\alpha)/\T$ defined right before \eqref{eq:J_def}, which in both figures are indistinguishable from each other. By \eqref{eq:capacity_UB} and \eqref{eq:capacity_LB}, we have that
\begin{equation}
\frac{\underline{\I}(\T,\rho)}{\T} \leq C(\T,\rho) \leq \sup_{0\leq\alpha\leq\rho} \frac{J(\T,\alpha)}{\T}.
\end{equation}
We thus conclude that the error term $\K_{\underline{\I}}(\T,\rho)$ in \eqref{eq:comb_for_asymp} is negligible for the SNR values considered in the figures. Observe that the high-SNR normal approximation of $\R^*(\L,\T,\epsilon,\rho)$ is accurate already for $\rho=15$\,dB and $\L\geq 10$ when we use the exact expression \eqref{eq:sec:first_first_moment_closed} for $\underline{\I}(\T,\rho)$. For $\rho=25$\,dB and $\L\geq10$, the normal approximation is accurate even when we approximate $\underline{\I}(\T,\rho)$ using the simplified expression \eqref{eq:sec:first_first_moment}. Further observe that the normal approximation is pessimistic for $\rho=15$\,dB and optimistic for $\rho=25$\,dB.
As expected, the normal approximation \eqref{eq:NA_coh} of the coherent channel is strictly larger than the high-SNR normal approximation \eqref{eq:asymp} and the gap between the two normal approximations appears to be independent of~$\L$. This agrees with the intuition that the cost for estimating the channel mainly depends on the length $\T$ of the coherence interval. Finally observe that the DT lower bound on $\R^*(\L,\T,\epsilon,\rho)$, computed for USTM channel inputs, is close to the MC upper bound, which holds for any input distribution satisfying the power constraint \eqref{power_const}, provided that $\rho=15$\,dB and $\L\geq 5$ or $\rho=25$\,dB and $L\geq 2$. Thus, USTM channel inputs, which achieve the capacity asymptotically as the SNR tends to infinity, also give rise to lower bounds on $\R^*(\L,\T,\epsilon,\rho)$ that are close to optimal for moderate SNR values and short blocklengths. A similar observation was also made in \cite{finte_block_length_2012}.

\begin{figure}
	\begin{flushright}
        \includegraphics[width=\columnwidth] {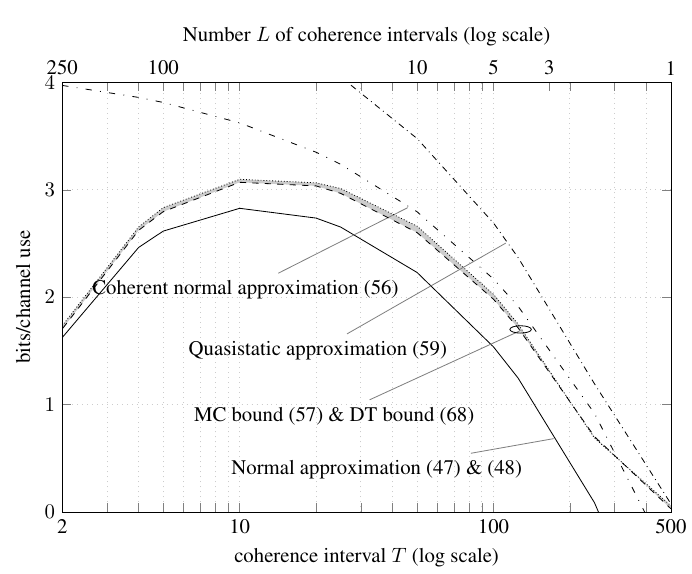}
		\end{flushright}
		\caption{Bounds on $\R^*(\L,\T,\epsilon,\rho)$ for $\L\T=500$, $\epsilon=10^{-3}$, $\rho=15$\,dB. The MC bound and the DT bound are almost indistinguishable. The shaded area indicates the area in which $\R^*(\L,\T,\epsilon,\rho)$ lies.}
				\label{fig:Rne_15db_DT}
\end{figure}

\begin{figure}
	\begin{flushright}
      \includegraphics[width=\columnwidth] {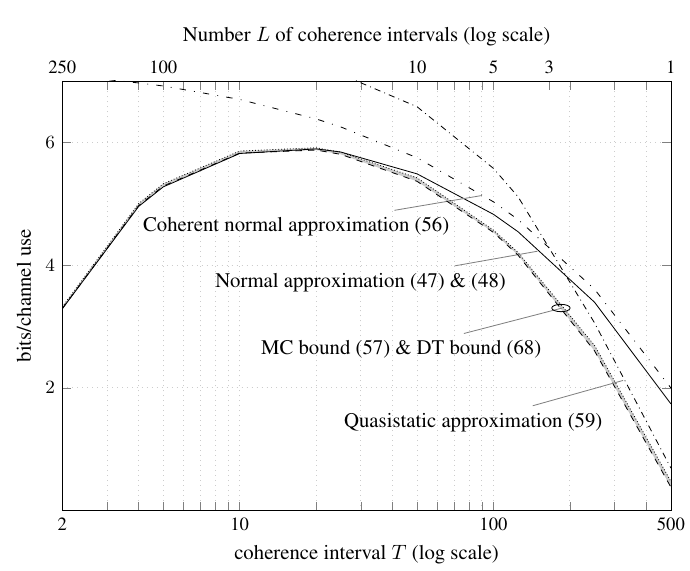}
       \end{flushright}
		\caption{Bounds on $\R^*(\L,\T,\epsilon,\rho)$ for $\L\T=500$, $\epsilon=10^{-3}$, $\rho=25$\,dB. The MC bound and the DT bound are almost indistinguishable. The shaded area indicates the area in which $\R^*(\L,\T,\epsilon,\rho)$ lies.}
		\label{fig:Rne_25db_DT}
\end{figure}

In Figs.~\ref{fig:Rne_15db_DT} and \ref{fig:Rne_25db_DT}, we show the high-SNR normal approximation \eqref{eq:asymp} (with $\underline{\I}(\T,\rho)/\T$ evaluated using the approximation \eqref{eq:sec:first_first_moment}) as a function of the coherence interval $\T$ for a fixed blocklength $\n$ (hence $\L$ is inversely proportional to $\T$). We further plot the normal approximation \eqref{eq:NA_coh} of the coherent channel. For comparison, we also show the DT bound (see \eqref{DT_bound_eq} below), evaluated for an USTM input distribution, and the weakened version of the MC bound~\eqref{eq:MC_numerical} evaluated by Monte Carlo simulations. Finally, we present the normal approximation that was proposed in \cite{Yang_2014} for quasistatic multiple-input multiple-output (MIMO) block-fading channels. To adapt the quasistatic MIMO block-fading channel to our system model, we replace $\mathbb{H}$ in \cite{Yang_2014} by an $\L\times \L$ diagonal matrix with diagonal entries $H_1,\ldots,H_\L$. Thus, specializing \cite[Eq. (95)]{Yang_2014} to our case, we obtain
\begin{equation}
	\epsilon\approx\mathbb{E}\Biggl[Q\Biggl(\frac{\C(\mathbb{H})-\L\,\R^*(\L,\T,\epsilon,\rho)}{\sqrt{\V(\mathbb{H})/\T}}\Biggr)\Biggr]\label{eq:quasistatic_approx_yang_et_al}
\end{equation}
where
\begin{subequations}\label{eq:quasi_eqs}
\begin{IEEEeqnarray}{lCl}
	\C(\mathbb{H}) & \triangleq & \sum_{j=1}^\L{\log(1+\rho|H_j|^2)} \\
	\V(\mathbb{H}) & \triangleq & \L-\sum_{j=1}^\L{\frac{1}{\log(1+\rho|H_j|^2)^2}}.
\end{IEEEeqnarray}
\end{subequations}

As already observed in Figs.~\ref{fig:Rne_10db} and \ref{fig:Rne_25db}, the high-SNR normal approximation is accurate for $\rho=15$~dB and $\L\geq 10$, and it is indistinguishable from the DT and MC bounds for $\rho=25$~dB and $\L\geq10$. The high-SNR normal approximation becomes less accurate as $\L$ decreases. Observe that the normal approximation of the coherent channel provides a good approximation  when $\T$ is large but becomes inaccurate when $\T\leq 100$. Further observe that the normal approximation for the quasistatic case \eqref{eq:quasistatic_approx_yang_et_al}, which is tailored towards the case where $\L$ is small, becomes accurate only for $\L\leq 3$ in both figures.
The figures show that, for a fixed blocklength $n=\L\T$, there is an optimal tradeoff between $\L$ and $\T$. This may be relevant, e.g., for the design of orthogonal frequency-division multiplexing (OFDM) systems, when the duration of a codeword is smaller than the coherence time, hence only frequency diversity is available. The system designer can then determine the number of diversity branches $\L$ available to each user by assigning OFDM symbols from different time and frequency slots. Figs.~\ref{fig:Rne_15db_DT} and \ref{fig:Rne_25db_DT} indicate the optimal value of $\L$ for $\n=500$, $\epsilon=10^{-3}$, and $\rho=\{15,25\}$\,dB. We refer to \cite{Ostman_HARQ_2018} for a more detailed discussion.

\begin{figure}
	\begin{flushright}
        \includegraphics[width=\columnwidth]  {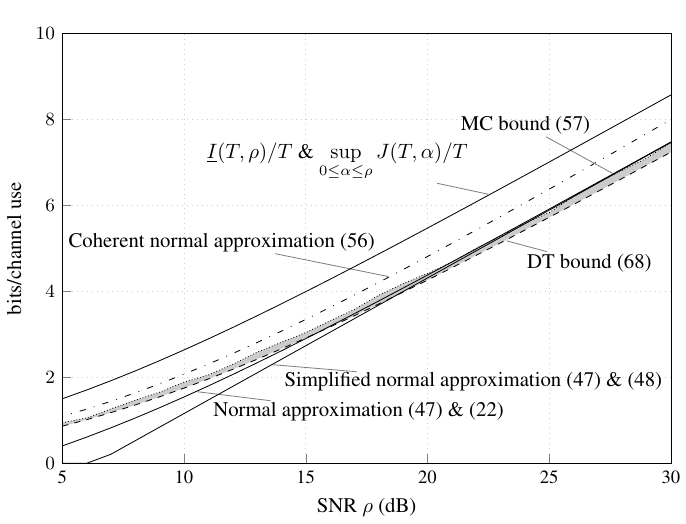}
     \end{flushright}
		\caption{Bounds on $\R^*(\L,\T,\epsilon,\rho)$ for $\T=20$, $\L=25$, $\epsilon=10^{-3}$. The shaded area indicates the area in which $\R^*(\L,\T,\epsilon,\rho)$ lies.}
		\label{fig:Rne_T20_L25}
\end{figure}

In Fig.~\ref{fig:Rne_T20_L25}, we plot the high-SNR normal approximation \eqref{eq:asymp}, evaluating $\underline{\I}(\T,\rho)$ using both \eqref{eq:sec:first_first_moment_closed} and \eqref{eq:sec:first_first_moment}, as a function of the SNR $\rho$ for fixed $\T$ and $\L$. We also plot the normal approximation \eqref{eq:NA_coh} of the coherent channel. For comparison, we further plot the DT bound (see \eqref{DT_bound_eq} below) evaluated for an USTM input distribution and the weakened version of the MC bound \eqref{eq:MC_numerical}.  Finally, we plot $\underline{\I}(\T,\rho)/\T$ given in \eqref{eq:sec:first_first_moment_closed} and $\sup_{0\leq\alpha\leq\rho} J(\T,\alpha)/\T$ defined right before \eqref{eq:J_def}, which in the figure are indistinguishable from each other. Recall that, by \eqref{eq:capacity_UB} and \eqref{eq:capacity_LB}, these terms bound the capacity $C(\T,\rho)$ from below and from above, so we conclude that the error term $\K_{\underline{\I}}(\T,\rho)$ in \eqref{eq:comb_for_asymp} is negligible for the SNR values considered in the figures. Observe that the normal approximation that uses~\eqref{eq:sec:first_first_moment_closed} becomes accurate already at SNR values of $15$ dB, while the normal approximation that uses \eqref{eq:sec:first_first_moment} is accurate for SNR values above $20$ dB. Further observe that the normal approximation is pessimistic for $\rho < 20$~dB and optimistic for $\rho \geq 20$~dB. As expected, the normal approximation \eqref{eq:NA_coh} of the coherent channel is strictly larger than the high-SNR normal approximation \eqref{eq:asymp}, but its gap to the nonasymptotic bounds decreases as $\rho$ becomes small. Intuitively, this is because, as $\rho$ decreases, knowledge of the fading coefficients becomes less important. Finally, we again observe that the DT lower bound on $\R^*(\L,\T,\epsilon,\rho)$ is close to the MC upper bound. Thus, USTM channel inputs, which achieve the capacity asymptotically as the SNR tends to infinity, are also close to optimal for all SNR values considered in the plot.

\begin{figure}
	\begin{center}
        \includegraphics[width=\columnwidth] {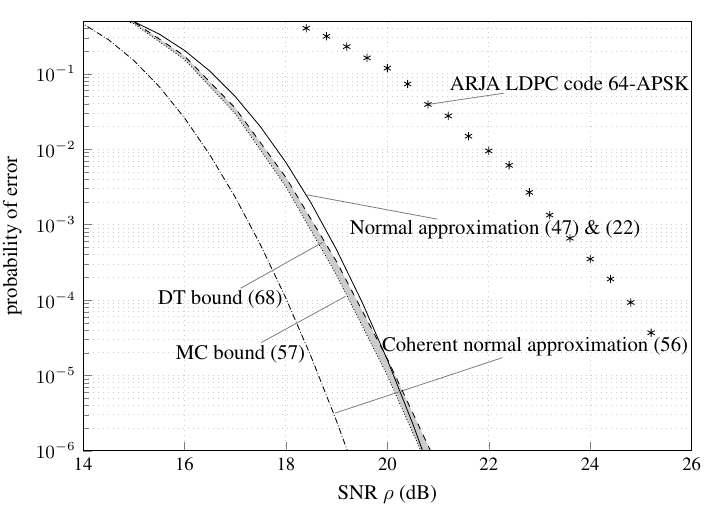}
    \end{center}
		\caption{Bounds on the probability of error $\epsilon$ for $R=4$, $\T=20$, $\L=25$. The shaded area indicates the area in which the true probability of error $\epsilon$ lies.}
		\label{fig:R4_T20_L25}
\end{figure}

In Fig.~\ref{fig:R4_T20_L25}, we plot the probability of error as a function of the SNR $\rho$ for $R=4$, $\T=20$, and $\L=25$. Specifically, we show the high-SNR normal approximation \eqref{eq:asymp} with $\underline{\I}(\T,\rho)$ evaluated using \eqref{eq:sec:first_first_moment_closed}, the normal approximation \eqref{eq:NA_coh} of the coherent channel, the DT bound evaluated for an USTM input distribution, and the weakened version of the MC bound \eqref{eq:MC_numerical}. For comparison, we further show the performance of an accumulate-repeat-jagged-accumulate (ARJA) low density parity check (LDPC) (3000,2000)-code combined with a \mbox{64-APSK} modulation, pilot-assisted transmission (2 pilot symbols per coherence block), and maximum likelihood channel estimation followed by mismatched nearest-neighbor decoding at the receiver (for details see \cite[Sec. 4]{Mustafa2019}). Observe that the high-SNR normal approximation is accurate for the whole range of SNR values evaluated. Further observe that the gap between the presented coding scheme and the rest of curves is substantial. This suggests that more sophisticated joint channel-estimation decoding procedures together with shaping techniques need to be adopted to close the gap; see also \cite{Coskun_2019}.
\subsection{Engineering Wisdom}
\label{sub:wisdom}
As argued, e.g., in \cite{Durisi_Koch_Popovski_2015}, the normal approximation can be used to analyze the performance of communication protocols. For example, let us consider the uplink scenario in \cite[Sec. IV-C]{Durisi_Koch_Popovski_2015}, where $\d$ devices intend to send $\k$ information bits to a base station within the time corresponding to $\n$ channel uses. The $\n$ channel uses are divided into $\s$ equally-sized slots of $\n_\s\triangleq n/\s$ channels uses and the devices apply a simple slotted-ALOHA protocol: each device picks randomly one of the $\s$ slots in the frame and sends its packet. If two or more devices pick the same slot, then a collision occurs and none of their packets is received correctly. If only one device picks a particular slot, then the error probability is calculated using the normal approximation. Specifically, in \cite[Sec. IV-C]{Durisi_Koch_Popovski_2015} the normal approximation for the AWGN channel was considered, i.e.,\footnote{For the AWGN channel, the $\mathcal{O}(\log\n/\n)$ in \eqref{eq:rate_ppv} can be replaced by $(\log\n)/(2n) + \mathcal{O}(1/\n)$ \cite{Polyanskiy_Poor_Verdu,Tan_Tomamichel_2015}.}
\begin{equation}\label{eq:rate_AWGN}
	\R^*(\n,\epsilon) \approx \C_{\textnormal{AWGN}}(\rho)-\sqrt{\frac{\V_{\textnormal{AWGN}}(\rho)}{\n}}Q^{-1}(\epsilon)+\frac{1}{2}\frac{\log \n}{\n}
\end{equation}
where
\begin{subequations}
\begin{IEEEeqnarray}{lCl}
\C_{\textnormal{AWGN}}(\rho) & = & \log(1+\rho) \\
\V_{\textnormal{AWGN}}(\rho) & = & \rho \frac{2+\rho}{(1+\rho)^2}.
\end{IEEEeqnarray}
\end{subequations}
By solving \eqref{eq:rate_AWGN} for $\epsilon$, we obtain an approximation of the packet error probability as a function of the packet length~$\n$, the number of information bits $\k=\n \R$ to be conveyed in a packet, and the SNR $\rho$, i.e.,
\begin{equation}\label{eq:error_AWGN}
	\epsilon^*(\k,\n,\rho) \approx Q\Biggl(\frac{\n \C_{\textnormal{AWGN}}(\rho)-\k\log2+(\log\n)/2}{\sqrt{\n \V_{\textnormal{AWGN}}(\rho)}}\Biggr).
\end{equation}
By replacing \eqref{eq:rate_AWGN} by our high-SNR normal approximation \eqref{eq:asymp}, we obtain the following approximation for the packet error probability when packets are transmitted over a noncoherent, single-antenna, Rayleigh block-fading channel of coherence interval $\T$:
\begin{equation}\label{eq:error_BFCh}
	\epsilon^*(\k,\n,\rho) \approx Q\Biggl(\frac{\n\underline{\I}(\T,\rho)-\k\T\log2}{\sqrt{\n\T\tilde{\U}(\T)}}\Biggr).
\end{equation}
Likewise, replacing \eqref{eq:rate_AWGN} by the normal approximation for the coherent Rayleigh block-fading channel \cite{Polyanskiy_2011,Collins_Polyanskiy_2018}, we obtain
\begin{equation}\label{eq:error_BFCh_coh}
	\epsilon^*(\k,\n,\rho) \approx Q\Biggl(\frac{\n \C_{c}(\rho)-\k\log2}{\sqrt{\n\T\V_{c}(\T,\rho)}}\Biggr)
\end{equation}
where
\begin{IEEEeqnarray}{rCl}\label{eq:capacity_var_coh}
\C_{c}(\rho) &\triangleq& \mathsf{E}\bigl[\log(1+\rho Z_1)\bigr]\IEEEyesnumber\IEEEyessubnumber \\
\V_{c}(\T,\rho) &\triangleq& \text{Var}\bigl[\log(1+\rho Z_1)\bigr] + \frac{1}{\T}-\frac{1}{\T}\mathsf{E}\biggl[\frac{1}{1+\rho Z_1}\biggr]^2.\,\IEEEeqnarraynumspace\IEEEyessubnumber
\end{IEEEeqnarray}

The probability of successful transmission is given by \cite[Eq. (24)]{Durisi_Koch_Popovski_2015}
\begin{equation}\label{eq:prob_success}
P_{\text{success}} = \frac{\d}{\s}\biggl(1-\frac{1}{\s}\biggr)^{\d-1}\bigl(1-\epsilon^*(\k,\n_\s,\rho)\bigr)
\end{equation} 
where $(\d/\s)(1-1/\s)^{\d-1}$ is the probability that only one device transmits in a given slot \cite[Sec.~5.3.2]{Kurose:2012:CNT:2584507}. Our goal is to choose $\s$ such that the probability of successful transmission is maximized given $\d$, $\k$, $\n$, and $\rho$. This problem entails a tradeoff between the probability of collision and the number of channel uses available for each packet, which affects the achievable error probability in a singleton slot. 
 
\begin{table}[t!]
\centering
\caption{Optimal number of slots for different channel models and $\n=\L\T=480$, $\k=256$, $\d=12$.}
\label{tab:aloha}
\begin{tabular}{|c|c||c|c|c|c|}
\hline
\multirow{2}{*}{\vspace{-3.5em}\textbf{SNR}}       & \multirow{2}{*}{\vspace{-3.5em} $T$ } & \multicolumn{4}{c|}{\textbf{optimal number of slots} $\s$}                                                                                                                                                                                          \\ \cline{3-6} 
                                &                                               & \begin{tabular}[c]{@{}c@{}}noncoherent\\ Rayleigh \\ block-\\fading\end{tabular} & \begin{tabular}[c]{@{}c@{}}coherent\\ Rayleigh \\ block-\\fading\end{tabular} & AWGN   & \begin{tabular}[c]{@{}c@{}}classic\\ slotted-\\ALOHA\end{tabular} \\ \hline \hline

\multirow{2}{*}{$\rho = 15$ dB} & $5$  & $\s=4$ & $\s=6$ & $\s=8$ & $\s=12$ \\ \cline{2-6} 
                                & $20$ & $\s=6$ & $\s=6$ & $\s=8$ & $\s=12$ \\ \hline \hline
\multirow{2}{*}{$\rho = 25$ dB} & $5$  & $\s=8$ & $\s=12$& $\s=12$& $\s=12$ \\ \cline{2-6} 
                                & $20$ & $\s=8$ & $\s=8$ & $\s=12$& $\s=12$ \\ \hline 
\end{tabular}
\end{table}

As a concrete example, we consider the case where $\n=480$, $\d=12$, and $\k=256$.\footnote{The fact that $\n$ is fixed implies that the number of coherence intervals $\L$ changes inversely proportional to $\T$ for the block-fading cases.} In Table~\ref{tab:aloha}, we show the optimal number of slots $\s$ for the noncoherent Rayleigh block-fading channel (with $\epsilon^*(\k,\n_\s,\rho)$ approximated by \eqref{eq:error_BFCh}), the coherent Rayleigh block-fading channel (with $\epsilon^*(\k,\n_\s,\rho)$ approximated by \eqref{eq:error_BFCh_coh}), the AWGN channel (with $\epsilon^*(\k,\n_\s,\rho)$ approximated by \eqref{eq:error_AWGN}), and the classic slotted-ALOHA protocol ($\epsilon^*(\k,\n_\s,\rho)=0$) for the SNR values $\rho=15$~dB and $\rho=25$~dB and coherence intervals $\T=5$ and $\T=20$. To be consistent with our system model, for the Rayleigh block-fading channel (both coherent and noncoherent) we only consider slot sizes $\n_\s$ that are integer multiples of $\T$. Observe that the optimal number of slots $\s$ depends critically on the SNR, the coherence interval, and the considered channel model. For example, for the classic slotted-ALOHA protocol, the optimal number of slots is $\s=12$, which coincides with the total number of devices $\d=12$. In contrast, for the AWGN channel, the optimal number of slots is $\s=8$ for $\rho=15$~dB and coincides with the one of the classic slotted-ALOHA for $\rho=25$~dB. In most cases, the optimal number of slots $\s$ for the Rayleigh block-fading channel (both coherent and noncoherent) is yet again smaller and depends both on the SNR and the coherence interval $\T$. When $\T=20$, the optimal number of slots $\s$ for the noncoherent Rayleigh block-fading channel coincides with that for the coherent channel. This agrees with the intuition that, when $\T$ is sufficiently large, the fading coefficients can be learned with little training overhead. In general, the optimal number of slots $\s$ decreases as the channel becomes less favorable. Intuitively, larger codes are required to combat the impairments due to AWGN and fading. Hence, the packet length $\n_\s$ must be increased or, equivalently, the number of slots $\s=\n/\n_\s$ must be reduced.
 
\section{Proof of Theorem~\ref{th:high_SNR}}\label{sec:Proofs}
\label{sec:proof}
The proof of Theorem \ref{th:high_SNR} is based on a lower bound on $\R^*(\L,\T,\epsilon,\rho)$, given in Section~\ref{sec:DT}, and on an upper bound on $\R^*(\L,\T,\epsilon,\rho)$, given in Section~\ref{sec:MC}. Since these bounds coincide up to the error terms $\K_{\textnormal{DT}}(L,\T,\rho)$ and $\K_{\textnormal{MC}}(L,\T,\rho)$ (whose difference is of order $\log \L/\L$ uniform in $\rho$) and up to the terms $\K_{\bar{\J}}(\T,\rho)$ and $\K_{\bar{\V}}(\T,\rho)$ given in \eqref{eq:barJ} and \eqref{eq:barV} (which are independent of $L$ and vanish as $\rho\to\infty$), they prove \eqref{eq:comb_for_asymp}.

\subsection{Dependence Testing Lower Bound}\label{sec:DT}
To obtain a lower bound on $\R^*(\L,\T,\epsilon,\rho)$, we evaluate the DT bound \cite[Th. 22]{Polyanskiy_Poor_Verdu} for the USTM input distribution defined in Section~\ref{sec:preliminaries}. Thus, assume that $\XL\sim \text{P}_{\XL}^{(\text{U})}$, which implies $\YL\sim\text{q}_{\mathbf{Y}^\L}^{(\text{U})}$. One can show (see \cite[App.~A]{Short_Packets_Durisi_Koch_TCOM_2015}) that the cumulative distribution function $\mathsf{P}[i(\mathbf{x}^\L;\boldsymbol{\tilde{Y}}^\L)\leq\pow]$ does not depend on $\mathbf{x}^\L$. Furthermore, the USTM input distribution satisfies the power constraint \eqref{power_const} with probability one. A lower bound on $\R^*(\L,\T,\epsilon,\rho)$ follows therefore from the DT bound (maximum probability of error) \cite[Th. 22]{Polyanskiy_Poor_Verdu}, which, after a standard change of measure, can be stated as follows: there exists a code with $M$ codewords, blocklength $\L\T$, and maximum probability of error $\epsilon$ not exceeding
\begin{multline}
		\epsilon \leq (M-1)\mathsf{E}\Bigl[e^{-i(\XL;\YL)}\text{I}{\{i(\XL;\YL)>\log(M-1)}\}\Bigr]
		\\{}+\mathsf{P}\bigl[i(\XL;\YL)\leq\log(M-1)\bigr]. \label{DT_bound_eq}
\end{multline}
To show that \eqref{DT_bound_eq} yields the lower bound
\begin{equation}
	\label{max_DT}
	\R^*(\L,\T,\epsilon,\rho) \geq \frac{\I(\T,\rho)}{\T}-\sqrt{\frac{\U(\T,\rho)}{\L\T^2}}Q^{-1}(\epsilon)+\K_{\textnormal{DT}}(L,\T,\rho)
\end{equation}
where for every $\T>2$ and some $\A$, $L_0$, and $\rho_0$ independent of $L$, and $\rho$,
\begin{equation}
\sup_{\rho\geq\rho_0}\bigl|\K_{\textnormal{DT}}(L,\T,\rho)\bigr|\leq \frac{\A}{L},\quad L\geq\L_0
\end{equation}
we follow almost verbatim the steps in \cite[Eqs. (258)--(267)]{Polyanskiy_Poor_Verdu} (with $\gamma$ in \cite{Polyanskiy_Poor_Verdu} replaced by $M-1$). The main difference is that, in our case, $\U(\T,\rho)$ defined in \eqref{eq:second_def} and $\B(\T,\rho)$ defined as (cf.\ \cite[Eq. (254)]{Polyanskiy_Poor_Verdu})
\begin{equation}
\label{B_DT_bound}
\B(\T,\rho)\triangleq \frac{6\mathsf{E}\Bigl[\bigl|i_{\ell}(\T,\rho)-\I(\T,\rho)\bigr|^3\Bigr]}{\U(\T,\rho)^{3/2}}
\end{equation}
depend on $\rho$. To ensure that the term $\K_{\textnormal{DT}}(L,\T,\rho)$ in \eqref{max_DT} is uniform in $\rho$, we will show that both $\U(\T,\rho)$ and $\B(\T,\rho)$ are bounded in $\rho$. We then apply the Berry-Esseen theorem \cite[Ch. XVI.5]{Feller} to obtain \cite[Eq. (259)]{Polyanskiy_Poor_Verdu} with $\B(\T,\rho)$ replaced by an upper bound $\B(\T,\rho_0)$ that holds for all $\rho\geq\rho_0$ and a sufficiently large $\rho_0$, followed by \cite[Eqs. (261)--(265)]{Polyanskiy_Poor_Verdu}, which gives
\begin{IEEEeqnarray}{lCl}
		\R^*(\L,\T,\epsilon,\rho) &\geq& \frac{\I(\T,\rho)}{\T}-\sqrt{\frac{\U(\T,\rho)}{\L\T^2}}Q^{-1}(\tau)\IEEEyesnumber\label{max_DT_firststep}
\end{IEEEeqnarray}
where 
\begin{equation}\label{eq:tau_def}
		 \tau = \epsilon-\biggl(\frac{2\log 2}{\sqrt{2\pi}}+5\B(\T,\rho_0)\biggr)\frac{1}{\sqrt{\L}}.
\end{equation}  
A Taylor-series expansion of $Q^{-1}(\tau)$ around $\epsilon$ yields then that
\begin{equation}\label{eq:KQ_big_O}
\sup_{\rho\geq\rho_0}\bigl|Q^{-1}(\tau) - Q^{-1}(\epsilon)\bigr|\leq \frac{\A}{\sqrt{L}},\quad L\geq\L_0
\end{equation}
for every $\T>2$ and some $\A$, $L_0$, and $\rho_0$ independent of $L$ and $\rho$. Combining \eqref{eq:KQ_big_O} with \eqref{max_DT_firststep}, we obtain \eqref{max_DT}.

To show that $\U(\T,\rho)$ and $\B(\T,\rho)$ are bounded in $\rho$, we resort to the following lemmas:

\begin{lemma}\label{lm:var_bounded_away0}
Let $0\leq\delta\leq 1/2$ and let $\bar{\V}_\rho(\T,\pow)$ be defined in \eqref{eq:second_def_bar}. For every $\rho(1-\delta)\leq\pow\leq\rho$, we have
\begin{equation}\label{eq:var_bounded_away_lemma}
\bar{\V}_\rho(\T,\pow)\geq \biggl(\frac{\T\rho}{1+\T\rho}\biggr)^2 (\T-1)-\Xi(\T)\delta + \K_{\bar{\V}}(\T,\rho)
\end{equation}
where $\K_{\bar{\V}}(\T,\rho)$ is a function of $\T$ and $\rho$ that satisfies
\begin{equation}\label{eq:KV_small_o}
\lim_{\rho\to\infty}\K_{\bar{\V}}(\T,\rho) = 0, \quad\T>2
\end{equation}
and $\Xi(\T)$ is a positive constant that only depends on $\T$.
\end{lemma}
\begin{IEEEproof}
See Appendix \ref{app:bounded_away0}.
\end{IEEEproof}
\begin{lemma}\label{lm:var_bounded}
	For every $\rho_0>0$ and $\T>2$, we have
	 \begin{IEEEeqnarray}{rCl}
	 \sup_{\begin{subarray}{l} \pow\geq 0,\\ \rho\geq\rho_0 \end{subarray}}{\bar{\V}_\rho(\T,\pow)}&<& \infty\label{eq:bar_var_bounded}\\
		 \sup_{\rho\geq\rho_0}{\U(\T,\rho)}&<& \infty.\label{eq:var_bounded}		 
	\end{IEEEeqnarray}
\end{lemma}
\begin{IEEEproof}
See Appendix \ref{app:second_moment}.
\end{IEEEproof}
\begin{lemma}\label{lm:third_moment_bounded}
	For every $\rho_0>0$ and $\T>2$, we have
\begin{IEEEeqnarray}{lCl}
	 \sup_{\begin{subarray}{l} \pow\geq 0,\\ \rho\geq\rho_0 \end{subarray}}{\mathsf{E}\Bigl[\bigl|\bar{j}_\ell(\T,\pow)-\bar{\J}(\T,\pow)\bigr|^3\Bigr]} & < & \infty\label{eq:bar_third_bounded}\\
		 \sup_{\rho\geq \rho_0}{\mathsf{E}\Bigl[\bigl|i_{\ell}(\T,\rho)-\I(\T,\rho)\bigr|^3\Bigr]} & <& \infty.\label{eq:third_bounded}
\end{IEEEeqnarray}
\end{lemma}
\begin{IEEEproof}
See Appendix~\ref{app:third_moment}.
\end{IEEEproof}

For $\delta=0$, Lemma~\ref{lm:var_bounded_away0} yields
 \begin{equation}
 \bar{\V}(\T,\rho) \geq \biggl(\frac{\T\rho}{1+\T\rho}\biggr)^2 (\T-1)+\K_{\bar{\V}}(\T,\rho)
   \end{equation} 
where $\K_{\bar{\V}}(\T,\rho)$ satisfies \eqref{eq:var_bounded_away_lemma}. Together with \eqref{eq:underU}, \eqref{eq:barV}, and \eqref{eq:former_36}, this implies that
 \begin{equation}
\label{eq:inf_dens_variance_lower_bound_DT_bound}
\U(\T,\rho) \geq \biggl(\frac{\T\rho_0}{1+\T\rho_0}\biggr)^2 \frac{\T-1}{2}, \quad \rho\geq\rho_0 
\end{equation}
for a sufficiently large $\rho_0$. Furthermore,  Lemma \ref{lm:var_bounded} implies that, for every $\rho_0>0$, there exists an $\U_{\text{UB}}(\T,\rho_0)$ that is independent of $\rho$ and that satisfies
\begin{equation}
\label{eq:U_upper_bound}
\U(\T,\rho) \leq \U_{\text{UB}}(\T,\rho_0), \quad \rho\geq\rho_0.
\end{equation}
Finally, Lemma~\ref{lm:third_moment_bounded} implies that for every $\rho_0>0$ there exists an $\S(\T,\rho_0)$ that is independent of $\rho$ and that satisfies
\begin{equation}
\mathsf{E}\Bigl[\bigl|i_{\ell}(\T,\rho)-\I(\T,\rho)\bigr|^3\Bigr] \leq \S(\T,\rho_0), \quad \rho\geq\rho_0. \label{eq:Srho_bounded}
\end{equation}
Combining \eqref{eq:inf_dens_variance_lower_bound_DT_bound} and \eqref{eq:Srho_bounded}, it follows that for a sufficiently large $\rho_0>0$ there exists a $\B(\T,\rho_0)$ that is independent of $\rho$ and that satisfies
\begin{equation}\label{B_DT_bound_bounded}
 		\B(\T,\rho) \leq \frac{6 \S(\T,\rho_0)}{\Bigl(\frac{\T\rho_0}{1+\T\rho_0}\Bigr)^3\bigl(\frac{\T-1}{2}\bigr)^{3/2}}\triangleq \B(\T,\rho_0), \quad \rho\geq \rho_0.
 \end{equation}
 This concludes the proof of the lower bound \eqref{max_DT}.


\subsection{Meta Converse Upper Bound}\label{sec:MC}
 An upper bound on $\R^*(\L,\T,\epsilon,\rho)$ follows from the MC bound \cite[Th. 31]{Polyanskiy_Poor_Verdu} computed for the auxiliary pdf $\text{q}_{\mathbf{Y}^\L}^{(\text{U})}$, i.e.,
 \begin{equation}
 \label{eq:MC}
 \R^*(\L,\T,\epsilon,\rho) \leq \frac{1}{LT} \sup_{\boldsymbol{\pow}\in[0,\rho]^\L}\log\Biggl(\frac{1}{\beta(\boldsymbol{\pow},\text{q}_{\mathbf{Y}^\L}^{(\text{U})})}\Biggr).
 \end{equation}
 Here, $\boldsymbol{\pow}=(\alpha_1,\ldots,\alpha_\L)$ denotes the vector of power allocations, and $\beta(\boldsymbol{\pow},\text{q}_{\mathbf{Y}^\L}^{(\text{U})})$ denotes the minimum probability of error under hypothesis $\text{q}_{\mathbf{Y}^\L}^{(\text{U})}$ if the probability of error under hypothesis $\text{p}_{\YL|\XL=\boldsymbol{x}^\L}$ does not exceed $\epsilon$~\cite[Eq.~(100)]{Polyanskiy_Poor_Verdu}. Note that, by \eqref{eq:distributions_cond_x}--\eqref{eq:norm_output_Q}, $\beta(\boldsymbol{\pow},\text{q}_{\mathbf{Y}^\L}^{(\text{U})})$ depends on $\boldsymbol{x}^\L$ only via $\boldsymbol{\pow}$ (recall that  $\|\X_l\|^2=\T\pow_{\ell}$).
 
 For $0<\delta<1$, let $\L_{\delta}(\boldsymbol{\pow})$ denote the number of $\pow_\ell$'s in $\boldsymbol{\pow}$ that satisfy $\rho(1-\delta)\leq\alpha_{\ell}\leq\rho$. The following lemma demonstrates that we can assume without loss of optimality that $\L_{\delta}(\boldsymbol{\pow}) \geq \L/2$, i.e., in at least half of the coherence intervals $\pow_\ell$ is larger than $\rho(1-\delta)$.
   \begin{lemma}\label{lm:opt_high_SNR}
 Let
   \begin{equation}
   \mathcal{A}_{\rho,\delta}\triangleq \{\boldsymbol{\pow}\in[0,\rho]^\L\colon \L_{\delta}(\boldsymbol{\pow}) \geq \L/2\}.
   \end{equation}
 For every $0<\delta<1$, $\T>2$, and $0<\epsilon<1/2$, there exists a pair $(\L_0,\rho_0)$ independent of $\L$ and $\rho$ such that, for $\L\geq L_0$ and $\rho\geq \rho_0$, the supremum in \eqref{eq:MC} can be replaced without loss of optimality by a supremum over $\boldsymbol{\pow}\in\mathcal{A}_{\rho,\delta}$.
    \end{lemma}
   \begin{IEEEproof}
    See Appendix~\ref{app:proof_lm_opt_high_SNR_A}.
   \end{IEEEproof}
 
In the following, we implicitly assume that $\L\geq \L_0$ and $\rho\geq\rho_0$ for some sufficiently large $\L_0$ and $\rho_0$ so that Lemma~\ref{lm:opt_high_SNR} holds. Applying Lemma~\ref{lm:opt_high_SNR} to \eqref{eq:MC}, and upper-bounding the right-hand side (RHS) of \eqref{eq:MC} using \cite[Eq. (106)]{Polyanskiy_Poor_Verdu} and \eqref{eq:j_u_bound}, we obtain
   \begin{multline}
   \label{eq:MC_logM}
   	\R^*(\L,\T,\epsilon,\rho) \leq\sup_{\boldsymbol{\pow}\in\mathcal{A}_{\rho,\delta}}\Biggl\{\frac{\log\xi(\boldsymbol{\pow})}{\L\T}\\
   	{} - \frac{\log\bigl(1-\epsilon-\mathsf{P}\bigl[\sum_{\ell=1}^\L{\bar{j}_\ell(\T,\pow_\ell)}\geq\log\xi(\boldsymbol{\pow})\bigr]\bigr)}{\L\T}\Biggr\}
   \end{multline}
  for every $\xi\colon [0,\rho]^\L \to (0,\infty)$.
  
  Let
  \begin{equation}\label{eq:B_bounds_first}
  	\bar{\B}(\T,\boldsymbol{\pow})\triangleq\frac{6\sum_{\ell=1}^\L\mathsf{E}\Bigl[\bigl|\bar{j}_\ell(\T,\pow_\ell)-\bar{\J}(\T,\pow_\ell)\bigr|^3\Bigr]}{\Bigl(\sum_{\ell=1}^\L\bar{\V}_\rho(\T,\pow_\ell)\Bigr)^{3/2}}.
  	\end{equation}
  By Lemma~\ref{lm:third_moment_bounded}, the expectation $\mathsf{E}\bigl[|\bar{j}_{\ell}(\T,\pow)-\bar{\J}(\T,\pow)|^3\bigr]$ can be upper-bounded by a constant $\bar{\S}(\T,\rho_0)$ that is independent of $\pow$ and $\rho$. Furthermore, by the nonnegativity of $\bar{\V}_\rho(\T,\pow_\ell)$,
  \begin{equation}
  	\sum\limits_{\ell=1}^\L\bar{\V}_\rho(\T,\pow_\ell) \geq \sum\limits_{\ell\in\mathscr{L}_{\delta}(\boldsymbol{\pow})}\bar{\V}_\rho(\T,\pow_\ell)\label{eq:var_low_bound_sets}
  \end{equation}
  where $\mathscr{L}_{\delta}(\boldsymbol{\pow})\triangleq\{\ell=1,\dots,\L:\,\pow_\ell\geq\rho(1-\delta)\}$. Lemma \ref{lm:var_bounded_away0} demonstrates that, for $\pow\geq\rho(1-\delta)$, 
  \begin{equation}
  \bar{\V}_\rho(\T,\pow)\geq \biggl(\frac{\T\rho}{1+\T\rho}\biggr)^2 (\T-1)-\Xi(\T) \delta +\K_{\bar{\V}}(\T,\rho).
  \end{equation}
Thus, for
\begin{equation}
\label{eq:delta}
\delta = \left(\frac{\T\rho_0}{1+\T\rho_0}\right)^2 \frac{\T-1}{3\Xi(\T)}
\end{equation}
and $\rho_0$ sufficiently large, we have
  \begin{IEEEeqnarray}{lCl}
  \sum\limits_{\ell=1}^\L\bar{\V}_\rho(\T,\pow_\ell) &\geq & \L_{\delta}(\boldsymbol{\pow}) \biggl(\frac{\T\rho_0}{1+\T\rho_0}\biggr)^2 \frac{\T-1}{2}, \quad \rho\geq\rho_0.\IEEEeqnarraynumspace
  \end{IEEEeqnarray}
It follows that, for every $\boldsymbol{\pow}\in\mathcal{A}_{\rho,\delta}$ and $\delta$ as chosen in \eqref{eq:delta},
  \begin{equation}
  \label{eq:Tobi_barB}
  \bar{\B}(\T,\boldsymbol{\pow})\leq\frac{6\L\bar{\S}(\T,\rho_0)}{\Bigl(\frac{(\T-1)\L}{4}\Bigr)^{3/2}\Bigl(\frac{\T\rho_0}{1+\T\rho_0}\Bigr)^3}\triangleq \frac{\bar{\B}(\T,\rho_0)}{\sqrt{\L}}.
  \end{equation}
  
  Let
  \begin{equation}
  \lambda= Q^{-1}\biggl(\epsilon + \frac{2\bar{\B}(\T,\rho_0)}{\sqrt{\L}}\biggr)
  \end{equation}
  and
  \begin{equation}
  	\log\xi(\boldsymbol{\pow}) = \sum\limits_{\ell=1}^\L\bar{\J}(\T,\pow_\ell) -\lambda\sqrt{\sum\limits_{\ell=1}^\L \bar{\V}_\rho(\T,\pow_\ell)}.
  \end{equation}
  With this choice, the Berry-Esseen theorem and \eqref{eq:Tobi_barB} imply that, for every $\boldsymbol{\pow}\in\mathcal{A}_{\rho,\delta}$,
  \begin{IEEEeqnarray}{lCl}
  \Biggl|\mathsf{P}\Biggl[\sum\limits_{\ell=1}^\L{\bar{j}_\ell(\T,\pow_\ell)}\leq \log\xi(\boldsymbol{\pow}) \Biggr] - Q(\lambda)\Biggr| &\leq&\frac{\bar{\B}(\T,\rho_0)}{\sqrt{\L}}.\IEEEeqnarraynumspace
  \end{IEEEeqnarray}
  Thus, for such $\boldsymbol{\pow}$'s,
  \begin{IEEEeqnarray}{lCl}
  	\mathsf{P}\Biggl[\sum\limits_{\ell=1}^\L{\bar{j}_\ell(\T,\pow_\ell)}\leq \log\xi(\boldsymbol{\pow}) \Biggr] \geq \epsilon + \frac{\bar{\B}(\T,\rho_0)}{\sqrt{\L}}.\IEEEyesnumber\label{second_Berry_Essen_MC}
  \end{IEEEeqnarray}
  Substituting \eqref{second_Berry_Essen_MC} into the upper bound \eqref{eq:MC_logM}, we obtain
  \begin{IEEEeqnarray}{lCl}
  	\IEEEeqnarraymulticol{3}{l}{\R^*(\L,\T,\epsilon,\rho)}\nonumber\\
	\quad & \leq &  \sup_{\boldsymbol{\pow}\in\mathcal{A}_{\rho,\delta}}\Biggl\{\frac{\sum_{\ell=1}^\L\bar{\J}(\T,\pow_\ell)}{\L\T}\nonumber\\
	& &\qquad\quad {} - \sqrt{\frac{\sum_{\ell=1}^\L \bar{\V}_\rho(\T,\pow_\ell)}{\L^2\T^2}}Q^{-1}\biggl(\epsilon + \frac{2\bar{\B}(\T,\rho_0)}{\sqrt{\L}}\biggr)\Biggr\} \nonumber\\
	& & {} - \frac{\log\bar{\B}(\T,\rho_0)}{\L\T}+ \frac{1}{2}\frac{\log \L}{\L\T}. \label{eq:mc_with_L_terms}
  \end{IEEEeqnarray}
By the assumption $0<\epsilon<\frac{1}{2}$, the inverse $Q$-function on the RHS of \eqref{eq:mc_with_L_terms} is positive for sufficiently large $\L$. It follows by the concavity of the square-root function and Jensen's inequality that \eqref{eq:mc_with_L_terms} can be further upper-bounded as
  \begin{IEEEeqnarray}{lCl}
  	\IEEEeqnarraymulticol{3}{l}{\R^*(\L,\T,\epsilon,\rho)} \nonumber\\
	\quad & \leq & \frac{1}{\L}\sum\limits_{\ell=1}^\L \sup_{0\leq\alpha_{\ell}\leq\rho}\biggl\{\frac{\bar{\J}(\T,\pow_{\ell})}{\T}\nonumber\\
  	&& \qquad\qquad\qquad {} -\sqrt{\frac{ \bar{\V}_\rho(\T,\pow_{\ell})}{\L\T^2}}Q^{-1}\biggl(\epsilon + \frac{2\bar{\B}(\T,\rho_0)}{\sqrt{\L}}\biggr)\biggr\} \nonumber\\
  	&&{}- \frac{\log\bar{\B}(\T,\rho_0)}{\L\T}+ \frac{1}{2}\frac{\log \L}{\L\T} \nonumber\\
  	& = & \sup_{0\leq\alpha\leq\rho}\biggl\{\frac{\bar{\J}(\T,\pow)}{\T} -\sqrt{\frac{ \bar{\V}_\rho(\T,\pow)}{\L\T^2}}Q^{-1}\biggl(\epsilon + \frac{2\bar{\B}(\T,\rho_0)}{\sqrt{\L}}\biggr)\biggr\}\nonumber\\
  	&&{} - \frac{\log\bar{\B}(\T,\rho_0)}{\L\T}+ \frac{1}{2}\frac{\log \L}{\L\T} \label{eq:mc_with_L_terms_2}
  \end{IEEEeqnarray}
  where the second step follows because the channel is blockwise i.i.d., so the terms inside the curly brackets do not depend on $\ell$.
  
  Performing a Taylor-series expansion of the inverse \mbox{$Q$-function} around $\epsilon$, we obtain
  \begin{equation}
  	\sup_{\rho\geq\rho_0}\biggl|Q^{-1}\biggl(\epsilon + \frac{2\bar{\B}(\T,\rho_0)}{\sqrt{\L}}\biggr) - Q^{-1}(\epsilon)\biggr| \leq \frac{\A}{\sqrt{L}}, \quad \L\geq \L_0
  	\end{equation}
  	for every $\T>2$ and some $\A$, $\L_0$, and $\rho_0$ independent of $\L$ and $\rho$. Further using that, by Lemma~\ref{lm:var_bounded}, $\bar{\V}_\rho(\T,\pow)$ is bounded in $\rho$ and $\alpha$, and collecting terms of order $\log \L/\L$, we can rewrite \eqref{eq:mc_with_L_terms_2} as 
  \begin{IEEEeqnarray}{lCl}\label{eq:MC_ub_alpha}
  	\R^*(\L,\T,\epsilon,\rho)&\leq& \sup_{0\leq\pow\leq\rho}\biggl\{\frac{\bar{\J}(\T,\pow)}{\T} - \sqrt{\frac{ \bar{\V}_\rho(\T,\pow)}{\L\T^2}}Q^{-1}(\epsilon)\biggr\}\nonumber\\
  	&&{}+\K_{1}(L,\T,\rho)
  	\end{IEEEeqnarray}
  	where $\K_{1}(\L,\T,\rho)$ is a function of $\L$, $\T$, and $\rho$ that satisfies
  	\begin{equation}
  	\sup_{\rho\geq\rho_0}\bigl|\K_{1}(L,\T,\rho)\bigr|\leq \A\frac{\log L}{L},\quad L\geq\L_0
  	\end{equation}
  	for every $\T>2$ and some $\A$, $L_0$, and $\rho_0$ independent of $L$ and $\rho$.
 
 We next show that 
  \begin{multline} \label{eq:diff_upp_bounds}
    	\sup_{0\leq\pow\leq\rho}\biggl\{\frac{\bar{\J}(\T,\pow)}{\T} - \sqrt{\frac{ \bar{\V}_\rho(\T,\pow)}{\L\T^2}}Q^{-1}(\epsilon)\biggr\}\\=\frac{\bar{\J}(\T,\rho)}{\T}-\sqrt{\frac{\bar{\V}(\T,\rho)}{\L\T^2}}Q^{-1}(\epsilon)+\K_{2}(L,\T,\rho)
    	  	\end{multline}
    	  	where $\K_{2}(L,\T,\rho)$ is a function of $\L$, $\T$, and $\rho$ that satisfies
    	  	\begin{equation}\label{eq:K2_bigO}
    	  	\sup_{\rho\geq\rho_0}\bigl|\K_{2}(L,\T,\rho)\bigr|\leq \frac{\A}{L},\quad L\geq\L_0
    	  	\end{equation}
    	  	for every $\T>2$ and some $\A$, $L_0$, and $\rho_0$ independent of $L$ and $\rho$.
  By using \eqref{eq:barJ} and \eqref{eq:barV}, we finally obtain the desired upper bound
  \begin{multline}\label{eq:MC_ub_final}
  	\R^*(\L,\T,\epsilon,\rho)\leq \frac{\underline{\I}(\T,\rho)+\K_{\bar{\J}}(\T,\rho)}{\T} \\
  	{}- \sqrt{\frac{\tilde{\U}(\T)+\K_{\bar{\V}}(\T,\rho)}{\L\T^2}}Q^{-1}(\epsilon)+\K_{\textnormal{MC}}(L,\T,\rho)
  	  	\end{multline}
  	  	where $\K_{\textnormal{MC}}(L,\T,\rho)$ is a function of $\L$, $\T$, and $\rho$ that satisfies
  	  	\begin{equation}
  	  	\sup_{\rho\geq\rho_0}\bigl|\K_{\textnormal{MC}}(L,\T,\rho)\bigr|\leq \A\frac{\log L}{L},\quad L\geq\L_0
  	  	\end{equation}
  	  	for every $\T>2$ and some $\A$, $L_0$, and $\rho_0$ independent of $L$ and $\rho$; and $\K_{\xi}(\T,\rho)$, $\xi=\{\bar{\J},\bar{\V}\}$ are functions of $\T$ and $\rho$ that satisfy
  	  	\begin{equation}
  	  	\lim_{\rho\to\infty}\K_\xi(\T,\rho) = 0, \quad \T>2.
  	  	\end{equation}
   
 To prove \eqref{eq:diff_upp_bounds}, we first present the following auxiliary results.
 
 \begin{lemma}\label{lm:opt_high_SNR_2}
 \begin{enumerate}
 \item Assume that $\T>2$. For sufficiently large $\rho$, we have
 \begin{equation}\label{eq:optimum_J}
 	 \sup_{0\leq\pow\leq\rho}{\bar{\J}(\T,\pow)}=\bar{\J}(\T,\rho).
 \end{equation}
 \item Assume that $\T>2$ and $0<\epsilon<\frac{1}{2}$. Consider the supremum on the left-hand side (LHS) of \eqref{eq:diff_upp_bounds}. For sufficiently large $\L$ and $\rho$, we can assume without loss of optimality that $\pow\in[\rho(1-\frac{\mathsf{K}(T)}{\L}),\rho]$ for some nonnegative constant $\mathsf{K}(\T)$ that is independent of $\L$, $\rho$, and $\pow$.
 \end{enumerate}
 \end{lemma}
 \begin{IEEEproof}
  See Appendix~\ref{app:proof_lm_opt_high_SNR_B}.
 \end{IEEEproof}
 
 We next set out to prove \eqref{eq:diff_upp_bounds}. By Part~2) of Lemma~\ref{lm:opt_high_SNR_2}, we can assume without loss of optimality that
 \begin{equation}\label{eq:optimal_alpha_start}
 	\pow\geq \rho\biggl(1-\frac{\mathsf{K}(\T)}{\L}\biggr).
 \end{equation}
Furthermore, we show in Appendix \ref{app:bound_U_alpha} that 
\begin{equation}
\bar{\V}_\rho(\T,\pow)\geq\bar{\V}(\T,\rho) - \Upsilon(\T)\delta, \quad \rho(1-\delta) \leq\pow\leq\rho
\end{equation}
where $\Upsilon(\T)$ is a positive constant that only depends on $\T$. Particularizing this bound for $\delta=\mathsf{K}(\T)/\L$, we obtain
\begin{equation}\label{eq:Tobi_optimal_alpha}
\bar{\V}_\rho(\T,\pow)\geq\bar{\V}(\T,\rho) - \Upsilon(\T)\frac{\mathsf{K}(\T)}{\L}, \quad \rho\biggl(1-\frac{\mathsf{K}(\T)}{\L}\biggr) \leq\pow\leq\rho.
\end{equation}
Combining \eqref{eq:Tobi_optimal_alpha} with Part~1) of Lemma~\ref{lm:opt_high_SNR_2}, and using that, by the assumption $0<\epsilon<\frac{1}{2}$ we have $Q^{-1}(\epsilon)>0$, we obtain
 \begin{IEEEeqnarray}{lCl}
 \IEEEeqnarraymulticol{3}{l}{
 	\sup_{0\leq\pow\leq\rho}\biggl\{\frac{\bar{\J}(\T,\pow)}{\L\T} - \sqrt{\frac{ \bar{\V}_\rho(\T,\pow)}{\L\T^2}}Q^{-1}(\epsilon)\biggr\}}\nonumber\\\qquad
 	&\leq&\frac{\bar{\J}(\T,\rho)}{\T}-\sqrt{\frac{\bar{\V}(\T,\rho)-\frac{\Upsilon(\T)\mathsf{K}(\T)}{\L}}{\L\T^2}}Q^{-1}(\epsilon)\IEEEnonumber\\
 	&=&\frac{\bar{\J}(\T,\rho)}{\T}-\sqrt{\frac{\bar{\V}(\T,\rho)}{\L\T^2}}Q^{-1}(\epsilon)
 +\K_{2}(L,\T,\rho)\label{eq:iter0_UB_rates}\IEEEeqnarraynumspace
  \end{IEEEeqnarray}
 where $\K_{2}(L,\T,\rho)$ is as in \eqref{eq:diff_upp_bounds}. This proves \eqref{eq:diff_upp_bounds} and concludes the proof of the upper bound.
 
\section{Conclusion}
\label{sec:conclusion}
We presented a high-SNR normal approximation for the maximum coding rate $\R^*(\L,\T,\epsilon,\rho)$ achievable over noncoherent, single-antenna, Rayleigh block-fading channels using an error-correcting code that spans $\L$ coherence intervals of length $\T$, has a block-error probability no larger than $\epsilon$, and satisfies the power constraint $\rho$. The high-SNR normal approximation is roughly equal to the normal approximation one obtains by transmitting one pilot symbol per coherence block to estimate the fading coefficient, and by then transmitting $\T-1$ symbols per coherence block over a coherent fading channel. This suggests the heuristic that, at high SNR, one pilot symbol per coherence block should be transmitted to achieve both the capacity and the channel dispersion. While the approximation was derived under the assumption that the number of coherence intervals $\L$ and the SNR $\rho$ tend to infinity, numerical analyses suggest that it becomes accurate already at SNR values of $15$~dB and for $10$ coherence intervals or more.

The obtained normal approximation is useful in two ways. First, it complements the nonasymptotic bounds provided in \cite{finte_block_length_2012, Ostman_2014, Short_Packets_Durisi_Koch_TCOM_2015}, whose evaluation is computationally demanding. Second, it lays the foundation for analytical studies that analyze the behavior of the maximum coding rate as a function of system parameters such as SNR, number of coherence intervals, or blocklength. An example of such a study was illustrated in Section~\ref{sub:wisdom} concerning the optimal design of a simple slotted-ALOHA protocol.
\appendices

\section{Proof of Lemma~\ref{lm:inc_gamma}}\label{app:inc_gamma}
The left-most inequality in \eqref{eq:part1_inc_gamma} follows because the regularized lower incomplete gamma function is no larger than~$1$. For the right-most inequality in \eqref{eq:part1_inc_gamma}, consider the following bound by Alzer \cite[Th. 1]{Alzer} (see also \cite[Eq.~(5.4)]{Gautschi})
\begin{IEEEeqnarray}{lCl}\label{eq:Alzer}
 		\tilde{\gamma}(a,x) & > & \bigl(1-e^{-s_a x}\bigr)^a, \quad (x\geq 0, \, a>0, \, a\neq1) \IEEEeqnarraynumspace
 	\end{IEEEeqnarray}
where
\begin{equation}
	s_a
	= \Biggl\{
 	\begin{array}{ll}
 	1, & \text{ if } 0<a<1\\
 	\Gamma(a+1)^{-\frac{1}{a}}, & \text{ if } a>1. \end{array} 
\end{equation}
In order to obtain the right-most inequality in \eqref{eq:part1_inc_gamma}, we first lower-bound $\tilde{\gamma}(\cdot,\cdot)$ using \eqref{eq:Alzer}
\begin{IEEEeqnarray}{lCl}
 	\log\frac{1}{\tilde{\gamma}(\T-1,x)} & \leq & (\T-1)
 	\log\Biggl(\frac{1}{1-e^{-x\Gamma(\T)^{-\frac{1}{\T-1}}}}\Biggr) \nonumber\\
	& = & (\T-1)\log\Biggl(1+\frac{1}{e^{x\Gamma(\T)^{-\frac{1}{\T-1}}}-1}\Biggr) \IEEEeqnarraynumspace
\end{IEEEeqnarray}
where the second step follows by simple algebraic manipulations. Since $e^z\geq1+z$, this can be further upper-bounded as	
\begin{IEEEeqnarray}{lCl}
	\log\frac{1}{\tilde{\gamma}(\T-1,x)} &\leq& (\T-1) \log\Biggl(1+\frac{\Gamma(\T)^{\frac{1}{\T-1}}}{x}\Biggr).\IEEEeqnarraynumspace
\end{IEEEeqnarray}
This proves Lemma~\ref{lm:inc_gamma}.

\section{Proof of Lemma \ref{lm:var_bounded_away0}}\label{app:bounded_away0}
To lower-bound $\bar{\V}_\rho(\T,\pow)$, we begin by lower-bounding
\begin{IEEEeqnarray}{lCl}
\IEEEeqnarraymulticol{3}{l}{\bigl(\bar{j}_{\ell}(\T,\pow)-\bar{\J}(\T,\pow)\bigr)^2} \IEEEnonumber*\\
\quad & = & \bigg(-\frac{\T\rho-\T\pow}{1+\T\rho}(Z_1-1)-\frac{\T\rho}{1+\T\rho}(Z_2-(\T-1))\\
	&&{}+(\T-1)\log\bigl((1+\T\pow)Z_1+Z_2+\beta(\T,\rho)\bigr)\\
	&&{}-(\T-1)\mathsf{E}\bigl[\log\bigl((1+\T\pow)Z_1+Z_2+\beta(\T,\rho)\bigr)\bigr]\biggr)^2 \\
	& \geq & \biggl(\frac{\T\rho-\T\pow}{1+\T\rho}(Z_1-1)+\frac{\T\rho}{1+\T\rho}(Z_2-(\T-1))\biggr)^2 \\
	& & {} -2 \biggl(\frac{\T\rho-\T\pow}{1+\T\rho}(Z_1-1)+\frac{\T\rho}{1+\T\rho}(Z_2-(\T-1))\biggr) \\
	& & \quad {}\times\bigg((\T-1)\log\bigl((1+\T\pow)Z_1+Z_2\bigr)\\
	& & \quad\qquad {} -(\T-1)\mathsf{E}\bigl[\log\bigl((1+\T\pow)Z_1+Z_2\bigr)\bigr]\\
	& &\quad\qquad {}+(\T-1)\log\biggl(1+\frac{\beta(\T,\rho)}{(1+\T\pow)Z_1+Z_2}\biggr)\\
	& & \quad\qquad {} -(\T-1)\mathsf{E}\biggl[\log\biggl(1+\frac{\beta(\T,\rho)}{(1+\T\pow)Z_1+Z_2}\biggr)\biggr]\bigg). \\\IEEEyesnumber\label{eq:this_is_my_bound}
\end{IEEEeqnarray}
We next note that
\begin{IEEEeqnarray}{rCl}
	\mathsf{E}\biggl[(Z_1-1)\log\biggl(1+\frac{\beta(\T,\rho)}{(1+\T\pow)Z_1+Z_2}\biggr)\biggr]&\leq& 0\label{eq:first_cov_term}\\
	\mathsf{E}\biggl[(Z_2-(\T-1))\log\biggl(1+\frac{\beta(\T,\rho)}{(1+\T\pow)Z_1+Z_2}\biggr)\biggr]&\leq& 0. \IEEEeqnarraynumspace\label{eq:sec_cov_term}
\end{IEEEeqnarray}
This follow from the inequalities
		\begin{IEEEeqnarray}{lCl}
		\IEEEeqnarraymulticol{3}{l}{(Z_1-1)\log\biggl(1+\frac{\beta(\T,\rho)}{(1+\T\pow)Z_1+Z_2}\biggr)} \nonumber\\
		\quad & \leq & (Z_1-1)\log\biggl(1+\frac{\beta(\T,\rho)}{(1+\T\pow)+Z_2}\biggr)
		\end{IEEEeqnarray}
		and
		\begin{IEEEeqnarray}{lCl}
	\IEEEeqnarraymulticol{3}{l}{(Z_2-(\T-1))\log\biggl(1+\frac{\beta(\T,\rho)}{(1+\T\pow)Z_1+Z_2}\biggr)}\nonumber\\
	\,\,\, & \leq & (Z_2-(\T-1))\log\biggl(1+\frac{\beta(\T,\rho)}{(1+\T\pow)Z_1+(\T-1)}\biggr) \IEEEeqnarraynumspace
		  \end{IEEEeqnarray}
whose RHSs are zero mean because $Z_1$ and $Z_2$ are independent and have mean $1$ and $T-1$, respectively.

Computing the expected value of \eqref{eq:this_is_my_bound}, and using \eqref{eq:first_cov_term} and~\eqref{eq:sec_cov_term}, we can lower-bounded $\bar{\V}_\rho(\T,\pow)$ as
\begin{IEEEeqnarray}{lCl}
\IEEEeqnarraymulticol{3}{l}{
	\bar{\V}_\rho(\T,\pow)}\IEEEnonumber*\\
	\quad & \triangleq & \mathsf{E}\bigl[\bigl(\bar{j}_{\ell}(\T,\pow)-\bar{\J}(\T,\pow)\bigr)^2\bigr] \\
	&\geq&\biggl(\frac{\T\rho-\T\pow}{1+\T\rho}\biggr)^2+\biggl(\frac{\T\rho}{1+\T\rho}\biggr)^2(\T-1)\\
	&&{}-\kappa_{\T,\rho} \left(1-\frac{\pow}{\rho}\right)\mathsf{E}\biggl[(Z_1-1)\log\biggl(Z_1+\frac{Z_2}{1+\T\pow}\biggr)\biggr]\\
	&&{}-\kappa_{\T,\rho}\mathsf{E}\biggl[(Z_2-(\T-1))\log\biggl(Z_1+\frac{Z_2}{1+\T\rho}\biggr)\biggr]\\
	& & {} + \kappa_{\T,\rho}\mathsf{E}\biggl[\bigl(Z_2-(\T-1)\bigr)\log\biggl(\frac{(1+\T\rho)Z_1+Z_2}{(1+\T\pow)Z_1+Z_2}\biggr)\biggr]\\\IEEEyesnumber\label{eq:low_var_first_app_low_a}
\end{IEEEeqnarray}
where $\kappa_{\T,\rho}\triangleq 2(\T-1)\T\rho/(1+\T\rho)$.

The first term on the RHS of \eqref{eq:low_var_first_app_low_a} is nonnegative, so discarding it yields a lower bound. The third term on the RHS of \eqref{eq:low_var_first_app_low_a} can be lower-bounded by upper-bounding 
\begin{IEEEeqnarray}{lCl}
	\IEEEeqnarraymulticol{3}{l}{\kappa_{\T,\rho} \left(1-\frac{\pow}{\rho}\right)\mathsf{E}\biggl[(Z_1-1)\log\biggl(Z_1+\frac{Z_2}{1+\T\pow}\biggr)\biggr]}\IEEEnonumber*\\
 &\leq&\kappa_{\T,\rho} \left(1-\frac{\pow}{\rho}\right)\sqrt{\mathsf{E}\bigl[(Z_1-1)^2\bigr]\mathsf{E}\biggl[\log^2\biggl(Z_1+\frac{Z_2}{1+\T\pow}\biggr)\biggr]}\\
	&\leq&2(\T-1)\delta\sqrt{\biggl(\frac{\pi^2}{6}+\gamma^2+\psi^2(\T)+\zeta(2,\T)\biggr)}\IEEEyesnumber\IEEEeqnarraynumspace\label{eq:Tobi_bounded_away_first}
\end{IEEEeqnarray}
for $\rho(1-\delta)\leq\pow\leq\rho$. Here, the first inequality follows from the Cauchy-Schwarz inequality, and the last inequality follows because $\kappa_{\T,\rho}\leq 2(\T-1)$, because $Z_1$ has variance $1$, and because
\begin{IEEEeqnarray}{lCl}
\IEEEeqnarraymulticol{3}{l}{
	\mathsf{E}\biggl[\log^2\biggl(Z_1+\frac{Z_2}{1+\T\pow}\biggr)\biggr]}\nonumber\\
	\quad	&\leq&\mathsf{E}\bigl[\log^2\bigl(Z_1+Z_2\bigr)\bigr]+\mathsf{E}\bigl[\log^2(Z_1)\bigr]\IEEEnonumber*\\
	&=&\frac{\pi^2}{6}+\gamma^2+\zeta(2,\T)+\psi^2(\T)\IEEEyesnumber\IEEEeqnarraynumspace\label{eq:log_square_bounded_away}
\end{IEEEeqnarray}
where we have evaluated the expected values using \cite[Sec. 4.335-1]{table_integrals} and \cite[Sec. 4.358-2]{table_integrals}, respectively. The first inequality in \eqref{eq:log_square_bounded_away} follows by treating the cases $Z_1+Z_2/(1+T\pow)\leq 1$ and $Z_1+Z_2/(1+T\pow)>1$ separately, and by lower-bounding in the former case $Z_1+Z_2/(1+T\pow)$ by $Z_1$ and upper-bounding in the latter case $Z_1+Z_2/(1+T\pow)$ by $Z_1+Z_2$. Hence
\begin{IEEEeqnarray}{lCl}
\log^2\left(Z_1+\frac{Z_2}{1+T\pow}\right) & \leq & \log^2(Z_1)\nonumber\\
& \leq& \log^2(Z_1)+\log^2(Z_1+Z_2) \IEEEyesnumber\IEEEeqnarraynumspace
\end{IEEEeqnarray}
if $Z_1+Z_2/(1+T\pow)\leq 1$, and
\begin{IEEEeqnarray}{lCl}
\log^2\left(Z_1+\frac{Z_2}{1+T\pow}\right) & \leq & \log^2(Z_1+Z_2)\nonumber\\
 & \leq & \log^2(Z_1)+\log^2(Z_1+Z_2)\IEEEeqnarraynumspace
\end{IEEEeqnarray}
if $Z_1+Z_2/(1+T\pow)> 1$, which yields the desired bound.

The fifth term on the RHS \eqref{eq:low_var_first_app_low_a} can be lower-bounded by upper-bounding
\begin{IEEEeqnarray}{lCl}
\IEEEeqnarraymulticol{3}{l}{
	\biggl|\mathsf{E}\biggl[(Z_2-(\T-1))\log\biggl(\frac{(1+\T\rho)Z_1+Z_2}{(1+\T\pow)Z_1+Z_2}\biggr)\biggr]\biggr|}\nonumber\\
\quad &\leq& \mathsf{E}\biggl[\bigl|Z_2-(\T-1)\bigr|\log\biggl(\frac{(1+\T\rho)Z_1+Z_2}{(1+\T\pow)Z_1+Z_2}\biggr)\biggr]\IEEEnonumber*\\
	&\leq&\mathsf{E}\bigl[|Z_2-(\T-1)|\bigr]\log\Bigl(\frac{\rho}{\pow}\Bigr)\\
	&\leq&\mathsf{E}\bigl[|Z_2-(\T-1)|\bigr]\log\biggl(\frac{1}{1-\delta}\biggr)\IEEEyesnumber\label{eq:Tobi_bounded_away_last}
\end{IEEEeqnarray}
for $\rho(1-\delta)\leq\pow\leq\rho$. Combining \eqref{eq:Tobi_bounded_away_first}--\eqref{eq:Tobi_bounded_away_last} with \eqref{eq:low_var_first_app_low_a}, and upper-bounding $\kappa_{\T,\rho}$ by $2(\T-1)$, we obtain the lower bound
\begin{IEEEeqnarray}{lCl}
	\IEEEeqnarraymulticol{3}{l}{\bar{\V}_\rho(\T,\pow)} \nonumber\\
	\quad & \geq & \biggl(\frac{\T\rho}{1+\T\rho}\biggr)^2(\T-1)\nonumber\\
	&&{} -2(\T-1)\delta\sqrt{\biggl(\frac{\pi^2}{6}+\gamma^2+\psi^2(\T)+\zeta(2,\T)\biggr)}\nonumber\\
	&&{}-\kappa_{\T,\rho}\mathsf{E}\biggl[\bigl(Z_2-(\T-1)\bigr)\log\biggl(Z_1+\frac{Z_2}{1+\T\rho}\biggr)\biggr] \nonumber\\
	&&{} - 2(\T-1)\mathsf{E}\bigl[|Z_2-(\T-1)|\bigr]\log\biggl(\frac{1}{1-\delta}\biggr).\IEEEyesnumber\IEEEeqnarraynumspace\label{eq:low_var_first_app_low_b}
\end{IEEEeqnarray}
Only the second and fourth term on the RHS of \eqref{eq:low_var_first_app_low_b} depend on $\delta$. The former term is linear in $\delta$, the latter term can be upper-bounded by a linear term by using that, for $0\leq\delta\leq1/2$,
\begin{equation}
\label{eq:watchawaiting}
\log\biggl(\frac{1}{1-\delta}\biggr) \leq \frac{\delta}{1-\delta} \leq 2\delta.
\end{equation}
Hence, there exists a positive constant $\Xi(\T)$ that only depends on $\T$ such that
\begin{multline}
\bar{\V}_\rho(\T,\pow)\geq\biggl(\frac{\T\rho}{1+\T\rho}\biggr)^2 (\T-1) - \Xi(\T) \delta \\ -\kappa_{\T,\rho}\mathsf{E}\biggl[\bigl(Z_2-(\T-1)\bigr)\log\biggl(Z_1+\frac{Z_2}{1+\T\rho}\biggr)\biggr].
\end{multline}

We conclude the proof of Lemma~\ref{lm:var_bounded_away0} by demonstrating that,  for every $\T$,
\begin{IEEEeqnarray}{lCl}\label{eq:DCT_var_bounded_away_alpha_final}
	\lim_{\rho\to\infty}\mathsf{E}\biggl[(Z_2-(\T-1))\log\biggl(Z_1+\frac{Z_2}{1+\T\rho}\biggr)\biggr]=0.\IEEEeqnarraynumspace
\end{IEEEeqnarray}
This is a direct consequence of the dominated convergence theorem \cite[Section 1.26]{Rudin}, which can be applied because
\begin{IEEEeqnarray}{lCl}
	\IEEEeqnarraymulticol{3}{l}{\biggl|(Z_2-(\T-1))\log\biggl(Z_1+\frac{Z_2}{1+\T\rho}\biggr)\biggr|}\nonumber\\\quad
	&\leq&\bigl|(Z_2-(\T-1))\bigr|\sqrt{\log^2(Z_1)+\log^2(Z_1+Z_2)}\label{eq:upper_bound_Z2-T_log}\IEEEeqnarraynumspace
\end{IEEEeqnarray}
which follows from the same steps as the first inequality in \eqref{eq:log_square_bounded_away}. Using the Cauchy-Schwarz inequality, the expected value of the RHS of \eqref{eq:upper_bound_Z2-T_log} can be upper-bounded by
\begin{equation*}
\sqrt{\mathsf{E}\Bigl[(Z_2-(\T-1))^2\Bigr]\mathsf{E}\Bigl[\log^2(Z_1)+\log^2(Z_1+Z_2)\Bigr]}
\end{equation*}
which is finite by \eqref{eq:log_square_bounded_away} and because $Z_2$ has finite variance.

\section{Proof of Lemma \ref{lm:var_bounded}}\label{app:second_moment}
We shall first prove \eqref{eq:bar_var_bounded}. Using the definitions of $\bar{j}_{\ell}(\T,\pow)$ and $\bar{\J}(\T,\pow)$ in \eqref{eq:j_upper_def} and \eqref{eq:J_def}, respectively, we upper-bound $\bar{\V}_\rho(\T,\pow)$ as
\begin{IEEEeqnarray}{lCl}
\IEEEeqnarraymulticol{3}{l}{\bar{\V}_\rho(\T,\pow)} \nonumber\\
\quad & = & \mathsf{E}\Biggl[\Biggl(\frac{\T\rho-\T\pow}{1+\T\rho}(1-Z_1)+\frac{\T\rho}{1\,+\T\rho}(\T-1-Z_2) \nonumber\\
& & \qquad {} + (\T-1) \log\biggl(Z_1+\frac{Z_2}{1+\T\pow}\biggr)\nonumber\\
&& \qquad {} - (\T-1) \mathsf{E}\biggl[\log\biggl(Z_1+\frac{Z_2}{1+\T\pow}\biggr)\biggr]\nonumber\\
& &  \qquad {} + (\T-1) \log\biggl(1+\frac{\beta(\T,\rho)}{(1+\T\pow)Z_1+Z_2}\biggr) \nonumber\\
&& \qquad {} - (\T-1) \mathsf{E}\biggl[\log\biggl(1+\frac{\beta(\T,\rho)}{(1+\T\pow)Z_1+Z_2}\biggr)\biggr]\Biggr)^2\Biggr] \nonumber\\
& \leq & \ceta_{4,2}\biggl(\frac{\T\rho-\T\pow}{1+\T\rho}\biggr)^2\mathsf{E}\bigl[(Z_1-1)^2\bigr]\nonumber\\
&& {} + \ceta_{4,2}\biggl(\frac{\T\rho}{1\,+\T\rho}\biggr)^2 \mathsf{E}\bigl[(Z_2-\T+1)^2\bigr] \nonumber\\
&&{} + \ceta_{4,2}(\T-1)^2\mathsf{E}\biggl[\log^2\biggl(Z_1+\frac{Z_2}{1+\T\pow}\biggr)\biggr]\nonumber\\
& & {}+ \ceta_{4,2}(\T-1)^2 \mathsf{E}\biggl[\log^2\biggl(1+\frac{\beta(\T,\rho)}{(1+\T\pow)Z_1+Z_2}\biggr)\biggr]\IEEEeqnarraynumspace \label{eq:var_bound_def_V_bar_rho}
\end{IEEEeqnarray}
where we have used that  
\begin{equation}\label{eq:c_terms_2}
|a_1+ \dots + a_\eta|^\nu\leq \ceta_{\eta,\nu}(|a_1|^\nu + \dots + |a_\eta|^\nu), \quad \eta, \nu \in\mathbb{Z}^+
\end{equation} 
for some positive constant $\ceta_{\eta,\nu}$ that only depends on $\eta$ and $\nu$, and that $\mathsf{E}\bigl[(X-\mathsf{E}[X])^2\bigr]\leq\mathsf{E}\bigl[X^2\bigr]$ for every random variable~$X$.

We next show that each term on the RHS of \eqref{eq:var_bound_def_V_bar_rho} is bounded in $\rho$ and $\alpha$. Indeed, we have $\mathsf{E}\bigl[(Z_1-1)^2\bigr]=1$ and $\mathsf{E}\bigl[(Z_2-(\T-1))^2\bigr] = (\T-1)$. Furthermore, since $0\leq(\T\rho-\T\pow)/(1+\T\rho)\leq 1$ and $0\leq \T\rho/(1\,+\T\rho)\leq 1$, the first two terms on the RHS of \eqref{eq:var_bound_def_V_bar_rho} are bounded in $\rho$ and $\alpha$. The third term on the RHS of \eqref{eq:var_bound_def_V_bar_rho} can be upper-bounded by (see \eqref{eq:log_square_bounded_away})
\begin{multline}
(\T-1)^2\mathsf{E}\biggl[\log^2\biggl(Z_1+\frac{Z_2}{1+\T\pow}\biggr)\biggr]\\ \leq (\T-1)^2 \mathsf{E}\bigl[\log^2(Z_1+Z_2)\bigr]+(\T-1)^2 \mathsf{E}\bigl[\log^2(Z_1)\bigr].
\label{eq:var_bound_sec_Bbar_first}
\end{multline}
Finally, for every $\rho_0>0$ and $\rho\geq\rho_0$, the fourth term on the RHS of \eqref{eq:var_bound_def_V_bar_rho} can be upper-bounded by
\begin{IEEEeqnarray}{lCl}
\IEEEeqnarraymulticol{3}{l}{
	\mathsf{E}\biggl[(\T-1)^2\log^2\biggl(1+\frac{\beta(\T,\rho)}{(1+\T\pow)Z_1+Z_2}\biggr)\biggr]}\nonumber\\\quad
	 &\leq & (\T-1)^2\mathsf{E}\biggl[\log^2\biggl(1+\frac{\beta(\T,\rho)}{Z_1+Z_2}\biggr)\biggr] \nonumber\\
	& \leq & (\T-1)^2\mathsf{E}\biggl[\log^2\biggl(1+\frac{\beta(\T,\rho_0)}{Z_1+Z_2}\biggr)\biggr] \label{eq:log_square_gamma_b2_bar}\end{IEEEeqnarray}		
where the second inequality follows because $\rho\mapsto\beta(\T,\rho)$ is monotonically decreasing in $\rho$. Since the RHSs of \eqref{eq:var_bound_sec_Bbar_first} and \eqref{eq:log_square_gamma_b2_bar} are finite, this proves \eqref{eq:bar_var_bounded}.

The proof of \eqref{eq:var_bounded} follows along similar lines. Indeed, using the definitions of $i_{\ell}(\T,\rho)$ and $\I(\T,\rho)$ in \eqref{eq:i_def_2} and \eqref{eq:I_def}, respectively, we can upper-bound $\U(\T,\rho)$ as
\begin{IEEEeqnarray}{lCl}
\IEEEeqnarraymulticol{3}{l}{\U(\T,\rho)} \nonumber\\
\quad & = & \mathsf{E}\Biggl[\Biggl(\frac{\T\rho}{1\,+\T\rho}(\T-1-Z_2)\nonumber\\
&&\qquad {} + (\T-1) \log\biggl(Z_1+\frac{Z_2}{1+\T\rho}\biggr)\nonumber\\
&& \qquad {}- (\T-1) \mathsf{E}\biggl[\log\biggl(Z_1+\frac{Z_2}{1+\T\rho}\biggr)\biggr] \nonumber\\
& & \qquad {} - \log\tilde{\gamma}\biggl(\T-1,\frac{\T\rho((1+\T\rho)Z_1+Z_2)}{1+\T\rho}\biggr) \nonumber\\
&& \qquad {}+ \mathsf{E}\biggl[\log\tilde{\gamma}\biggl(\T-1,\frac{\T\rho((1+\T\rho)Z_1+Z_2)}{1+\T\rho}\biggr)\biggr]\Biggr)^2\Biggr] \nonumber\\
& \leq & \ceta_{3,2}\biggl(\frac{\T\rho}{1\,+\T\rho}\biggr)^2\mathsf{E}\bigl[(Z_2-\T+1)^2\bigr]\nonumber\\
&& {} + \ceta_{3,2}(\T-1)^2 \mathsf{E}\biggl[\log^2\biggl(Z_1+\frac{Z_2}{1+\T\rho}\biggr)\biggr]\nonumber\\
&&{}+ \ceta_{3,2}\mathsf{E}\biggl[\log^2\tilde{\gamma}\biggl(\T-1,\frac{\T\rho((1+\T\rho)Z_1+Z_2)}{1+\T\rho}\biggr)\biggr].  \nonumber\\ \label{eq:var_bound_def_U_rho}
\end{IEEEeqnarray}

We next show that each summand is bounded in $\rho$. Indeed, as shown before, the first and the second term on the RHS of \eqref{eq:var_bound_def_U_rho} are bounded in $\rho$. As for the third term on the RHS of \eqref{eq:var_bound_def_U_rho}, we use Lemma \ref{lm:inc_gamma} to obtain
\begin{multline}
\mathsf{E}\biggl[\log^2\tilde{\gamma}\biggl(\T-1,\frac{\T\rho((1+\T\rho)Z_1+Z_2)}{1+\T\rho}\biggr)\biggr]\\ \leq (\T-1)^2\mathsf{E}\biggl[\log^2\biggl(1+\frac{\beta(\T,\rho)}{(1+\T\rho)Z_1+Z_2}\biggr)\biggr].\label{eq:loggamma_app2_2}
\end{multline}
By the monotonicity of $\rho\mapsto \beta(\T,\rho)$, it follows that for every $\rho_0>0$ and $\rho\geq\rho_0$, the third term on the RHS of \eqref{eq:var_bound_def_U_rho} is upper-bounded by
\begin{multline}
\mathsf{E}\biggl[\log^2\tilde{\gamma}\biggl(\T-1,\frac{\T\rho((1+\T\rho)Z_1+Z_2)}{1+\T\rho}\biggr)\biggr]\\ \leq (\T-1)^2\mathsf{E}\biggl[\log^2\biggl(1+\frac{\beta(\T,\rho_0)}{Z_1+Z_2}\biggr)\biggr].\label{eq:loggamma_app2_3}
\end{multline}
Combining the above steps with \eqref{eq:var_bound_def_U_rho} yields \eqref{eq:var_bounded}.

\section{Proof of Lemma \ref{lm:third_moment_bounded}}\label{app:third_moment}
We shall first prove \eqref{eq:bar_third_bounded}. Using the definitions of $\bar{j}_{\ell}(\T,\pow)$ and $\bar{\J}(\T,\pow)$ in \eqref{eq:j_upper_def} and \eqref{eq:J_def}, respectively, we can upper-bound the third moment of $\bar{j}_{\ell}(\T,\pow)$ as
\begin{IEEEeqnarray}{lCl}
\IEEEeqnarraymulticol{3}{l}{
	\mathsf{E}\Bigl[\bigl|\bar{j}_{\ell}(\T,\pow)-\bar{\J}(\T,\pow)\bigr|^3\Bigr]}\nonumber\\\quad
	&=&\mathsf{E}\Biggl[\biggl|\frac{\T\rho-\T\pow}{1+\T\rho}(1-Z_1)+\frac{\T\rho}{1+\T\rho}(\T-1-Z_2) \nonumber\\
	& & \qquad {} +(\T-1)\log\biggl(Z_1+\frac{Z_2}{1+\T\pow}\biggr)\nonumber\\
	&&\qquad{}-(\T-1)\mathsf{E}\biggl[\log\biggl(Z_1+\frac{Z_2}{1+\T\pow}\biggr)\biggr] \nonumber\\
	& & \qquad {} -(\T-1)\log\biggl(1+\frac{\beta(\T,\rho)}{(1+\T\pow)Z_1+Z_2}\biggr) \nonumber\\
	& & \qquad {} +(\T-1)\mathsf{E}\biggl[\log\biggl(1+\frac{\beta(\T,\rho)}{(1+\T\pow)Z_1+Z_2}\biggr)\biggr]\biggr|^3\Biggr] \nonumber\\
		& \leq & \ceta_{6,3}\biggl|\frac{\T\rho-\T\pow}{1+\T\rho}\biggr|^3 \mathsf{E}\bigl[|Z_1-1|^3\bigr]\nonumber\\
		&&{} + \ceta_{6,3}\biggl|\frac{\T\rho}{1+\T\rho}\biggr|^3 \mathsf{E}\bigl[|Z_2-\T+1|^3\bigr] \nonumber\\
		&& {} +2\ceta_{6,3}(\T-1)^3\mathsf{E}\Biggl[\biggl|\log\biggl(Z_1+\frac{Z_2}{1+\T\pow}\biggr)\biggr|^3\Biggr] \nonumber\\
		& &  {} + 2\ceta_{6,3}(\T-1)^3\mathsf{E}\Biggl[\biggl|\log\biggl(1+\frac{\beta(\T,\rho)}{(1+\T\pow)Z_1+Z_2}\biggr)\biggr|^3\Biggr] \nonumber\\ 
		 \label{eq:bar_third_terms_result}\label{eq:bar_third_terms}
\end{IEEEeqnarray}
where we have used \eqref{eq:c_terms_2} and that $\mathsf{E}[|X|^3]\geq |\mathsf{E}[X]|^3$ for every random variable $X$.

We next show that each term on the RHS of \eqref{eq:bar_third_terms_result} is bounded in $\rho$ and $\alpha$. Indeed, the first two terms on the RHS of \eqref{eq:bar_third_terms_result} are bounded because the third central moments of the Gamma-distributed random variables $Z_1$ and $Z_2$ are bounded, and because $0\leq(\T\rho-\T\pow)/(1+\T\rho)\leq 1$ and $0\leq \T\rho/(1\,+\T\rho)\leq 1$.
The third term on the RHS of \eqref{eq:bar_third_terms_result} can be upper-bounded by using that
\begin{equation}
\label{eq:111}
\biggl|\log\biggl(Z_1+\frac{Z_2}{1+\T\pow}\biggr)\biggr|\leq|\log Z_1|+|\log(Z_1+Z_2)|
\end{equation}
which follows from similar steps as the first inequality in \eqref{eq:log_square_bounded_away}. Hence, by \eqref{eq:c_terms_2}
\begin{multline}
\label{expectation_sec_B_third_ineq}
	\mathsf{E}\Biggl[\biggl|\log\biggl(Z_1+\frac{Z_2}{1+\T\pow}\biggr)\biggr|^3\Biggr]\\\leq c_{2,3}\bigl(\mathsf{E}\bigl[|\log Z_1|^3\bigr]+\mathsf{E}\bigl[|\log(Z_1+Z_2)|^3\bigr]\bigr).
\end{multline}
Finally, the fourth term on the RHS of \eqref{eq:bar_third_terms_result} can be upper-bounded as
\begin{IEEEeqnarray}{lCl}
\IEEEeqnarraymulticol{3}{l}{(\T-1)^3\mathsf{E}\Biggl[\biggl|\log\biggl(1+\frac{\beta(\T,\rho)}{(1+\T\pow)Z_1+Z_2}\biggr)\biggr|^3\Biggr]} \nonumber \\
\quad & \leq & (\T-1)^3\mathsf{E}\Biggl[\biggl|\log\biggl(1+\frac{\beta(\T,\rho)}{Z_1+Z_2}\biggr)\biggr|^3\Biggr] \nonumber\\
& \leq & (\T-1)^3\mathsf{E}\Biggl[\biggl|\log\biggl(1+\frac{\beta(\T,\rho_0)}{Z_1+Z_2}\biggr)\biggr|^3\Biggr] \label{eq:blablabla}
\end{IEEEeqnarray}
for every $\rho_0>0$ and $\rho\geq\rho_0$. Here, the second inequality follows from the monotonicity of $\rho\mapsto \beta(\T,\rho)$. Since the RHSs of \eqref{expectation_sec_B_third_ineq} and \eqref{eq:blablabla} are finite, this proves \eqref{eq:bar_third_bounded}.

We establish \eqref{eq:third_bounded} along similar lines. Using the definitions of $i_{\ell}(\T,\rho)$ and $\I(\T,\rho)$ in \eqref{eq:i_def_2} and \eqref{eq:I_def}, respectively, we can upper-bound the third moment of $i_{\ell}(\T,\rho)$ as
\begin{IEEEeqnarray}{lCl}
	\IEEEeqnarraymulticol{3}{l}{\mathsf{E}\Bigl[\bigl|i_{\ell}(\T,\rho)-\I(\T,\rho)\bigr|^3\Bigr]}\nonumber\\
\,\,	 & = & \mathsf{E}\Biggl[\biggl|\frac{\T\rho}{1+\T\rho}(\T-1-Z_2)+(\T-1)\log\biggl(Z_1+\frac{Z_2}{1+\T\rho}\biggr)\nonumber\\
	&&{} \qquad-(\T-1)\mathsf{E}\biggl[\log\biggl(Z_1+\frac{Z_2}{1+\T\rho}\biggr)\biggr]\nonumber\\
		&& \qquad {}-\log\tilde{\gamma}\biggl(\T-1,\frac{\T\rho((1+\T\rho)Z_1+Z_2)}{1+\T\rho}\biggr)\nonumber\\
		&&\qquad{}+\mathsf{E}\biggl[\log\tilde{\gamma}\biggl(\T-1,\frac{\T\rho((1+\T\rho)Z_1+Z_2)}{1+\T\rho}\biggr)\biggr]\biggr|^3\Biggr]\nonumber\\
		&\leq & \ceta_{5,3} \biggl|\frac{\T\rho}{1+\T\rho}\biggr|^3 \mathsf{E}\bigl[|Z_2-\T+1|^3\bigr] \nonumber\\
		& & + 2\ceta_{5,3}(\T-1)^3 \mathsf{E}\Biggl[\biggl|\log\biggl(Z_1+\frac{Z_2}{1+\T\rho}\biggr)\biggr|^3\Biggr]\nonumber\\ 
		&&{} + 2\ceta_{5,3} \mathsf{E}\Biggl[\biggl|\log\tilde{\gamma}\biggl(\T-1,\frac{\T\rho((1+\T\rho)Z_1+Z_2)}{1+\T\rho}\biggr)\biggr|^3\Biggr]\nonumber\\ \label{eq:third_terms}
\end{IEEEeqnarray}
where we have used \eqref{eq:c_terms_2} and that $\mathsf{E}[|X|^3]\geq |\mathsf{E}[X]|^3$ for every random variable $X$.

As shown before, the first two terms on the RHS of \eqref{eq:third_terms} are bounded in $\rho$. As for the third term, we first use Lemma~\ref{lm:inc_gamma} to obtain
\begin{multline}
\mathsf{E}\Biggl[\biggl|\log\tilde{\gamma}\biggl(\T-1,\frac{\T\rho((1+\T\rho)Z_1+Z_2)}{1+\T\rho}\biggr)\biggr|^3\Biggr] \\ \leq (\T-1)^3 \mathsf{E}\biggl[\log^3\biggl(1+\frac{\beta(\T,\rho)}{(1+\T\rho)Z_1+Z_2}\biggr)\biggr].
\end{multline}
By the monotonicity of $\rho\mapsto \beta(\T,\rho)$, it follows that for every $\rho_0>0$ and $\rho\geq\rho_0$, the third term on the RHS of \eqref{eq:third_terms} is upper-bounded by
\begin{multline}
\mathsf{E}\Biggl[\biggl|\log\tilde{\gamma}\biggl(\T-1,\frac{\T\rho((1+\T\rho)Z_1+Z_2)}{1+\T\rho}\biggr)\biggr|^3\Biggr] \\ \leq (\T-1)^3 \mathsf{E}\biggl[\log^3\biggl(1+\frac{\beta(\T,\rho_0)}{(1+\T\rho)Z_1+Z_2}\biggr)\biggr]. \label{eq:heybaby}
\end{multline} 
Since the RHS of \eqref{eq:heybaby} is finite, this proves \eqref{eq:third_bounded}.

\section{Proof of Lemma \ref{lm:opt_high_SNR}}\label{app:proof_lm_opt_high_SNR_A}
Consider the upper bound  \eqref{eq:MC}, namely,
\begin{equation}
 \label{eq:MC_app}
 \R^*(\L,\T,\epsilon,\rho) \leq \sup_{\boldsymbol{\pow}\in[0,\rho]^\L}\log\Biggl(\frac{1}{\beta(\boldsymbol{\pow},\text{q}_{\mathbf{Y}^\L}^{(\text{U})})}\Biggr).
 \end{equation}
 In the following, we show that, for sufficiently large $\L$ and $\rho$, we can assume without loss of optimality that $\boldsymbol{\pow}\in\mathcal{A}_{\rho,\delta}$. To this end, we demonstrate that for all $\boldsymbol{\pow}\notin\mathcal{A}_{\rho,\delta}$ and sufficiently large $\L$ and $\rho$, we can find a lower bound on $\R^*(\L,\T,\epsilon,\rho)$ that exceeds an upper bound on \eqref{eq:MC_app}. Hence, such $\boldsymbol{\pow}$ cannot be optimal.

A lower bound on $\R^*(\L,\T,\epsilon,\rho)$ follows from \eqref{max_DT}, and by bounding $\I(\T,\rho)$ and $\U(\T,\rho)$ using \eqref{eq:low_bound_I} and \eqref{eq:U_upper_bound}, respectively:
	\begin{IEEEeqnarray}{lCl}
		\R^*(\L,\T,\epsilon,\rho) &\geq& \frac{\underline{\I}(\T,\rho)}{\T}-\sqrt{\frac{\U_{\text{UB}}(\T,\rho_0)}{\L\T^2}}Q^{-1}(\tau)\nonumber\\
		& \triangleq & \frac{\R_{\text{LB}}(\T,\rho)}{\T}, \quad \rho\geq\rho_0 \label{max_DT_secstep}
	\end{IEEEeqnarray}
with $\tau$ defined in \eqref{eq:tau_def}. Recall that, by the assumption ${0<\epsilon<\frac{1}{2}}$, we have $Q^{-1}(\tau)>0$  for $\L$ sufficiently large.
 
It follows from \cite[Eq. (106)]{Polyanskiy_Poor_Verdu} and \eqref{eq:j_u_bound} that the RHS of~\eqref{eq:MC_app} can be upper-bounded as 
 \begin{IEEEeqnarray}{lCl}
 \label{eq:MC_logM_app}
 	&&\sup_{\boldsymbol{\pow}\in[0,\rho]^\L}\log\Biggl(\frac{1}{\beta(\boldsymbol{\pow},\text{q}_{\mathbf{Y}^\L}^{(\text{U})})}\Biggr) \leq\sup_{\boldsymbol{\pow}\in[0,\rho]^\L}\Biggl\{\frac{\log\xi(\boldsymbol{\pow})}{\L\T}\nonumber\\
 	&&\qquad{} - \frac{\log\Bigl(1-\epsilon-\mathsf{P}\bigl[\sum_{\ell=1}^\L{\bar{j}_\ell(\T,\pow_\ell)}\geq\log\xi(\boldsymbol{\pow})\bigr]\Bigr)}{\L\T}\Biggr\}\IEEEeqnarraynumspace
 \end{IEEEeqnarray}
for every $\xi\colon [0,\rho]^\L \to (0,\infty)$. By Lemma~\ref{lm:var_bounded}, for every $\rho_0>0$ there exists a $\bar{\V}_{\text{UB}}(\T,\rho_0)$ that is independent of $\pow$ and $\rho$ and that satisfies
\begin{equation}
   \label{eq:Vbar}
   \bar{\V}_\rho(\T,\pow) \leq \bar{\V}_{\text{UB}}(\T,\rho_0), \quad \pow\geq 0, \,\rho\geq \rho_0.
\end{equation}
Let
 \begin{equation}
	 \log\xi(\boldsymbol{\pow})=\sum_{\ell=1}^\L{\bar{\J}(\T,\pow_\ell)}+\sqrt{\frac{\L\bar{\V}_{\text{UB}}(\T,\rho_0)}{(1-\epsilon)-\frac{1}{\sqrt{\L}}}}.
\end{equation}
By Chebyshev's inequality \cite[Ch. V.7]{Feller} and \eqref{eq:Vbar}, we obtain
\begin{IEEEeqnarray}{lCl}
\IEEEeqnarraymulticol{3}{l}{\mathsf{P}\Biggl[\sum_{\ell=1}^\L{\bar{j}_\ell(\T,\pow_{\ell})}\geq\log\xi(\boldsymbol{\pow})\Biggr]} \nonumber\\
\quad &\leq&\frac{\sum_{\ell=1}^\L{\bar{\V}_\rho(\T,\pow_\ell)}}{\L\bar{\V}_{\text{UB}}(\T,\rho_0)}\biggl(1-\epsilon-\frac{1}{\sqrt{\L}}\biggr)\nonumber\\
&\leq& 1-\epsilon-\frac{1}{\sqrt{\L}}, \quad \rho\geq\rho_0. \label{eq:cheby}
 \end{IEEEeqnarray}
Combining \eqref{eq:cheby} with \eqref{eq:MC_logM_app} then yields
\begin{IEEEeqnarray}{lCl}
	\R^{*}(\L,\epsilon,\rho)&\leq &\sup_{\boldsymbol{\pow}\in[0,\rho]^\L}\frac{\sum_{\ell=1}^\L{\bar{\J}(\T,\pow_\ell)}}{\L\T}\nonumber\\
	&&{}+\sqrt{\frac{\bar{\V}_{\text{UB}}(\T,\rho_0)}{\L\T^2(1-\epsilon)-\T^2\sqrt{\L}}}+\frac{\log \L}{2\L\T}\nonumber\\
	&\triangleq & \sup_{\boldsymbol{\pow}\in[0,\rho]^\L} \frac{1}{\L}\sum_{\ell=1}^\L{\frac{\R_{\text{UB}}(\T,\pow_\ell)}{\T}}, \quad \rho\geq\rho_0. \IEEEeqnarraynumspace\label{eq:max_rate_cheby}
\end{IEEEeqnarray}
The $\boldsymbol{\pow}$'s for which $\frac{1}{\L}\sum_{\ell=1}^\L{\R_{\text{UB}}(\T,\pow_\ell)/\T}$ is smaller than \eqref{max_DT_secstep} can be discarded without loss of optimality, since the upper bound can never be smaller than the lower bound. We next use this argument to show that the fraction of $\pow_{\ell}$'s in $\boldsymbol{\pow}$ that satisfy $\pow_\ell\geq \rho(1-\delta)$ tends to $1$ as $\L$ and $\rho$ tend to infinity. Specifically, we consider the difference
\begin{IEEEeqnarray}{lCl}
\IEEEeqnarraymulticol{3}{l}{
	\frac{1}{\L}\sum_{\ell=1}^\L\bigl[\R_{\text{LB}}(\T,\rho) -	\R_{\text{UB}}(\T,\pow_\ell)\bigr]}\IEEEnonumber\\\,\,\,
	&=&\frac{1}{\L}\sum_{\ell=1}^\L\Biggl[\frac{\T\rho-\T\pow_\ell}{1+\T\rho}+\log\biggl(\frac{1+\T\pow_\ell}{1+\T\rho}\biggr)\nonumber\\
	&&{}+(\T-1)\mathsf{E}\biggl[\log\frac{(1+\T\rho)Z_{1,\ell}+Z_{2,\ell}+\beta(\T,\rho)}{(1+\T\pow_\ell)Z_{1,\ell}+Z_{2,\ell}+\beta(\T,\rho)}\biggr]\IEEEnonumber*\\
	&&{}-(\T-1)\mathsf{E}\biggl[\log\biggl(1+\frac{\beta(\T,\rho)}{(1+\T\rho)Z_{1,\ell}+Z_{2,\ell}}\biggr)\biggr]\\
	&&{}-\sqrt{\frac{\U_{\text{UB}}(\T,\rho_0)}{\L}}Q^{-1}(\tau) -\sqrt{\frac{\bar{\V}_{\text{UB}}(\T,\rho_0)}{\L(1-\epsilon)-\sqrt{\L}}}-\frac{\log \L}{2\L}\Biggr]\\\IEEEyesnumber\label{eq:diff_J}
\end{IEEEeqnarray}
where we have evaluated $\R_{\text{LB}}(\T,\rho)$ and $\R_{\text{UB}}(\T,\pow_{\ell})$ using \eqref{eq:sec:first_first_moment_closed} and \eqref{eq:J_def}. We next fix a sufficiently large $\rho_0$ and assume that $\rho\geq\rho_0$. Since $\rho\mapsto\beta(\T,\rho)$ is decreasing in $\rho$, we can lower-bound the third-term on the RHS of \eqref{eq:diff_J} by replacing $\beta(\T,\rho)$ by $\beta(\T,\rho_0)$. We can further lower-bound \eqref{eq:diff_J} by omitting the first term on the RHS of \eqref{eq:diff_J}, which is nonnegative since $\pow_\ell \leq\rho$. This yields
\begin{IEEEeqnarray}{lCl}
\IEEEeqnarraymulticol{3}{l}{
	\frac{1}{\L}\sum_{\ell=1}^\L\bigl[\R_{\text{LB}}(\T,\rho) -\R_{\text{UB}}(\T,\pow_\ell)\bigr]}\nonumber\\\,\,\,
	 &\geq& {}\frac{1}{\L}\sum_{\ell=1}^\L\Biggl[\log\biggl(\frac{1+\T\pow_\ell}{1+\T\rho}\biggr)\nonumber\\
	 &&{}+(\T-1)\mathsf{E}\biggl[\log\frac{(1+\T\rho)Z_{1,\ell}+\T-1+\beta(\T,\rho_0)}{(1+\T\pow_\ell)Z_{1,\ell}+\T-1+\beta(\T,\rho_0)}\biggr]\nonumber\\
	&& {}-(\T-1)\mathsf{E}\biggl[\log\biggl(1+\frac{\beta(\T,\rho)}{(1+\T\rho)Z_{1,\ell}+Z_{2,\ell}}\biggr)\biggr]\nonumber\\
	&& {} -\sqrt{\frac{\U_{\text{UB}}(\T,\rho_0)}{\L}}Q^{-1}(\tau)-\sqrt{\frac{\bar{\V}_{\text{UB}}(\T,\rho_0)}{\L(1-\epsilon)-\sqrt{\L}}}-\frac{\log \L}{2\L}\Biggr]\nonumber\\
	&\triangleq & \frac{1}{\L}\sum_{\ell=1}^\L \Delta_{\L,\T,\rho}(\pow_\ell), \quad \rho\geq\rho_0.\label{eq:analysis_diff_rates}
\end{IEEEeqnarray}
In the following, we analyze the behaviour of the function $\pow_\ell\mapsto \Delta_{\L,\T,\rho}(\pow_\ell)$. Let 
\begin{multline}
	g_{\T,\rho}(\pow_\ell)\triangleq\log\biggl(\frac{1+\T\pow_\ell}{1+\T\rho}\biggr)\\+(\T-1)\mathsf{E}\biggl[\log\frac{(1+\T\rho)Z_{1,\ell}+\T-1+\beta(\T,\rho_0)}{(1+\T\pow_\ell)Z_{1,\ell}+\T-1+\beta(\T,\rho_0)}\biggr] \label{eq:diff_low_1}
\end{multline}
and
\begin{multline}\label{eq:omega_L_rho_def}
	\omega_{\L,\T,\rho}\triangleq (\T-1)\mathsf{E}\biggl[\log\biggl(1+\frac{\beta(\T,\rho)}{(1+\T\rho)Z_1+Z_2}\biggr)\biggr]\\+\sqrt{\frac{\U_{\text{UB}}(\T,\rho_0)}{\L}}Q^{-1}(\tau)+\sqrt{\frac{\bar{\V}_{\text{UB}}(\T,\rho_0)}{\L(1-\epsilon)-\sqrt{\L}}}+\frac{\log \L}{2\L}.
\end{multline}
Thus, $\Delta_{\L,\T,\rho}(\pow_\ell)=g_{\T,\rho}(\pow_\ell)-\omega_{\L,\T,\rho}$. Note that $\frac{\partial}{\partial\pow_\ell}g_{\T,\rho}(\pow_\ell)=\frac{\partial}{\partial\pow_\ell}\Delta_{\L,\T,\rho}(\pow_\ell)$, since $\omega_{\L,\T,\rho}$ does not depend on $\pow_\ell$. Further note that 
\begin{IEEEeqnarray}{lCl}
	\IEEEeqnarraymulticol{3}{l}{\lim_{\begin{subarray}{l} \L\to\infty,\\ \rho\to\infty\end{subarray}}{\omega_{\L,\T,\rho}}} \nonumber\\
\quad &=&\lim_{\rho\to\infty}{(\T-1)\mathsf{E}\biggl[\log\biggl(1+\frac{\beta(\T,\rho)}{(1+\T\rho)Z_1+Z_2}\biggr)\biggr]}\nonumber\\
	&&{}+\lim_{\L\to\infty}\sqrt{\frac{\U_{\text{UB}}(\T,\rho_0)}{\L}}Q^{-1}(\tau)\nonumber\\
	&&{}+\lim_{\L\to\infty}\sqrt{\frac{\bar{\V}_{\text{UB}}(\T,\rho_0)}{\L(1-\epsilon)-\sqrt{\L}}}+\lim_{\L\to\infty}\frac{\log \L}{2\L}\nonumber\\
	& = & 0 \label{eq:double_limit}
\end{IEEEeqnarray} 
where the second line in \eqref{eq:double_limit} is zero by the dominated convergence theorem. The following lemma discusses the behavior of $\pow_\ell\mapsto g_{\T,\rho}(\pow_\ell)$.
\begin{lemma}\label{lm:g_alpha}
	The function $\pow\mapsto g_{\T,\rho}(\pow)$ has the following properties: 
\begin{enumerate}
	\item The derivative of $\pow\mapsto g_{\T,\rho}(\pow)$ is either strictly positive, strictly negative, or changes its sign once from positive to negative. This implies that $g_{\T,\rho}(\pow)$, $0\leq\pow\leq\rho$  is minimized at the boundary of $[0,\rho]$, and it has a unique maximizer.
	\item The derivative of $\pow\mapsto g_{\T,\rho}(\pow)$ with respect to $\pow$, denoted by $\pow\mapsto g'_{\T,\rho}(\pow)$, does not depend on $\rho$.
	\item It holds that $g_{\T,\rho}(\rho)=0$.  Furthermore, $g_{\T,\rho}(0)\to\infty$ as $\rho\to\infty$ for $\T>2$.
	\item Let $\pow^*$ denote the unique maximizer of $\pow\mapsto g_{\T,\rho}(\pow)$, which by Part~2) does not depend on $\rho$. For $\T>2$ and every $\pow'>\pow^*$ independent of $\rho$, we have
	\begin{equation}\label{eq:part4_lm_g}
	\sup_{\rho\geq\pow'}\sup_{\pow'\leq\pow\leq\rho} \rho  g'_{\T,\rho}(\pow) < 0.
	\end{equation}
\end{enumerate}
\end{lemma}
\begin{IEEEproof}
	See Appendix \ref{app:about_g}.
\end{IEEEproof}

We next study those $\boldsymbol{\pow}$'s for which $\sum_{\ell=1}^{\L}\Delta_{\L,\T,\rho}(\pow_\ell)\geq 0$, since they can be discarded without loss of optimality. Let
\begin{equation}
\mathscr{L}_{\delta}(\boldsymbol{\pow})\triangleq\{\ell=1,\dots,\L:\,\pow_\ell\geq\rho(1-\delta)\}
\end{equation}
and let $\L_{\delta}(\boldsymbol{\pow})$ denote the number of $\pow_\ell$'s in $\boldsymbol{\pow}$ that satisfy $\rho(1-\delta)\leq\alpha_{\ell}\leq\rho$. (Thus, $\L_{\delta}(\boldsymbol{\pow})$ is the cardinality of $\mathscr{L}_{\delta}(\boldsymbol{\pow})$.) Further let
\begin{equation}
\Delta_{\L,\T,\rho}^*(\delta) \triangleq \inf_{0\leq\pow\leq\rho(1-\delta)}\Delta_{\L,\T,\rho}(\pow).
\end{equation}
The sum of $\Delta_{\L,\T,\rho}(\pow_{\ell})$ in \eqref{eq:analysis_diff_rates} can be expressed as
\begin{IEEEeqnarray}{lCl}
	\IEEEeqnarraymulticol{3}{l}{\sum_{\ell=1}^\L \Delta_{\L,\T,\rho}(\pow_\ell)} \nonumber\\
	\quad & = & \sum_{\mathscr{L}_{\delta}(\boldsymbol{\pow})} \Delta_{\L,\T,\rho}(\pow_\ell) + \sum_{\mathscr{L}^\mathsf{c}_{\delta}(\boldsymbol{\pow})} \Delta_{\L,\T,\rho}(\pow_\ell). \label{eq:Tobi_analysis_diff}
\end{IEEEeqnarray}
By Parts 1) and 3) of Lemma~\ref{lm:g_alpha},
\begin{IEEEeqnarray}{rl}
	 \Delta_{\L,\T,\rho}(\pow_\ell)\geq -\omega_{\L,\T,\rho}, \quad 0 \leq \pow_\ell \leq \rho
\end{IEEEeqnarray}
for $\T>2$ and $\rho$ sufficiently large. Thus, we can lower-bound the first sum on the RHS of \eqref{eq:Tobi_analysis_diff} by $-\L_{\delta}(\boldsymbol{\pow})\omega_{\L,\T,\rho}$ and the second sum on the RHS of \eqref{eq:Tobi_analysis_diff} by $(\L-\L_{\delta}(\boldsymbol{\pow}))\Delta_{\L,\T,\rho}^*(\delta)$. This yields
\begin{IEEEeqnarray}{lCl}
\IEEEeqnarraymulticol{3}{l}{\sum_{\ell=1}^\L \Delta_{\L,\T,\rho}(\pow_\ell)} \nonumber\\
\quad & \geq & (\L-\L_{\delta}(\boldsymbol{\pow}))\Delta_{\L,\T,\rho}^*(\delta) -\L_{\delta}(\boldsymbol{\pow})\omega_{\L,\T,\rho}.\label{eq:Delta_bound_explicit}
\end{IEEEeqnarray}
It follows that we can discard without loss of optimality every $\boldsymbol{\pow}$ for which
\begin{equation}
\label{eq:bound_sum_Deltas}
\L \Delta_{\L,\T,\rho}^*(\delta) \geq \L_{\delta}(\boldsymbol{\pow})[\omega_{\L,\T,\rho}+\Delta_{\L,\T,\rho}^*(\delta)]
\end{equation}
since for such $\boldsymbol{\pow}$'s the RHS of \eqref{eq:Delta_bound_explicit}, and hence also \eqref{eq:diff_J}, is nonnegative.  We conclude that an $\boldsymbol{\pow}$ maximizing \eqref{eq:MC_app} must satisfy
\begin{equation}
\label{eq:L1_lim}
\frac{\L_{\delta}(\boldsymbol{\pow})}{\L} > 1 - \frac{\omega_{\L,\T,\rho}}{\omega_{\L,\T,\rho}+\Delta_{\L,\T,\rho}^*(\delta)}.
\end{equation}
As we shall show below, for every $0<\delta<1$ we have
\begin{equation}\label{eq:min_Delta}
	 \omega_{\L,\T,\rho}+\Delta_{\L,\T,\rho}^*(\delta) \geq -\delta \sup_{\rho\geq\pow'} \sup_{\pow'\leq\pow\leq\rho} \rho  g'_{\T,\rho}(\pow)
\end{equation}
for some $0<\pow'<\rho(1-\delta)$ that is independent of $\rho$. We further show that the RHS of \eqref{eq:min_Delta} is independent of $\L$ and $\rho$ and strictly positive. It follows that
\begin{equation}
\frac{\L_{\delta}(\boldsymbol{\pow})}{\L} > 1 - \frac{\omega_{\L,\T,\rho}}{-\delta \sup_{\rho\geq\pow'} \sup_{\pow'\leq\pow\leq\rho} \rho  g'_{\T,\rho}(\pow)}
\end{equation}
which, by \eqref{eq:double_limit}, tends to one as $\rho$ and $\L$ tend to infinity. Thus, for every $0<\delta<1$, there exist sufficiently large $\L_0$ and $\rho_0$ such that 
\begin{equation}
\L_{\delta}(\boldsymbol{\pow}) \geq \L/2, \quad \L\geq \L_0, \, \rho\geq\rho_0.
\end{equation}
This proves Lemma~\ref{lm:opt_high_SNR}. 

It remains to show \eqref{eq:min_Delta}. Let $\pow_{\min}=\rho(1-\delta)$.  By Part~1) of Lemma \ref{lm:g_alpha}, $\pow\mapsto g_{\T,\rho}(\pow)$ has exactly one maximizer, which we shall denote by $\pow^*$. Since $\omega_{\L,\T,\rho}$ does not depend on $\pow$, it follows that $\pow^*$ is also the maximizer of $\pow\mapsto \Delta_{\L,\T,\rho}(\pow)$. Furthermore, the infimum of $\Delta_{\L,\T,\rho}(\pow)$ over $0\leq\pow\leq\pow_{\min}$, denoted by $\Delta_{\L,\T,\rho}^*(\delta)$, is either achieved at $\pow=0$ or at $\pow_{\min}$.

By Part~3) of Lemma \ref{lm:g_alpha} and \eqref{eq:double_limit}, we have
\begin{equation}
\lim_{\begin{subarray}{l} \L\to\infty,\\ \rho\to\infty \end{subarray}} \Delta_{\L,\T,\rho}(0) = \infty.
\end{equation}
We next show that
\begin{equation}
\label{eq:Tobi_min_Delta}
\Delta_{\L,\T,\rho}(\pow_{\min}) + \omega_{\L,\T,\rho} \geq -\delta \sup_{\rho\geq\pow'} \sup_{\pow'\leq\pow\leq\rho} \rho  g'_{\T,\rho}(\pow).
\end{equation}
If $\pow_{\min}\leq \pow^*$, then this is clearly satisfied, since in this case $\Delta_{\L,\T,\rho}(\pow_{\min})\geq \Delta_{\L,\T,\rho}(0)$ and $\Delta_{\L,\T,\rho}(0)$ tends to infinity as $\L\to\infty$ and $\rho\to\infty$. However, in general this case does not occur for large $\rho$ and $\L$, since $\pow_{\min}$ tends to infinity as $\rho\to\infty$ and, by Part 2) of Lemma \ref{lm:g_alpha}, $\pow^*$ does not depend on $\rho$, which implies that $\pow_{\min} > \pow^*$ for $\rho$ sufficiently large. We thus focus on the case where $\pow_{\min} > \pow^*$. Note that
\begin{equation}
\Delta_{\L,\T,\rho}(\rho)-\Delta_{\L,\T,\rho}(\pow_{\min})=-\omega_{\L,\T,\rho}-\Delta_{\L,\T,\rho}(\pow_{\min})
\end{equation}
since $g_{\T,\rho}(\rho)=0$. Thus, by the mean value theorem \cite[Th. 5.10]{rudin-principles}, there exists an $x_0\in[\pow_{\min},\rho]$ such that
\begin{IEEEeqnarray}{lCl}
	\IEEEeqnarraymulticol{3}{l}{-\omega_{\L,\T,\rho}-\Delta_{\L,\T,\rho}(\pow_{\min})} \nonumber\\ 
	\quad & = & \int_{\pow_{\min}}^{\rho}{\Delta_{\L,\T,\rho}'(\pow)d\pow} = \rho\delta \Delta_{\L,\T,\rho}'(x_0) \label{eq:MVT}
\end{IEEEeqnarray}
where $\Delta_{\L,\T,\rho}'(\cdot)$ denotes the derivative of $\pow\mapsto\Delta_{\L,\T,\rho}(\pow)$. We can therefore lower-bound
\begin{IEEEeqnarray}{lCl}
\Delta_{\L,\T,\rho}(\pow_{\min}) + \omega_{\L,\T,\rho}& \geq &   - \delta \sup_{\pow_{\min}\leq \alpha\leq\rho}\rho \Delta_{\L,\T,\rho}'(\alpha)\nonumber\\
& \geq & -\delta \sup_{\rho\geq\pow'} \sup_{\pow'\leq\pow\leq\rho} \rho  g'_{\T,\rho}(\pow)\IEEEeqnarraynumspace\label{eq:delta_def_app_A}
\end{IEEEeqnarray}
for every $\pow'\in (\pow^*,\pow_{\min})$ independent of $\L$ and $\rho$.\footnote{Since $\pow^*$ is independent of $\rho$ and $\pow_{\min}\to\infty$ as $\rho\to\infty$, it follows that such an $\pow'$ exists.} In \eqref{eq:delta_def_app_A}, the second inequality follows by noting that \mbox{$\Delta_{\L,\T,\rho}'=g_{\T,\rho}'$} and by further optimizing over $\rho$. It remains to show that the RHS of \eqref{eq:delta_def_app_A} is independent of $\L$ and $\rho$ and strictly positive. To this end, we first note that $\alpha\mapsto g_{\T,\rho}(\alpha)$ is independent of $\L$. Furthermore, by optimizing over $\rho\geq\alpha'$, the RHS of \eqref{eq:delta_def_app_A} becomes also independent of $\rho$. Finally, by Part 4) of Lemma~\ref{lm:g_alpha},
\begin{equation}
\sup_{\rho\geq\pow'} \sup_{\pow'\leq\pow\leq\rho} \rho  g'_{\T,\rho}(\pow)<0, \quad \T>2,\,\rho\geq\pow'
\end{equation}
for every $\pow'\in (\pow^*,\pow_{\min})$ independent of $\L$ and $\rho$. Thus, the claim \eqref{eq:min_Delta} follows, which concludes the proof of Lemma~\ref{lm:opt_high_SNR}.

\section{Proof of Lemma \ref{lm:g_alpha}}\label{app:about_g}
The derivative of $\pow\mapsto g_{\T,\rho}(\pow)$ can be expressed as

\begin{IEEEeqnarray}{lCl}
	g_{\T,\rho}'(\pow)&=&\frac{\T}{1+\T\pow}\nonumber\\
	&&{}-(\T-1)\mathsf{E}\biggl[\frac{\T Z_1}{(1+\T\pow)Z_1+(\T-1)+\beta(\T,\rho_0)}\biggr]\IEEEnonumber*\\
	&=&\T\Biggl[\frac{1}{1+\T\pow}-\frac{\T-1}{1+\T\pow}\nonumber\\
	&&\quad{}+\frac{\T-1}{1+\T\pow}\varphi\biggl(\frac{1+\T\pow}{\T-1+\beta(\T,\rho_0)}\biggr)\Biggr]\\
	&=&{}\frac{\T}{1+\T\pow}\Biggl[-(\T-2) \nonumber\\
	& &\quad\qquad {} +(\T-1)\varphi\biggl(\frac{1+\T\pow}{\T-1+\beta(\T,\rho_0)}\biggr)\Biggr] \IEEEyesnumber\IEEEeqnarraynumspace\label{eq:derivative_g}
\end{IEEEeqnarray}
where
\begin{equation}
\varphi(x) \triangleq \frac{1}{x}e^{\frac{1}{x}}\text{E}_1\biggl(\frac{1}{x}\biggr), \quad x>0.
\end{equation}
The first equality follows because, by \cite[App. A.9]{Durrett_Prob_theory}, we can swap derivative and expected value; the second equality follows by solving the expected value using \cite[Sec. 3.353-5.$^7$]{table_integrals}. Note that the RHS of \eqref{eq:derivative_g} does not depend on $\rho$. Hence Part 2) of Lemma \ref{lm:g_alpha} follows immediately.

We next prove Part~1) of Lemma \ref{lm:g_alpha}. Because $\T/(1+\T\pow)$ in~\eqref{eq:derivative_g} is nonnegative, the sign of $\pow\mapsto g_{\T,\rho}'(\pow)$ is determined by the terms inside the square brackets. Note that $x\mapsto\varphi(x)$ is strictly decreasing since, by \cite[Sec. 3.353-3]{table_integrals},
  \begin{equation}\label{eq:monton_expression_with_exp_int_1}
	\frac{1}{x}e^{\frac{1}{x}}\text{E}_1\biggl(\frac{1}{x}\biggr)=1-\int_0^1{e^{-\frac{t}{(1-t)x}}\mathrm{d} t}
 \end{equation}
and $x\mapsto  e^{-\frac{t}{(1-t)x}}$ is strictly positive and strictly increasing in $x$. Hence, the function inside the squared brackets is strictly decreasing in $\pow$. This implies that $\pow\mapsto g_{\T,\rho}'(\pow)$ is either strictly positive, strictly negative, or changes its sign once from positive to negative.
 
We next prove Part~3) of Lemma \ref{lm:g_alpha} by showing that $\lim_{\rho\to\infty} g_{\T,\rho}(0)=\infty$ for $\T>2$. To this end, we express $g_{\T,\rho}(0)$ as
\begin{IEEEeqnarray}{lCl}
	g_{\T,\rho}(0)&=& (\T-2)\mathsf{E}\biggl[\log\biggl(1+\frac{\T\rho Z_1}{Z_1+(\T-1)+\beta(\T,\rho_0)}\biggr)\biggr]\nonumber\\
	&&{} + \mathsf{E}\biggl[\log\biggl(Z_1+\frac{\T-1+\beta(\T,\rho_0)}{1+\T\rho}\biggr)\biggr] \nonumber\\
	& & {} - \mathsf{E}\bigl[\log\bigl(Z_1 + \T-1 + \beta(\T,\rho_0)\bigr)\bigr].\label{eq:zeta_def}
	\end{IEEEeqnarray}
The first expected value on the RHS of \eqref{eq:zeta_def} tends to infinity as $\rho\to\infty$, whereas the other expected values are bounded in $\rho$. For $\T>2$, it follows that the RHS of \eqref{eq:zeta_def} tends to infinity as $\rho\to\infty$. Hence the claim follows.

We finally prove Part 4) of Lemma \ref{lm:g_alpha} by analyzing $\rho g_{\T,\rho}'(\pow)$. It follows from \eqref{eq:derivative_g} that
\begin{multline}
 	\rho g_{\T,\rho}'(\pow)=\frac{\T\rho}{1+\T\pow}\biggl[-(\T-2) \\
	{} +(\T-1)\varphi\biggl(\frac{1+\T\pow}{\T-1+\beta(\T,\rho_0)}\biggr)\biggr].\label{eq:p_deriv_g}
 \end{multline}
Observe that the function inside the square brackets is independent of $\L$ and $\rho$. Further note that, as argued above, it is strictly decreasing in $\pow$. Hence, its supremum over $\pow'\leq\pow\leq\rho$ is achieved for $\pow=\pow'$. Furthermore, this function is strictly negative for $\T>2$ and $\pow'>\pow^*$. As for the term outside the curly brackets, we have for every $\pow'>\pow^*$
\begin{equation}
\inf_{\rho\geq\pow'} \inf_{\pow'\leq\pow\leq\rho} \frac{\T\rho}{1+\T\pow} = \frac{\T\pow'}{1+\T\pow'} > 0.
\end{equation}
Combining these two results, we conclude that
\begin{equation}
\sup_{\rho\geq\pow'} \sup_{\pow'\leq\pow\leq \rho} \rho g_{\T,\rho}'(\pow) < 0, \quad \T>2,\,\pow'>\pow^*.
\end{equation}
This proves Part 4) of Lemma \ref{lm:g_alpha} and concludes the proof of Lemma~\ref{lm:g_alpha}.

\section{Proof of Lemma \ref{lm:opt_high_SNR_2}}\label{app:proof_lm_opt_high_SNR_B}
\subsubsection{Part~1)}
The difference between $\bar{\J}(\T,\pow)$ and $\bar{\J}(\T,\rho)$ can be lower-bounded by
\begin{equation}
\bar{\J}(\T,\rho) - \bar{\J}(\T,\pow) \geq g_{\T,\rho}(\pow)
\end{equation}
where the function $\pow\mapsto g_{\T,\rho}(\pow)$ was defined in \eqref{eq:diff_low_1}. By Parts~1) and 3) of Lemma~\ref{lm:g_alpha} (Appendix~\ref{app:proof_lm_opt_high_SNR_A}), $\pow\mapsto g_{\T,\rho}(\pow)$ is nonnegative for sufficiently large $\rho$. It follows that, for such~$\rho$,
\begin{equation}\label{eq:sup_J_alpha_J_p}
	 \sup_{0\leq\pow\leq\rho}{\bar{\J}(\T,\pow)}=\bar{\J}(\T,\rho).
\end{equation}
This proves Part~1) of Lemma~\ref{lm:opt_high_SNR_2}.
 
\subsubsection{Part~2)}
To study
\begin{IEEEeqnarray}{lCl}
	\sup_{0\leq\pow\leq\rho}\Biggl\{\frac{\bar{\J}(\T,\pow)}{\T} - \sqrt{\frac{ \bar{\V}_\rho(\T,\pow)}{\L\T^2}}Q^{-1}(\epsilon)\Biggr\}
\end{IEEEeqnarray}
we consider the difference 
\begin{IEEEeqnarray}{lCl}
		\IEEEeqnarraymulticol{3}{l}{\bar{\J}(\T,\rho)-\bar{\J}(\T,\pow)-\Biggl(\sqrt{\frac{\bar{\V}(\T,\rho)}{\L}}-\sqrt{\frac{ \bar{\V}_\rho(\T,\pow)}{\L}}\Biggr)Q^{-1}(\epsilon)} \nonumber\\
		\quad &\geq& g_{\T,\rho}(\pow)-\Biggl(\sqrt{\frac{\bar{\V}(\T,\rho)}{\L}}-\sqrt{\frac{ \bar{\V}_\rho(\T,\pow)}{\L}}\Biggr)Q^{-1}(\epsilon).\IEEEeqnarraynumspace\label{eq:diff_up}
\end{IEEEeqnarray}
Clearly, every $\pow$ for which the RHS of \eqref{eq:diff_up} is nonnegative is suboptimal and can be discarded without loss of optimality. We continue by lower-bounding $\bar{\V}_\rho(\T,\pow) \geq 0$ and by using that $\bar{\V}(\T,\rho)\leq\bar{\V}_\text{UB}(\T,\rho_0)$, $\rho\geq\rho_0$ for sufficiently large $\rho_0$ and for some constant $\bar{\V}_\text{UB}(\T,\rho_0)$ that is independent of $\rho$ (Lemma~\ref{lm:var_bounded}). Since by the assumption $0<\epsilon<\frac{1}{2}$ we have $Q^{-1}(\epsilon)>0$, this yields
\begin{IEEEeqnarray}{lCl}
		\IEEEeqnarraymulticol{3}{l}{g_{\T,\rho}(\pow)-\sqrt{\frac{\bar{\V}(\T,\rho)}{\L}}Q^{-1}(\epsilon)+\sqrt{\frac{ \bar{\V}_\rho(\T,\pow)}{\L}}Q^{-1}(\epsilon)}\IEEEnonumber*\\
		\qquad &\geq&g_{\T,\rho}(\pow)-\sqrt{\frac{\bar{\V}_\text{UB}(\T,\rho_0)}{\L}}Q^{-1}(\epsilon)\\
		&\triangleq& f_{\L,\T,\rho}(\pow).\IEEEyesnumber\label{eq:diff_up_def}
\end{IEEEeqnarray}
Again, the values of $\pow$ for which $f_{\L,\T,\rho}(\pow)\geq 0$ are suboptimal and can be discarded without loss of optimality. 

Let us write $f_{\L,\T,\rho}$ as $f_{\L,\T,\rho}(\pow)\triangleq g_{\T,\rho}(\pow)-\omega_{\L,\T}$, where
\begin{equation}
\omega_{\L,\T}\triangleq\sqrt{\frac{ \bar{\V}_\text{UB}(\T,\rho_0)}{\L}}Q^{-1}(\epsilon).
\end{equation}
Note that $\Delta_{\L,\T,\rho}$ defined in \eqref{eq:analysis_diff_rates} and $f_{\L,\T,\rho}$ only differ in terms that do not depend on $\pow$ (namely, $\omega_{\L,\T,\rho}$ and $\omega_{\L,\T}$), so they have the same behavior with respect to $\pow$ as summarized in Lemma~\ref{lm:g_alpha}. Let $\delta_{\L,\T}\triangleq 1-\pow_{0}/\rho$, where $\pow_{0}$ is the unique real root of $\pow\mapsto f_{\L,\T,\rho}(\pow)$. Indeed, we know that $\pow\mapsto f_{\L,\T,\rho}(\pow)$ has at least one root because $\omega_{\L,\T}\geq0$ and $\omega_{\L,\T}\to 0$ as $\L\to\infty$, so $f_{\L,\T,\rho}(\rho)=-\omega_{\L,\T}\leq0$ and ${f_{\L,\T,\rho}(0)}>0$ for $\L$ and $\rho$ sufficiently large. Furthermore, we have $f_{\L,\T,\rho}'=g_{\T,\rho}'$ and $\pow \mapsto g_{\T,\rho}'(\pow)$ is either strictly positive, strictly negative, or changes its sign once from positive to negative (Part 1) of Lemma~\ref{lm:g_alpha}). Consequently, $f_{\L,\T,\rho}(\pow)$, $0\leq\pow\leq\rho$ is minimized at an endpoint of $[0,\rho]$ and it has a unique maximizer, so its root is unique. By the same line of arguments, we also conclude that all $\pow$'s between $0$ and $\rho(1-\delta_{\L,\T})$ can be discarded without loss of optimality, since for such $\pow$'s the function $f_{\L,\T,\rho}(\pow)$ is nonnegative.

To study the behavior of $\delta_{\L,\T}$, we next note that
\begin{equation}
\omega_{\L,\T}=-(f_{\L,\T,\rho}(\rho)-f_{\L,\rho}(\pow_{0})).
\end{equation}
It follows then by similar steps as in \eqref{eq:MVT}--\eqref{eq:delta_def_app_A} that
\begin{equation} \label{eq:bound_omega_1}
	\omega_{\L,\T} \geq -\delta_{\L,\T}\sup_{\pow_0\leq \pow \leq \rho}{\rho f_{\L,\T,\rho}'(\pow)}.
\end{equation} 
Let $\pow^*$ denote the unique maximizer of $\pow\mapsto f_{\L,\T,\rho}(\pow)$. Recall that, by Part~2) of Lemma \ref{lm:g_alpha}, $\pow^*$ does not depend on $\rho$. We next show that we can find an $\tilde{\pow}$ independent of $\L$ and $\rho$ such that $\pow^*<\tilde{\pow}<\pow_0$. Indeed, by Lemma \ref{lm:g_alpha}, we have that $g_{\T,\rho}(\pow^*)>0$ for sufficiently large $\rho$. This in turn implies that
\begin{equation}\label{eq:inf_alpha_star_neg_2}
	\varliminf_{\begin{subarray}{l} \L\to\infty,\\ \rho\to\infty \end{subarray}}f_{\L,\T,\rho}(\pow^*)>0
\end{equation} 
since $\omega_{\L,\T}\to 0$ as $\L\to\infty$. We next note that, for every $\tilde{\pow}$,
\begin{IEEEeqnarray}{lCl}
\IEEEeqnarraymulticol{3}{l}{\varliminf_{\begin{subarray}{l} \L\to\infty,\\ \rho\to\infty \end{subarray}}f_{\L,\T,\rho}(\tilde{\pow})} \nonumber \\
\quad & \geq & \varliminf_{\begin{subarray}{l} \L\to\infty,\\ \rho\to\infty \end{subarray}}f_{\L,\T,\rho}(\pow^*)-\bigl|f_{\L,\T,\rho}(\tilde{\pow})-f_{\L,\T,\rho}(\pow^*)\bigr| \IEEEeqnarraynumspace
\end{IEEEeqnarray} 
where the difference
\begin{IEEEeqnarray}{lCl}
\IEEEeqnarraymulticol{3}{l}{
	f_{\L,\T,\rho}(\tilde{\pow})-f_{\L,\T,\rho}(\pow^*)}\nonumber\\\quad
	&=&g_{\T,\rho}(\tilde{\pow})-g_{\T,\rho}(\pow^*)\IEEEnonumber\\
	&=& \log\biggl(\frac{1+\T\tilde{\pow}}{1+\T\pow^*}\biggr)\nonumber\\
	&&{}+(\T-1)\mathsf{E}\biggl[\log\frac{(1+\T\pow^*)Z_1+\T-1+\beta(\T,\rho_0)}{(1+\T\tilde{\pow})Z_1+\T-1+\beta(\T,\rho_0)}\biggr]\nonumber\\\label{eq:bounds_inf_alpha_tilde_2}
\end{IEEEeqnarray}
is independent of $\L$ and $\rho$. By the continuity of $\pow\mapsto g_{\T,\rho}(\pow)$, this difference can be made arbitrarily small by choosing $\tilde{\pow}$ sufficiently close to $\pow^*$. It thus follows from \eqref{eq:inf_alpha_star_neg_2}--\eqref{eq:bounds_inf_alpha_tilde_2} that there exists an $\tilde{\pow}\in(\pow^*,\rho]$ that is independent of $\L$ and $\rho$ and that satisfies
\begin{equation}
	\varliminf_{\begin{subarray}{l} \L\to\infty,\\ \rho\to\infty \end{subarray}}f_{\L,\T,\rho}(\tilde{\pow})>0.
\end{equation} 
In other words, if $\L$ and $\rho$ are sufficiently large, then we can find an $\tilde{\pow}\in(\pow^*, \pow_0)$ that is independent of $\L$ and $\rho$. In this case the RHS of \eqref{eq:bound_omega_1} can be further lower-bounded by
\begin{IEEEeqnarray}{lCl}
	\omega_{\L,\T}& \geq & -\delta_{\L,\T}\sup_{\tilde{\pow}\leq \pow\leq\rho}{\rho f_{\L,\T,\rho}'(\pow)} \nonumber\\
	& \geq & -\delta_{\L,\T} \sup_{\rho\geq\tilde{\pow}}\sup_{\tilde{\pow}\leq \pow\leq\rho}{\rho f_{\L,\T,\rho}'(\pow)}.\label{eq:omega_bound_last}
\end{IEEEeqnarray}
We next argue that the constant
\begin{equation}\label{eq:inf_K_lemma_opt_2}
\mathsf{F}(\T)\triangleq -\sup_{\rho\geq\tilde{\pow}}\sup_{\tilde{\pow}\leq \pow\leq\rho}{\rho f_{\L,\T,\rho}'(\pow)}
\end{equation}
is independent of $\L$ and $\rho$ and strictly positive. Indeed, we have that $f_{\L,\T,\rho}'=g_{\T,\rho}'$, which is independent of $\L$. Furthermore, by optimizing over $\rho\geq\tilde{\pow}$, the RHS of \eqref{eq:inf_K_lemma_opt_2} becomes independent of $\rho$. Finally, setting $\alpha'=\tilde{\pow}$ in \eqref{eq:part4_lm_g} (Part 4) of Lemma~\ref{lm:g_alpha}) yields
\begin{equation}\label{eq:inf_K_lemma_opt_1}
\sup_{\rho\geq\tilde{\pow}} \sup_{\tilde{\pow}\leq\pow\leq\rho}\rho g_{\T,\rho}'(\pow) <0, \quad \rho\geq\tilde{\pow}
\end{equation}
hence the claim follows. Consequently, we obtain from \eqref{eq:omega_bound_last} and the definition of $\omega_\L$ and $\mathsf{F}(\T)$ that, for sufficiently large $\L_0$ and $\rho_0$,
\begin{equation}\label{eq:delta_end_lemma_1}
	\delta_{\L,\T}\leq \frac{\sqrt{\bar{\V}_\text{UB}(\T,\rho_0)}Q^{-1}(\epsilon)}{\mathsf{F}(\T)}\frac{1}{\sqrt{\L}}, \quad \rho\geq\rho_0,\,\L\geq \L_0.
\end{equation}

We next tighten this bound on $\delta_{\L,\T}$. Indeed, using that without loss of optimality we can assume $\rho(1-\delta_{\L,\T})\leq\pow\leq\rho$, we can derive a tighter lower bound on \eqref{eq:diff_up} by lower-bounding $\bar{\V}_\rho(\T,\pow)$ using the lower bound given in Appendix~\ref{app:bound_U_alpha} instead of lower-bounding it by zero. Specifically, by \eqref{eq:LB_var_alpha} in Appendix~\ref{app:bound_U_alpha},
\begin{IEEEeqnarray}{lCl}\label{eq:iterative1}
	\sqrt{\frac{\bar{\V}_\rho(\T,\pow)}{\L}}&\geq& \sqrt{\frac{\bar{\V}(\T,\rho)-\Upsilon(\T)\delta_{\L,\T}}{\L}}\nonumber\\
	&\geq& \sqrt{\frac{\bar{\V}(\T,\rho)}{\L}}-\sqrt{\frac{\Upsilon(\T)\delta_{\L,\T}}{\L}}
\end{IEEEeqnarray}
for every $\rho(1-\delta_{\L,\T})\leq\pow\leq\rho$. We can thus lower-bound \eqref{eq:diff_up} as
\begin{IEEEeqnarray}{lCl}
		\IEEEeqnarraymulticol{3}{l}{\bar{\J}(\T,\rho)-\bar{\J}(\T,\pow) - \Biggl(\sqrt{\frac{\bar{\V}(\T,\rho)}{\L}}-\sqrt{\frac{ \bar{\V}_\rho(\T,\pow)}{\L}}\Biggr)Q^{-1}(\epsilon)} \IEEEnonumber*\\
		\quad &\geq & g_{\T,\rho}(\pow)-\sqrt{\frac{\Upsilon(\T)\delta_{\L,\T}}{\L}}Q^{-1}(\epsilon)\\
		&\triangleq & \tilde{f}_{\L,\T,\rho}(\pow),\quad\rho(1-\delta_{\L,\T})\leq\pow\leq\rho.\IEEEyesnumber\label{eq:diff_up_definition}
\end{IEEEeqnarray}
Again, the values of $\pow$ for which $\tilde{f}_{\L,\T,\rho}(\pow)\geq 0$ are suboptimal and can be discarded without loss of optimality.

Let us write $\tilde{f}_{\L,\T,\rho}(\pow)=g_{\T,\rho}(\pow)-\tilde{\omega}_{\L,\T}$, where
\begin{equation}
\tilde{\omega}_{\L,\T}\triangleq\sqrt{\frac{\Upsilon(\T)\delta_{\L,\T}}{\L}}Q^{-1}(\epsilon).
\end{equation}
 Further let $\tilde{\delta}_{\L,\T}\triangleq 1-\tilde{\pow}_{0}/\rho$, where $\tilde{\pow}_{0}$ is the unique real root of $\pow\mapsto \tilde{f}_{\L,\T,\rho}(\pow)$. As above, it can be shown that all $\pow$'s between $0$ and $\rho(1-\tilde{\delta}_{\L,\T})$ can be discarded without loss of optimality, since for such $\pow$'s the function $\tilde{f}_{\L,\T,\rho}(\pow)$ is nonnegative. By repeating the steps \eqref{eq:bound_omega_1}--\eqref{eq:delta_end_lemma_1} with $\omega_{\L,\T}$ replaced by $\tilde{\omega}_{\L,\T}$, we obtain that
\begin{IEEEeqnarray}{lCl}
	\tilde{\delta}_{\L,\T} &\leq& \frac{1}{\mathsf{F}(\T)}\sqrt{\frac{\Upsilon(\T)\delta_{\L,\T}}{\L}}Q^{-1}(\epsilon) \nonumber\\
	& \leq & \biggl(\frac{Q^{-1}(\epsilon)}{\mathsf{F}(\T)}\biggr)^{3/2}\sqrt{\Upsilon(\T) \sqrt{\bar{\V}_\text{UB}(\T,\rho_0)}} \frac{1}{\L^{3/4}} \IEEEeqnarraynumspace
\end{IEEEeqnarray}
for every $\rho\geq\rho_0, \, L\geq L_0$, and sufficiently large $\L_0$ and~$\rho_0$.
The last inequality follows by upper-bounding $\delta_{\L,\T}$ using~\eqref{eq:delta_end_lemma_1}.

If we perform the above steps $N$ times, then we obtain that, without loss of optimality,
\begin{equation}\label{eq:alpha_iterN}
	\pow\geq\rho\Bigl(1-\delta^{(N)}_{\L,\T}\Bigr)
\end{equation}
where $\delta^{(N)}_{\L,\T}$ satisfies
\begin{IEEEeqnarray}{lCl}
0 & \leq & \delta^{(N)}_{\L,\T} \nonumber\\
&\leq& \frac{\Bigl(\frac{Q^{-1}(\epsilon)\sqrt{\Upsilon(\T)}}{\mathsf{F}(\T)}\Bigr)^{2-2^{-N+1}}}{\L^{1-2^{-N}}}\bigg(\frac{\bar{\V}_\text{UB}(\T,\rho_0)}{\Upsilon(\T)}\biggr)^{2^{-N}} \label{eq:delta_iterN}
	\end{IEEEeqnarray}
	for every $\rho\geq\rho_0, \, L\geq L_0$, and sufficiently large $\L_0$ and~$\rho_0$.\footnote{While, in principle, $L_0$ and $\rho_0$ may depend on $N$, it can be shown that one can find pairs $(L_0,\rho_0)$ that are independent of $N$ and that satisfy \eqref{eq:delta_iterN} for every $N$.}
Thus, by letting $N$ tend to infinity, we conclude that we can assume without loss of optimality that
\begin{equation}\label{eq:alpha_iterN_inf}
	\pow\geq\rho\Bigl(1-\delta^{(\infty)}_{\L,\T}\Bigr)
\end{equation}
where $\delta^{(\infty)}_{\L,\T}$ satisfies
\begin{equation}
0\leq \delta^{(\infty)}_{\L,\T} \leq \frac{\Bigl(\frac{Q^{-1}(\epsilon)\sqrt{\Upsilon(\T)}}{\mathsf{F}(\T)}\Bigr)^2}{\L}.
\end{equation}
This concludes the proof of Part~2) of Lemma~\ref{lm:opt_high_SNR_2}.

\section{Lower Bound on $\bar{\V}_\rho(\T,\pow)$}\label{app:bound_U_alpha}
We show that for all $\rho(1-\delta) \leq \pow \leq \rho$, $0\leq\delta\leq1/2$, and $\rho \geq \rho_0$, we have
\begin{equation}\label{eq:LB_var_alpha}
\bar{\V}_\rho(\T,\pow)\geq \bar{\V}(\T,\rho)- \Upsilon(\T)\delta
\end{equation}
 where $\Upsilon(\T)$ is a positive constant that only depends on $\T$. Let $\Omega(\T,\pow)\triangleq \bar{j}_\ell(\T,\pow)-\bar{\J}(\T,\pow)$, i.e.,
\begin{IEEEeqnarray}{lCl}
	\Omega(\T,\pow)& = &-\frac{\T\rho-\T\pow}{1+\T\rho}(Z_1-1)-\frac{\T\rho}{1+\T\rho}(Z_2-(\T-1))\nonumber\\
	&&{}+(\T-1)\log\bigl((1+\T\pow)Z_1+Z_2+\beta(\T,\rho)\bigr)\IEEEnonumber*\\
	&&{} -(\T-1)\mathsf{E}\bigl[\log\bigl((1+\T\pow)Z_1+Z_2+\beta(\T,\rho)\bigr)\bigr].\nonumber\\\IEEEyesnumber
\end{IEEEeqnarray}
It follows that $\bar{\V}_\rho(\T,\pow)=\mathsf{E}[\Omega^2(\T,\pow)]$. We next analyze the difference
\begin{IEEEeqnarray}{lCl}
\IEEEeqnarraymulticol{3}{l}{
	\bar{\V}(\T,\rho)-\bar{\V}_\rho(\T,\pow)}\nonumber\\\quad
	&=&\mathsf{E}\bigl[\bigl(\Omega(\T,\rho)-\Omega(\T,\pow)\bigr)\bigl(\Omega(\T,\rho)+\Omega(\T,\pow)\bigr)\bigr]\IEEEnonumber\\
	&\leq&\sqrt{\mathsf{E}\bigl[(\Omega(\T,\rho)-\Omega(\T,\pow))^2\bigr]\mathsf{E}\bigl[(\Omega(\T,\rho)+\Omega(\T,\pow))^2\bigr]}\nonumber\\\label{eq:cauchy_schwarz_var}
\end{IEEEeqnarray}
where the inequality follows from the Cauchy-Schwarz inequality. On the one hand, using \eqref{eq:c_terms_2}, we have for every $\rho_0>0$,
\begin{IEEEeqnarray}{lCl}
\IEEEeqnarraymulticol{3}{l}{
	\sup_{\begin{subarray}{l} \pow>0,\\ \rho\geq\rho_0 \end{subarray}}{\mathsf{E}\Bigl[\bigl(\Omega(\T,\rho)+\Omega(\T,\pow)\bigr)^2\Bigr]}}\nonumber\\\quad
	 & \leq & \ceta_{2,2} \sup_{\rho\geq \rho_0}\mathsf{E}\bigl[\Omega^2(\T,\rho)\bigr] + \ceta_{2,2} \sup_{\begin{subarray}{l} \pow\geq 0,\\ \rho\geq\rho_0 \end{subarray}}\mathsf{E}\bigl[\Omega^2(\T,\pow)\bigr] \nonumber\\
	& = & \ceta_{2,2} \sup_{\rho\geq \rho_0}\bar{\V}(\T,\rho)+ \ceta_{2,2} \sup_{\begin{subarray}{l} \pow\geq 0,\\ \rho\geq\rho_0 \end{subarray}} \bar{\V}_\rho(\T,\pow) \label{eq:sum_var_const}
\end{IEEEeqnarray}
which, by Lemma~\ref{lm:var_bounded}, is bounded. On the other hand, using \eqref{eq:c_terms_2} and that $\mathsf{E}[(X-\mathsf{E}[X])^2]\leq\mathsf{E}[X^2]$ for every random variable $X$, we obtain
\begin{IEEEeqnarray}{lCl}
\IEEEeqnarraymulticol{3}{l}{
	\mathsf{E}\Bigl[\bigl(\Omega(\T,\rho)-\Omega(\T,\pow)\bigr)^2\Bigr]}\nonumber\\\quad
	&=&\mathsf{E}\Biggl[\biggl(\frac{\T\rho-\T\pow}{1+\T\rho}(Z_1-1)\nonumber\\
	&&{}+(\T-1)\log\biggl(\frac{(1+\T\rho) Z_1+Z_2+\beta(\T,\rho)}{(1+\T\pow) Z_1+Z_2+\beta(\T,\rho)}\biggr)\IEEEnonumber*\\
	&& {}-(\T-1)\mathsf{E}\biggl[\log\biggl(\frac{(1+\T\rho) Z_1+Z_2+\beta(\T,\rho)}{(1+\T\pow) Z_1+Z_2+\beta(\T,\rho)}\biggr)\biggr]\biggr)^2\Biggr]\\
	&\leq&\ceta_{2,2}(\T-1)^2\mathsf{E}\biggl[\log^2\biggl(\frac{(1+\T\rho) Z_1+Z_2+\beta(\T,\rho)}{(1+\T\pow) Z_1+Z_2+\beta(\T,\rho)}\biggr)\biggr]\nonumber\\
	&&{}+\ceta_{2,2}\biggl(\frac{\T\rho-\T\pow}{1+\T\rho}\biggr)^2.\IEEEyesnumber
	\end{IEEEeqnarray}
When  $\rho(1-\delta) \leq \pow \leq \rho$, this can be further upper-bounded as
\begin{IEEEeqnarray}{lCl}
\IEEEeqnarraymulticol{3}{l}{
	\mathsf{E}\Bigl[\bigl(\Omega(\T,\rho)-\Omega(\T,\pow)\bigr)^2\Bigr]}\nonumber\\\quad
	&\leq&\ceta_{2,2}(\T-1)^2\log^2\biggl(\frac{1}{1-\delta}\biggr)+\ceta_{2,2}\delta^2\nonumber\\
	&\leq& \ceta_{2,2}\bigl(4(\T-1)^2+1\bigr)\delta^2\IEEEyesnumber\label{eq:dif_var_decay}
\end{IEEEeqnarray}
where the last inequality follows from \eqref{eq:watchawaiting}. Combining \eqref{eq:sum_var_const} and \eqref{eq:dif_var_decay} with \eqref{eq:cauchy_schwarz_var} yields \eqref{eq:LB_var_alpha}.

\section{High-SNR Approximations of Information Rates}\label{app:first_asymp}
\begin{lemma}\label{lm:first_asy}
The quantities $\bar{\J}(\T,\rho)$, $\I(\T,\rho)$, and $\underline{\I}(\T,\rho)$ satisfy
	 \begin{IEEEeqnarray}{lCl}\label{eq:first_asymp_gen}
	 \lim_{\rho\to\infty}\bigl\{\bar{\J}(\T,\rho)-\bigl((\T-1)&&\log(\T\rho)-\log\Gamma(\T)\nonumber\\
	 &&{}-(\T-1)(1+\gamma)\bigr)\bigr\} =  0\label{eq:first_asymp_1}\\
	 	\lim_{\rho\to\infty}\bigl\{\I(\T,\rho)-\bigl((\T-1)&&\log(\T\rho)-\log\Gamma(\T)\nonumber\\
	 		 	&&{}-(\T-1)(1+\gamma)\bigr)\bigr\}  =  0\label{eq:first_asymp_2}\\
	 		 	\lim_{\rho\to\infty}\bigl\{\underline{\I}(\T,\rho)-\bigl((\T-1)&&\log(\T\rho)-\log\Gamma(\T)\nonumber\\
	 		 		 	&&{}-(\T-1)(1+\gamma)\bigr)\bigr\}  =  0 \IEEEeqnarraynumspace\label{eq:first_asymp_3}
	\end{IEEEeqnarray}
for every $\T>2$.
\end{lemma}
\begin{IEEEproof}
We can express $\bar{\J}(\T,\rho)$, $\I(\T,\rho)$, and $\underline{\I}(\T,\rho)$ as (see \eqref{eq:J_def}, \eqref{eq:I_def}, and \eqref{eq:sec:first_first_moment_closed})
\begin{IEEEeqnarray}{lCl}
	\bar{\J}(\T,\rho) & = &(\T-1)\log(\T\rho) -\log\Gamma(\T)-(\T-1)\frac{\T\rho}{1+\T\rho}\IEEEnonumber*\\
		& &{}+(\T-1)\mathsf{E}\biggl[\log\biggl(Z_1+\frac{Z_2}{(1+\T\rho)}\biggr)\biggr]\\
		&&{}+(\T-1)\mathsf{E}\biggl[\log\biggl(1+\frac{\beta(\T,\rho)}{(1+\T\rho)Z_1+Z_2}\biggr)\biggr]\\\IEEEyesnumber\label{eq:J_def_alt}\\
	\I(\T,\rho)& = & (\T-1)\log(\T\rho) -\log\Gamma(\T)-(\T-1)\frac{\T\rho}{1+\T\rho}\\
		& & {}+(\T-1)\mathsf{E}\biggl[\log\biggl(Z_1+\frac{Z_2}{(1+\T\rho)}\biggr)\biggr]\\
		&&{}-\mathsf{E}\biggl[\log\tilde{\gamma}\biggl(\T-1,\frac{\T\rho((1+\T\rho)Z_1+Z_2)}{1+\T\rho}\biggr)\biggr]\\\IEEEyesnumber\label{eq:I_def_alt}\\
	\underline{\I}(\T,\rho) & = &(\T-1)\log(\T\rho) -\log\Gamma(\T)-(\T-1)\frac{\T\rho}{1+\T\rho}\\
	&&{}+(\T-1)\mathsf{E}\biggl[\log\biggl(Z_1+\frac{Z_2}{(1+\T\rho)}\biggr)\biggr].\IEEEyesnumber\label{eq:I_low_def_alt}
\end{IEEEeqnarray}
Note that these expressions differ only in terms that vanish as $\rho\to\infty$. Indeed, we have  for every $\T>2$
\begin{IEEEeqnarray}{rCl}
	\lim_{\rho\to\infty}\biggl\{\mathsf{E}\biggl[\log\biggl(Z_1+\frac{Z_2}{(1+\T\rho)}\biggr)\biggr]-\gamma\biggr\} & = & 0\label{eq:first_DCT_1}\\
	\lim_{\rho\to\infty}\mathsf{E}\biggl[\log\biggl(1+\frac{\beta(\T,\rho)}{(1+\T\rho)Z_1+Z_2}\biggr)\biggr] & = & 0\label{eq:first_DCT_2}\\
	\lim_{\rho\to\infty}\mathsf{E}\biggl[\log\tilde{\gamma}\biggl(\T-1,\frac{\T\rho((1+\T\rho)Z_1+Z_2)}{1+\T\rho}\biggr)\biggr] & = & 0.\label{eq:first_DCT_3}\IEEEeqnarraynumspace
\end{IEEEeqnarray}	
We further have that
\begin{equation}
	\lim_{\rho\to\infty} (\T-1)\frac{\T\rho}{1+\T\rho}=(\T-1).
\end{equation}
Hence \eqref{eq:first_asymp_1}--\eqref{eq:first_asymp_3} follow.

It remains to prove \eqref{eq:first_DCT_1}--\eqref{eq:first_DCT_3}. Equation \eqref{eq:first_DCT_1} follows because, by the dominated convergence theorem,
\begin{IEEEeqnarray}{lCl}
	\IEEEeqnarraymulticol{3}{l}{\lim_{\rho\to\infty} \mathsf{E}\biggl[\log\biggl(Z_1+\frac{Z_2}{(1+\T\rho)}\biggr)\biggr]} \nonumber\\
	\quad & = & \mathsf{E}\biggl[\lim_{\rho\to\infty} \log\biggl(Z_1+\frac{Z_2}{(1+\T\rho)}\biggr)\biggr]
\end{IEEEeqnarray}
  and because $\mathsf{E}[\log Z_1]=-\gamma$. The dominated convergence theorem can be applied since (see \eqref{eq:111})
\begin{equation}
\biggl|\log\biggl(Z_1+\frac{Z_2}{1+\T\rho}\biggr)\biggr| \leq \bigl|\log(Z_1+Z_2)\bigr|+ \bigl|\log(Z_1)\bigr|\label{eq:first_asy_1}
\end{equation}
and $\mathsf{E}\bigl[\bigl|\log(Z_1+Z_2)\bigr|+ \bigl|\log(Z_1)\bigr|\bigr]<\infty$.

 Similarly, \eqref{eq:first_DCT_2} and \eqref{eq:first_DCT_3} follow by the dominated convergence theorem and by noting that the terms inside the expected values on the LHS of \eqref{eq:first_DCT_2} and \eqref{eq:first_DCT_3} vanish as $\rho\to\infty$. The dominated convergence theorem can be applied because, for every $\rho_0>0$ and $\rho\geq\rho_0$,
\begin{IEEEeqnarray}{lCl}
\IEEEeqnarraymulticol{3}{l}{
	\biggl|\log\tilde{\gamma}\biggl(\T-1,\frac{\T\rho((1+\T\rho)Z_1+Z_2)}{1+\T\rho}\biggr)\biggr| }\nonumber\\\quad
	&  \leq& (\T-1) \log\biggl(1+\frac{\beta(\T,\rho)}{(1+\T\rho)Z_1+Z_2}\biggr) \nonumber\\
 & \leq& {}(\T-1)\log\biggl(1+\frac{\beta(\T,\rho_0)}{Z_1+Z_2}\biggr) \label{eq:first_asy_2}
\end{IEEEeqnarray}
and because the expected value of the RHS of \eqref{eq:first_asy_2} is finite. Here, the first inequality follows from Lemma \ref{lm:inc_gamma}, and the second inequality follows because $\rho\mapsto\beta(\T,\rho)$ is monotonically decreasing in~$\rho$.
\end{IEEEproof}

\section{High-SNR Approximations of Dispersions}\label{app:second_asymp}
\begin{lemma}\label{lm:var_asy}
The quantities $\bar{\V}(\T,\rho)$ and $\U(\T,\rho)$ defined in \eqref{eq:barV} and \eqref{eq:underU}, respectively, satisfy
	 \begin{IEEEeqnarray}{lCl}
	 \lim_{\rho\to\infty}\bar{\V}(\T,\rho) & = & (\T-1)^2\frac{\pi^2}{6}+(\T-1)\label{eq:var_asymp_def_1}\\
	 \lim_{\rho\to\infty} \U(\T,\rho) & = & (\T-1)^2\frac{\pi^2}{6}+(\T-1)\label{eq:var_asymp_def_2} 
	\end{IEEEeqnarray}
for every $\T>2$.
\end{lemma}
\begin{IEEEproof}
We first prove \eqref{eq:var_asymp_def_1} by analyzing $\bar{\V}(\T,\rho)$ in the limit as $\rho\to\infty$. To this end, we first note that
\begin{IEEEeqnarray}{lCl}
\IEEEeqnarraymulticol{3}{l}{
\bar{j}_{\ell}(\T,\rho)-\bar{\J}(\T,\rho)}\nonumber\\\quad
 & = & \frac{\T\rho}{1\,+\T\rho}(\T-1-Z_2) + (\T-1) \log\biggl(Z_1+\frac{Z_2}{1+\T\rho}\biggr) \nonumber\\
&&{}- (\T-1) \mathsf{E}\biggl[\log\biggl(Z_1+\frac{Z_2}{1+\T\rho}\biggr)\biggr]\nonumber\\
& &  {} + (\T-1) \log\biggl(1+\frac{\beta(\T,\rho)}{(1+\T\rho)Z_1+Z_2}\biggr) \nonumber\\
&&{}- (\T-1) \mathsf{E}\biggl[\log\biggl(1+\frac{\beta(\T,\rho)}{(1+\T\rho)Z_1+Z_2}\biggr)\biggr]\quad
\end{IEEEeqnarray}
tends to
\begin{equation}
\label{eq:Tobi_appIII_lim}
\T-1-Z_2 + (\T-1) \log(Z_1)  - (\T-1)\mathsf{E}[\log Z_1]
\end{equation}
as $\rho\to\infty$. (To obtain $\mathsf{E}[\log Z_1]$, we interchange limit and expectation, which can be justified by the dominated convergence theorem.) Since $Z_1$ and $Z_2$ are independent, we have that
\begin{IEEEeqnarray}{lCl}
\IEEEeqnarraymulticol{3}{l}{\mathsf{E}\Bigl[\bigl(\T-1-Z_2 + (\T-1) \log(Z_1)  - (\T-1)\mathsf{E}[\log Z_1]\bigr)^2\Bigr]}\nonumber\\
\quad &= &  \mathsf{E}\bigl[(\T-1-Z_2)^2\bigr] \nonumber\\
& & {} + (\T-1)^2 \Bigl(\mathsf{E}\bigl[\log^2(Z_1)\bigr] - \mathsf{E}\bigl[\log Z_1 \bigr]^2\Bigl) \nonumber\\
& = & (\T-1) + (\T-1)^2 \frac{\pi^2}{6}.
\end{IEEEeqnarray}
It remains to show that we can swap limit and expectation. To this end, we next argue that the dominated convergence theorem applies. Indeed, proceeding similarly as in Appendix~\ref{app:second_moment}, we note that for every $\rho_0>0$ and $\rho\geq\rho_0$
\begin{IEEEeqnarray}{lCl}
\IEEEeqnarraymulticol{3}{l}{
\bigl(\bar{j}_{\ell}(\T,\rho)-\bar{\J}(\T,\rho)\bigr)^2}\nonumber\\\quad
 & \leq & \ceta_{5,2} \biggl(\frac{\T\rho}{1\,+\T\rho}\biggr)^2(Z_2-\T+1)^2 \nonumber\\
& & {} +  \ceta_{5,2} (\T-1)^2 \log^2\biggl(Z_1+\frac{Z_2}{1+\T\rho}\biggr)\nonumber\\
&&{} +  \ceta_{5,2}(\T-1)^2 \mathsf{E}\biggl[\log\biggl(Z_1+\frac{Z_2}{1+\T\rho}\biggr)\biggr]^2\nonumber\\
& & {} + \ceta_{5,2} (\T-1)^2 \log^2\biggl(1+\frac{\beta(\T,\rho)}{(1+\T\rho)Z_1+Z_2}\biggr) \nonumber\\
&&{} +  \ceta_{5,2}(\T-1)^2 \mathsf{E}\biggl[\log\biggl(1+\frac{\beta(\T,\rho)}{(1+\T\rho)Z_1+Z_2}\biggr)\biggr]^2\nonumber\\ 
& \leq & \ceta_{5,2}(Z_2-\T+1)^2 +(\T-1)^2 \log^2(Z_1) \nonumber\\
& & {} +  \ceta_{5,2}(\T-1)^2 \log^2(Z_1+Z_2) \nonumber\\  
&&{}+  \ceta_{5,2}(\T-1)^2 \mathsf{E}\bigl[\bigl|\log(Z_1+Z_2)\bigr|+ \bigl|\log(Z_1)\bigr|\bigr]^2 \nonumber\\
&& {} +  \ceta_{5,2}(\T-1)^2\log^2\biggl(1+\frac{\beta(\T,\rho_0)}{Z_1+Z_2}\biggr) \nonumber\\
& & {} +   \ceta_{5,2}(\T-1)^2 \mathsf{E}\biggl[\log\biggl(1+\frac{\beta(\T,\rho_0)}{Z_1+Z_2}\biggr)\biggr]^2.
\label{eq:var_bound_asy_V_bar_rho}
\end{IEEEeqnarray}
To obtain the second inequality in \eqref{eq:var_bound_asy_V_bar_rho}, we upper-bound the second term using that (see~\eqref{eq:log_square_bounded_away})
\begin{equation*}
\log^2\biggl(Z_1+\frac{Z_2}{1+\T\rho}\biggr) \leq\log^2(Z_1+Z_2)+\log^2(Z_1),\label{eq:var_asy_bar_second}
\end{equation*}
the third term using that (see \eqref{eq:111})
\begin{equation*}
\biggl|\log\biggl(Z_1+\frac{Z_2}{1+\T\rho}\biggr)\biggr| \leq \bigl|\log(Z_1+Z_2)\bigr|+ \bigl|\log(Z_1)\bigr|,\label{eq:var_asy_bar_3}
\end{equation*}
the fourth term using that, for every $\rho_0>0$ and $\rho\geq\rho_0$,
\begin{IEEEeqnarray*}{lCl}
	\log^2\biggl(1+\frac{\beta(\T,\rho)}{(1+\T\rho)Z_1+Z_2}\biggr) &\leq &\log^2\biggl(1+\frac{\beta(\T,\rho)}{Z_1+Z_2}\biggr) \nonumber\\
	& \leq & \log^2\biggl(1+\frac{\beta(\T,\rho_0)}{Z_1+Z_2}\biggr),
\end{IEEEeqnarray*}
and the fifth term using that, for every $\rho_0>0$ and $\rho\geq\rho_0$,
\begin{equation*}
\log\biggl(1+\frac{\beta(\T,\rho)}{(1+\T\rho)Z_1+Z_2}\biggr) \leq \log\biggl(1+\frac{\beta(\T,\rho_0)}{Z_1+Z_2}\biggr).
\end{equation*}
Since the expected value of the RHS of \eqref{eq:var_bound_asy_V_bar_rho} is finite, the dominated convergence theorem applies and \eqref{eq:var_asymp_def_1} follows.

To prove \eqref{eq:var_asymp_def_2} we proceed similarly. Indeed, by Lemma~\ref{lm:inc_gamma},
\begin{IEEEeqnarray}{lCl}
\IEEEeqnarraymulticol{3}{l}{
i_{\ell}(\T,\rho)-\I(\T,\rho)}\nonumber\\\quad
& = & \frac{\T\rho}{1\,+\T\rho}(\T-1-Z_2) + (\T-1) \log\biggl(Z_1+\frac{Z_2}{1+\T\rho}\biggr) \nonumber\\
&&{} - (\T-1) \mathsf{E}\biggl[\log\biggl(Z_1+\frac{Z_2}{1+\T\rho}\biggr)\biggr] \nonumber\\
 && {} - \log\tilde{\gamma}\biggl(\T-1,\frac{\T\rho((1+\T\rho)Z_1+Z_2)}{1+\T\rho}\biggr) \nonumber\\
 &&{} + \mathsf{E}\biggl[\log\tilde{\gamma}\biggl(\T-1,\frac{\T\rho((1+\T\rho)Z_1+Z_2)}{1+\T\rho}\biggr)\biggr]
\end{IEEEeqnarray}
tends to \eqref{eq:Tobi_appIII_lim} as $\rho$ tends to infinity. It remains to show that limit and expectation can be swapped. We next argue that this follows from dominated convergence theorem. Indeed, using \eqref{eq:c_terms_2}, we obtain for every $\rho_0>0$ and $\rho\geq\rho_0$ that
\begin{IEEEeqnarray}{lCl}
\IEEEeqnarraymulticol{3}{l}{
\bigl(i_{\ell}(\T,\rho)-\I(\T,\rho)\bigr)^2}\nonumber\\\quad
 & \leq & \ceta_{5,2} \biggl(\frac{\T\rho}{1\,+\T\rho}\biggr)^2(Z_2-\T+1)^2 \nonumber\\
 &&{} +  \ceta_{5,2}(\T-1)^2 \log^2\biggl(Z_1+\frac{Z_2}{1+\T\rho}\biggr) \nonumber\\
& & {} + \ceta_{5,2}(\T-1)^2 \mathsf{E}\biggl[\log\biggl(Z_1+\frac{Z_2}{1+\T\rho}\biggr)\biggr]^2 \nonumber\\
&&{} + \ceta_{5,2}\log^2\tilde{\gamma}\biggl(\T-1,\frac{\T\rho((1+\T\rho)Z_1+Z_2)}{1+\T\rho}\biggr)\nonumber\\
 & & {} + \ceta_{5,2}\mathsf{E}\biggl[\log\tilde{\gamma}\biggl(\T-1,\frac{\T\rho((1+\T\rho)Z_1+Z_2)}{1+\T\rho}\biggr)\biggr]^2\nonumber\\
 & \leq & \ceta_{5,2}(Z_2-\T+1)^2 \nonumber\\
 &&{}+ \ceta_{5,2}(\T-1)^2 \log^2(Z_1+Z_2)+(\T-1)^2 \log^2(Z_1) \nonumber\\
 & & {} + \ceta_{5,2}(\T-1)^2 \mathsf{E}\bigl[\bigl|\log(Z_1+Z_2)\bigr|+ \bigl|\log(Z_1)\bigr|\bigr]^2\nonumber\\
 &&{} + \ceta_{5,2}(\T-1)^2 \log^2\biggl(1+\frac{\beta(\T,\rho_0)}{Z_1+Z_2}\biggr) \nonumber\\
 & & {}  + \ceta_{5,2}(\T-1)^2 \mathsf{E}\biggl[\log\biggl(1+\frac{\beta(\T,\rho_0)}{Z_1+Z_2}\biggr)\biggr]^2. \label{eq:var_bound_asy_U_rho}
\end{IEEEeqnarray}
Here, we upper-bound the first three terms as in \eqref{eq:var_bound_asy_V_bar_rho}, and the fourth and fifth term using Lemma \ref{lm:inc_gamma} and the monotonicity of $\rho\mapsto \beta(\T,\rho)$.
Since the expected value of the RHS of \eqref{eq:var_bound_asy_U_rho} is finite, the dominated convergence theorem applies and \eqref{eq:var_asymp_def_2} follows.
\end{IEEEproof}

\section*{Acknowledgment}
The authors gratefully acknowledge fruitful discussions with Gonzalo Vazquez-Vilar. They further thank  Mustafa C. Co\c{s}kun for producing the performance curve of the ARJA LDPC code shown in Fig.~\ref{fig:R4_T20_L25}.

\bibliographystyle{IEEEtran}

\begin{thebibliography}{10}
\providecommand{\url}[1]{#1}
\csname url@samestyle\endcsname
\providecommand{\newblock}{\relax}
\providecommand{\bibinfo}[2]{#2}
\providecommand{\BIBentrySTDinterwordspacing}{\spaceskip=0pt\relax}
\providecommand{\BIBentryALTinterwordstretchfactor}{4}
\providecommand{\BIBentryALTinterwordspacing}{\spaceskip=\fontdimen2\font plus
\BIBentryALTinterwordstretchfactor\fontdimen3\font minus
  \fontdimen4\font\relax}
\providecommand{\BIBforeignlanguage}[2]{{%
\expandafter\ifx\csname l@#1\endcsname\relax
\typeout{** WARNING: IEEEtran.bst: No hyphenation pattern has been}%
\typeout{** loaded for the language `#1'. Using the pattern for}%
\typeout{** the default language instead.}%
\else
\language=\csname l@#1\endcsname
\fi
#2}}
\providecommand{\BIBdecl}{\relax}
\BIBdecl

\bibitem{Durisi_Koch_Popovski_2015}
G.~Durisi, T.~Koch, and P.~Popovski, ``Towards massive, ultra-reliable, and
  low-latency wireless communication with short packets,'' \emph{Proc. IEEE},
  vol. 104, no.~9, pp. 1711--1726, Sep. 2016.

\bibitem{Hayashi_2009}
M.~Hayashi, ``Information spectrum approach to second-order coding rate in
  channel coding,'' \emph{IEEE Trans. Inf. Theory}, vol.~55, no.~11, pp.
  4947--4966, Nov. 2009.

\bibitem{Polyanskiy_Poor_Verdu}
Y.~Polyanskiy, H.~V. Poor, and S.~Verd\'u, ``Channel coding rate in the finite
  blocklength regime,'' \emph{IEEE Trans. Inf. Theory}, vol.~56, no.~5, pp.
  2307--2359, May 2010.

\bibitem{Polyanskiy_2011}
Y.~Polyanskiy and S.~Verd\'u, ``Scalar coherent fading channel: Dispersion
  analysis,'' in \emph{Proc. IEEE Int. Symp. Inf. Theory (ISIT)}, St.
  Petersburg, Russia, Jul. 2011, pp. 2959--2963.

\bibitem{Collins_Polyanskiy_2014}
A.~Collins and Y.~Polyanskiy, ``Orthogonal designs optimize achievable
  dispersion for coherent {MISO} channels,'' in \emph{Proc. IEEE Int. Symp.
  Inf. Theory (ISIT)}, Honolulu, HI, USA, Jun. 2014, pp. 2524--2528.

\bibitem{Collins_Polyanskiy_MIMO_coh_BFC}
------, ``Dispersion of the coherent {MIMO} block-fading channel,'' in
  \emph{Proc. IEEE Int. Symp. Inf. Theory (ISIT)}, Barcelona, Spain, Jul. 2016,
  pp. 1068--1072.

\bibitem{Collins_Polyanskiy_2018}
------, ``Coherent multiple-antenna block-fading channels at finite
  blocklength,'' \emph{IEEE Trans. Inf. Theory}, vol.~65, no.~1, pp. 380--405,
  Jan. 2019.

\bibitem{Yang_LongTerm}
W.~Yang, G.~Caire, G.~Durisi, and Y.~Polyanskiy, ``Optimum power control at
  finite blocklength,'' \emph{IEEE Trans. Inf. Theory}, vol.~61, no.~9, pp.
  4598--4615, Sep. 2015.

\bibitem{Yang_2014}
W.~Yang, G.~Durisi, T.~Koch, and Y.~Polyanskiy, ``Quasi-static multiple-antenna
  fading channels at finite blocklength,'' \emph{IEEE Trans. Inf. Theory},
  vol.~60, no.~7, pp. 4232--4265, Jul. 2014.

\bibitem{MolavianJazi_2013}
E.~MolavianJazi and J.~N. Laneman, ``On the second-order coding rate of
  non-ergodic fading channels,'' in \emph{Proc. 51st Allerton Conf. Commun. Control
  Comput.}, Monticello, IL, USA, Oct. 2013, pp. 583--587.

\bibitem{Hoydis_2015}
J.~Hoydis, R.~Couillet, and P.~Piantanida, ``The second-order coding rate of
  the {MIMO} quasi-static {R}ayleigh fading channel,'' \emph{IEEE Trans. Inf.
  Theory}, vol.~61, no.~12, pp. 6591--6622, Dec. 2015.

\bibitem{finte_block_length_2012}
W.~Yang, G.~Durisi, T.~Koch, and Y.~Polyanskiy, ``Diversity versus channel
  knowledge at finite block-length,'' in \emph{Proc. IEEE Inf. Theory Workshop
  (ITW)}, Lausanne, Switzerland, Sep. 2012, pp. 572--576.

\bibitem{Ostman_2014}
J.~\"Ostman, W.~Yang, G.~Durisi, and T.~Koch, ``Diversity versus multiplexing
  at finite blocklength,'' in \emph{Proc. IEEE Int. Symp. on Wireless Commun.
  Syst. (ISWCS)}, Barcelona, Spain, Aug. 2014, pp. 702--706.

\bibitem{Short_Packets_Durisi_Koch_TCOM_2015}
G.~Durisi, T.~Koch, J.~\"Ostman, Y.~Polyanskiy, and W.~Yang, ``Short-packet
  communications over multiple-antenna {R}ayleigh-fading channels,'' \emph{IEEE
  Trans. Commun.}, vol.~64, no.~2, pp. 618--629, Feb. 2016.

\bibitem{Hochwald_2000}
B.~Hochwald and T.~Marzetta, ``Unitary space-time modulation for
  multiple-antenna communications in {R}ayleigh flat fading,'' \emph{IEEE
  Trans. Inf. Theory}, vol.~46, no.~2, pp. 543--564, Mar. 2000.

\bibitem{Zheng_2002}
L.~Zheng and D.~Tse, ``Communication on the {G}rassmann manifold: {A} geometric
  approach to the noncoherent multiple-antenna channel,'' \emph{IEEE Trans.
  Inf. Theory}, vol.~48, no.~2, pp. 359--383, Feb. 2002.

\bibitem{Yang_2013}
W.~Yang, G.~Durisi, and E.~Riegler, ``On the capacity of large-{MIMO}
  block-fading channels,'' \emph{IEEE J. Sel. Areas Commun.}, vol.~31, no.~2,
  pp. 117--132, Feb. 2013.

\bibitem{Abramowitz}
M.~Abramowitz, \emph{Handbook of Mathematical Functions, With Formulas, Graphs,
  and Mathematical Tables}.\hskip 1em plus 0.5em minus 0.4em\relax New York,
  NY, USA: Dover Publications, 1974.

\bibitem{table_integrals}
I.~S. Gradshteyn and I.~M. Ryzhik, \emph{Table of Integrals, Series, and
  Products}, 7th~ed.\hskip 1em plus 0.5em minus 0.4em\relax Amsterdam, The Netherlands: Elsevier/Academic
  Press, 2007.

\bibitem{table_Ei}
M.~Geller and E.~W.~Ng, ``A table of integrals of the exponential integral,''
  \emph{Journal of Research of the National Bureau of Standards - B,
  Mathematics and Mathematical Science}, vol. 738, no.~3, pp. 191--210,
  Jul.-Sep. 1969.

\bibitem{Biglieri}
E.~Biglieri, J.~Proakis, and S.~Shamai, ``Fading channels:
  Information-theoretic and communications aspects,'' \emph{IEEE Trans. Inf.
  Theory}, vol.~44, no.~6, pp. 2619--2692, Oct. 1998.

\bibitem{Lapidoth_Moser_duality}
A.~Lapidoth and S.~M. Moser, ``Capacity bounds via duality with applications to
  multiple-antenna systems on flat-fading channels,'' \emph{IEEE Trans. Inf.
  Theory}, vol.~49, no.~10, pp. 2426--2467, Oct. 2003.

\bibitem{Marzetta_99}
T.~Marzetta and B.~Hochwald, ``Capacity of a mobile multiple-antenna
  communication link in {R}ayleigh flat fading,'' \emph{IEEE Trans. Inf.
  Theory}, vol.~45, no.~1, pp. 139--157, Jan. 1999.

\bibitem{Ericson70}
T.~Ericson, ``A {Gaussian} channel with slow fading,'' \emph{IEEE Trans. Inf.
  Theory}, vol.~16, no.~3, pp. 353--355, May 1970.

\bibitem{Ostman_2018}
J.~\"Ostman, G.~Durisi, E.~G. Str\"om, M.~C. Co\c{s}kun, and G.~Liva, ``Short
  packets over block-memoryless fading channels: Pilot-assisted or noncoherent
  transmission?'' \emph{IEEE Trans. Commun.}, vol.~67, no.~2, pp. 1521--1536, Feb. 2019.

\bibitem{nagaoka2001}
H.~Nagaoka, ``Strong converse theorems in quantum information theory,'' in
  \emph{Proc. ERATO Workshop on Quantum Information Science}, Tokyo, Japan, Sep.
  2001, p. 6--8.

\bibitem{verdu_han}
S.~Verdu and T.~S. Han, ``A general formula for channel capacity,'' \emph{IEEE
  Trans. Inf. Theory}, vol.~40, no.~4, pp. 1147--1157, Jul. 1994.

\bibitem{hayashi-nagaoka}
M.~Hayashi and H.~Nagaoka, ``General formulas for capacity of classical-quantum
  channels,'' \emph{IEEE Trans. Inf. Theory}, vol.~49, no.~7, pp. 1753--1768,
  Jul. 2003.

\bibitem{Ostman_HARQ_2018}
J.~\"Ostman, R.~Devassy, G.~C. Ferrante, and G.~Durisi, ``Low-latency
  short-packet transmissions: Fixed length or {HARQ}?'' in \emph{Proc. IEEE Global
  Telecommun. Conf. (GLOBECOM)}, Abu Dhabi, United Arab Emirates, Dec. 2018,
  pp. 1--6.

\bibitem{Mustafa2019}
M.~C. Coşkun, G.~Durisi, T.~Jerkovits, G.~Liva, W.~Ryan, B.~Stein, and
  F.~Steiner, ``Efficient error-correcting codes in the short blocklength
  regime,'' \emph{Physical Communication}, vol.~34, pp. 66 -- 79, Jun. 2019.

\bibitem{Coskun_2019}
M.~C. Co\c{s}kun, G.~Liva, J.~\"Ostman, and G.~Durisi, ``Low-complexity joint
  channel estimation and list decoding of short codes,'' in \emph{Proc. 12th Int.
  ITG Conf. Sys. Commun. Coding (SCC)}, Rostock, Germany, Feb.~2019.

\bibitem{Tan_Tomamichel_2015}
V.~Y.~F. Tan and M.~Tomamichel, ``The third-order term in the normal
  approximation for the {AWGN} channel,'' \emph{IEEE Trans. Inf. Theory},
  vol.~61, no.~5, pp. 2430--2438, May 2015.

\bibitem{Kurose:2012:CNT:2584507}
J.~F. Kurose and K.~W. Ross, \emph{Computer Networking: A Top-Down Approach},
  6th~ed.\hskip 1em plus 0.5em minus 0.4em\relax New Jersey, USA: Pearson,
  2012.

\bibitem{Feller}
W.~Feller, \emph{An Introduction To Probability Theory And Its Applications},
  2nd~ed.\hskip 1em plus 0.5em minus 0.4em\relax New York, NY, USA: Wiley,
  1971, vol.~II.

\bibitem{Alzer}
H.~Alzer, ``On some inequalities for the incomplete gamma function,''
  \emph{Math Comp.}, vol.~66, no. 218, pp. 771--778, Apr. 1997.

\bibitem{Gautschi}
W.~Gautschi, ``The incomplete gamma functions since {T}ricomi,'' in \emph{Atti
  dei Convegni Lincei, 147}, Accademia Nazionale dei Lincei, Rome, 1998, pp.
  203--237.

\bibitem{Rudin}
W.~Rudin, \emph{Real and Complex Analysis}, 3rd~ed.\hskip 1em plus 0.5em minus
  0.4em\relax New York, NY, USA: McGraw-Hill, 1987.

\bibitem{rudin-principles}
------, \emph{Principles of Mathematical Analysis}, 3rd~ed.\hskip 1em plus
  0.5em minus 0.4em\relax New York, NY, USA: McGraw-Hill, 1976.

\bibitem{Durrett_Prob_theory}
R.~Durrett, \emph{Probability: Theory and Examples}, 3rd~ed.\hskip 1em plus
  0.5em minus 0.4em\relax Belmont, USA: Duxbury Advances Series, 2005.

\end{thebibliography}

\begin{IEEEbiographynophoto}{Alejandro Lancho} (S'14--M'19) received the B.E., M.Sc.\ and Ph.D.\ degrees in Electrical Engineering from the Universidad Carlos III de Madrid, Spain, in 2013, 2014 and 2019, respectively. During the fall of 2017, he was a visiting researcher at Chalmers University of Technology. Since October, 2019, he works at Chalmers University of Technology as a postdoctoral researcher. 

Alejandro Lancho was an FPU fellow (Spanish Ministerio de Educaci\'on Cultura y Deporte) during his Ph.D.\ studies. He was among the six finalists for the IEEE Jack Keil Wolf ISIT Student Paper Award at the 2017 IEEE International Symposium on Information Theory, Aachen, Germany, 2017.  His research interests are in the areas of information theory and wireless communications.
\end{IEEEbiographynophoto}

\vfill

\newpage

\begin{IEEEbiographynophoto}{Tobias Koch} (S'02--M'09--SM'16) is a Visiting Professor and Ram\'on y Cajal Research Fellow with the Signal Theory and Communications Department of Universidad Carlos III de Madrid (UC3M). He received the M.Sc.\ degree in electrical engineering (with distinction) in 2004 and the Ph.D.\ degree in electrical engineering in 2009, both from ETH Zurich, Switzerland. From June 2010 until May 2012 he was a Marie Curie Intra-European Research Fellow with the University of Cambridge, UK. He was also a research intern at Bell Labs, Murray Hill, NJ, USA in 2004, and the Universitat Pompeu Fabra (UPF), Barcelona, Spain, in 2007. He joined the Signal Processing Group of UC3M in June 2012. His research interests are in digital communication theory and information theory.

Dr.\ Koch received a Starting Grant from the European Research Council (ERC), a Ram\'on y Cajal Research Fellowship, a Marie Curie Intra-European Fellowship, a Marie Curie Career Integration Grant, and a Fellowship for Prospective Researchers from the Swiss National Science Foundation. He further received a medal of the 2018 Young Researchers Award ``Agust\'\i n de Betancourt y Molina" by the Real Academia de Ingenier\'\i a. In 2013--2016 he served as Vice Chair of the Spain Chapter of the IEEE Information Theory Society.
\end{IEEEbiographynophoto}

\begin{IEEEbiographynophoto}{Giuseppe Durisi} (S'02-M'06-SM'12) received the Laurea degree summa cum laude and the Doctor degree both from Politecnico di Torino, Italy, in 2001 and 2006, respectively. From 2002 to 2006, he was with Istituto Superiore Mario Boella, Torino, Italy. From 2006 to 2010 he was a postdoctoral researcher at ETH Zurich, Zurich, Switzerland. In 2010, he joined Chalmers University of Technology, Gothenburg, Sweden, where he is now professor with the Communication Systems Group. He is also co-director of Chalmers ICT Area of Advance, and of Chalmers AI Research Center.

Dr. Durisi is a senior member of the IEEE. He is the recipient of the 2013 IEEE ComSoc Best Young Researcher Award for the Europe, Middle East, and Africa Region, and is co-author of a paper that won a “student paper award” at the 2012 International Symposium on Information Theory, and of a paper that won the 2013 IEEE Sweden VT-COM-IT joint chapter best student conference paper award. In 2015, he joined the editorial board of the IEEE Transactions on Communications as associate editor. From 2011 to 2014, he served as publications editor for the IEEE Transactions on Information Theory. His research interests are in the areas of communication and information theory and machine learning.
\end{IEEEbiographynophoto}

\vfill

\end{document}